\documentclass{elsart}
 \usepackage{graphicx,amssymb,amsfonts,amsmath}

 \newcommand{\mlg}{{}^\lessgtr}                %less-greater
 \newcommand{\mgl}{{}^\gtrless}                %greater-less
 \newcommand{\myhat}[1]{\,\,\hat{\!\!#1}}
 \newcommand{\mynot}[1]{\!\not{\!#1}}
 \newcommand{\mmynot}[1]{\!\not{\!\!#1}}
 \newcommand{\mPi}{{\mathnormal\Pi}}
 \newcommand{\und}[1]{\underline{#1}}

 \newcommand{\beq}{\begin{equation}}
 \newcommand{\eeq}{\end{equation}}

 \renewcommand{\theequation}{\arabic{section}.\arabic{equation}}

\journal{Ann. Phys.}

\begin{document}

 \begin{frontmatter}
 \title{
  Kinetic Theory of Radiation in Nonequilibrium Relativistic Plasmas}
 \author[Moscow]{V.G. Morozov}
 \ead{vmorozov@orc.ru}\ and\
 \author[Rostock]{G. R\"opke\corauthref{cor}}
 \corauth[cor]{Corresponding author}
 \ead{gerd.roepke@uni-rostock.de}

  \address[Moscow]{Moscow State Institute of
   Radioengineering, Electronics, and Automation (Technical
   University),
 Vernadsky Prospect, 78, 119454\ Moscow, Russia }
 \address[Rostock]{University of Rostock, FB Physik,
   Universit\"atsplatz 3, D-18051\ Rostock Germany}

\begin{abstract}
 Many-particle QED is applied to kinetic
 theory of radiative  processes in many-component plasmas with
 relativistic electrons and nonrelativistic heavy particles.
  Within the framework of nonequilibrium Green's function technique,
  transport and mass-shell equations for fluctuations of the
  electromagnetic field are obtained. We show that the transverse field
  correlation functions can be decomposed
  into sharply peaked (non-Lorentzian) parts that describe
  resonant (propagating) photons
  and off-shell parts corresponding to virtual photons in plasmas. Analogous
  decompositions are found for the longitudinal field correlation functions
 and the correlation functions of relativistic electrons. As a novel result
  a kinetic equation for the resonant photons with a finite spectral width
  is derived.  The off-shell parts
  of the particle and field correlation functions are shown to be essential
  to calculate the local radiating power in relativistic
  plasmas and recover the results of vacuum QED.
  The influence of plasma effects and collisional
  broadening of the relativistic quasiparticle
  spectral function on radiative processes is discussed.
\end{abstract}
\end{frontmatter}

\section{Introduction}
Current developments in ultra-short high-intensity laser technology
 and recent progress in laser-plasma experiments have regenerated
interest in theory of relativistic plasmas. At focused intensities
of 10$^{19}$ - 10$^{20}$ W/cm$^2$, the electrons in the
laser channels are accelerated up
to multi-MeV energies \cite{Gordon98,Clayton98,Santala01,Mangles05}.
 It is expected that in near future the focal laser intensities will
  reach 10$^{21}$ -  10$^{22}$ W/cm$^2$, so that electron energies are
  predicted to exceed 1 GeV.

In the presence of dense laser-generated particle beams, the plasma is to
be considered as a many-component nonequilibrium medium composed of highly
relativistic beam electrons, relativistic (or nonrelativistic) electrons
in the plasma return current, and nonrelativistic heavy
particles\footnote{In laser-plasma experiments  protons and ions
 with energies in the MeV region can be
 produced~\cite{Clark00,Clark00a,Snavely00}, but their velocities
are much less than the speed of light.}.
Up to now theoretical investigations of such systems were concerned mainly
with semi-macroscopic mechanisms for acceleration of self-injected electrons in a
plasma channel~\cite{Rosenz91,Pukhov99,Tsung04}, collective stopping, ion
heating~\cite{Honda00}, and collective beam-plasma
instabilities~\cite{Bret04,Mendonca05,Tautz07}. Note, however, that
 a consistent
theory of many other transport phenomena should be based on
many-particle nonequilibrium QED which also
 takes account of characteristic plasma properties.
 In particular, a problem of great interest is photon kinetics,
  because radiative processes contribute to
 stopping power for highly relativistic electron beams, and they
 are interesting  in themselves  as an example of
 fundamental
 QED processes in a medium~\cite{Klein99}.
 It should also be noted that the measurement of angular
 distribution of $\gamma$ rays  in laser-plasma
experiments has been found to be a powerful diagnostic tool~\cite{Santala00}.

 It is important to
recognize that  the standard ``golden rule'' approach of relativistic
kinetic theory~\cite{DeGroot80} cannot be applied to the
 laser-plasma medium, since the very concepts of
 ``collisions'' and ``asymptotic free states'' are not straightforward due to
 collective plasma effects.
 There is also another non-trivial
 problem related to the kinetic description of radiation in
 a nonequilibrium plasma. The point is that transverse
   wave excitations may be associated with two quite different states
    of photon modes. First, \textit{virtual photons\/} are responsible for
    the interaction
between particles and cannot be detected as real photons outside the
system. Second, for sufficiently high frequencies, propagation of weakly
damped \textit{resonant photons\/} is possible. These photons behave as
well-defined quasiparticles and contribute to the plasma radiation.
Clearly, the above physical arguments must be
 supplemented by a consistent mathematical prescription for
 separating the contributions of virtual and resonant photons
 in transport equations.
 The main difficulty is that, in general, characteristics of both the virtual
 and resonant (propagating) photons involve medium corrections.
 A systematic treatment of the photon degrees of freedom in nonequilibrium
plasmas is still a challenging problem.

At present, the most highly developed methods for treating nonequilibrium
processes in many-particle systems from first principles are the density
matrix method~\cite{ZMR96,ZMR97,Bonitz98} and the real-time Green's
function method~\cite{KadBaym62,Danielew84,BotMalf90}. Of the two methods,
the former is especially suited for systems in which many-particle
correlations are important. On the other hand, the Green's function method
turns out to be efficient in the cases where the description of transport
 processes does not go far beyond the quasiparticle picture. Since in the
laser-plasma interaction experiments the plasma may be considered to be
weakly coupled and the dynamics of many-particle correlations (bound
states etc.)
 is of minor importance for the radiative
 phenomena considered here, the Green's function formalism
 seems to be the most direct way to
  transport equations for particles and radiation including
   relativistic and plasma effects.

 A systematic
 approach to nonequilibrium QED plasmas based on the Green's function
 technique was developed by Bezzerides and DuBois~\cite{BezDub72}.
  Within the weak coupling approximation,
 they were able to derive a covariant particle kinetic equation
 involving electron-electron  collisions and
 Cherenkov emission and absorption of plasmons.
 Note, however, that the kinetics of \textit{transverse\/} photons and
 radiative phenomena are mainly related to higher-order processes,
 like bremsstrahlung or Compton scattering. Although these
  fundamental processes have been studied in great
 detail within the framework of vacuum QED,
 they have not been studied to sufficient generality and clarity
 in many-particle nonequilibrium QED when plasma effects are
 of importance. Therefore, the purpose of the analysis
 in this paper is to perform the still missing formulation
 of photon kinetics which includes relativistic and
 plasma effects, and,
 where appropriate, recovers the conventional
  results of vacuum QED for radiative processes.

 The paper is outlined as follows. In Section~\ref{Sec:GF}
 the essential features of
 the real-time Green's function formalism for many-component relativistic
 plasmas are summarized. For the problems under consideration, there is a
 preferred frame of reference, the rest frame of the system, in which the
 heavy particles are nonrelativistic. It is thus convenient to choose the
 Coulomb gauge, allowing a description of electromagnetic
 fluctuations in terms of the longitudinal and transverse (photon) modes.
 The polarization matrix and the particle self-energies on the time-loop
 Schwinger-Keldysh contour are
  expressed in terms of Green's functions and
 the vertex functions which are found
  in the leading approximation for weakly coupled relativistic plasmas.

 In Section~\ref{Sec:KinEq} the kinetic description of the transverse
  wave fluctuations in plasmas is discussed.
  Dyson's equation for the transverse field Green's function is used to
  derive the Kadanoff-Baym equations and then the
   transport and mass-shell equations for the Wigner transformed
  correlation functions in the limit of slow macroscopic space-time
  variations. The analysis of the drift term in the transport equation
   shows that the transverse field correlation functions are
 naturally decomposed into  ``resonant'' and ``off-shell'' parts which,
  in the case of small damping, may be regarded as contributions from
  the propagating and virtual photons, respectively.
 The resonant parts are sharply peaked near the effective photon
  frequencies and have smaller wings than a Lorentzian.
  This observation allows the derivation of a kinetic equation for the
  local energy-momentum photon distribution
  which is related to the resonant parts of the transverse field correlation
  functions. An essential point is that the transport and mass-shell
  equations lead to the same photon kinetic equation. This fact
    attests to the self-consistency of the approach.
 In the zero width limit for the resonant spectral function, the
  kinetic equation derived in the present work reduces to the well-known
  kinetic equation for the ``on-shell'' photon distribution
   function~\cite{Dub67}.

Section~\ref{Sec:Electrons} is concerned with the relativistic electron
propagators and correlation functions which are required for calculating
the photon emission and absorption rates. As in the case of transverse
photons,
 the structure of the relativistic transport equation leads to the
 decomposition of the electron correlation functions into sharply peaked
  ``quasiparticle'' parts and ``off-shell'' parts.
 To leading order in the field fluctuations, the full electron correlation
 functions are expressed in terms of their quasiparticle parts. This
 representation is analogous to the so-called ``extended quasiparticle
 approximation'' that was previously discussed by several
 workers~\cite{KoelMalf93,SpLip94,SpLip95,BornKremp96}
 in a context of nonrelativistic kinetic theory.
 As shown later on, the off-shell parts of the electron correlation
 functions play an extremely important role in computing
 radiation effects.

In Section~5 we consider  the transverse polarization functions
 which are the key quantities to calculate
 the collision terms in the photon kinetic equation.
 By classification of diagrams for the polarization matrix
 on the Schwinger-Keldysh contour, relevant contributions to
 the photon emission and absorption rates in weakly coupled plasmas
 are selected. It is important to note that the final expressions for
 the transverse polarization functions contain terms coming from both the
 vertex corrections and the off-shell parts of the particle correlation
 functions.

The results of Section~5 are used in Section~6 to discuss the
contributions of various scattering processes to the local radiating power with
special attention paid to the influence of plasma effects. It is shown
that Cherenkov radiation processes, which
  are energetically forbidden in a collisionless plasma,
  occur if the collisional broadening of the quasiparticle spectral
  function is taken into account. The physical interpretation of
   radiative processes is closely related to the decomposition of
   the field correlation functions into resonant and off-shell parts.
 For instance, the contribution of the resonant parts of the transverse field
correlation functions to the radiating power may be interpreted as
 Compton scattering while the contribution of the off-shell
 parts corresponds to a relativistic  plasma effect --- the  scattering of electrons
 on current fluctuations. The contribution
  to the radiating power arising from interaction between
   the relativistic electrons and the longitudinal field fluctuations can
 be divided into several terms corresponding to the electron scattering
  by resonant plasmons (the Compton conversion effect),
 by the off-resonant electron charge fluctuations, and by
   ions (bremsstrahlung). It is important to note that the transition
   probabilities for bremsstrahlung and Compton scattering known from vacuum QED
    are recovered within the present many-particle approach if all plasma
    effects are removed.

Finally, in Section~7 we discuss the results and possible
 extensions of the theory. Some special questions are considered in
 the appendices.

Throughout the paper we use the system of units with $c=\hbar=1$ and the
Heaviside's units for electromagnetic field, i.e., the Coulomb force is
written as $qq'/4\pi r$. Although we work in the Coulomb gauge and
 in the rest frame of the system, many formulas are represented more
compactly in the relativistic four-notation.
 The signature of the metric tensor $g^{\mu\nu}$ is
$(+,-,-,-)$.  Summation over repeated Lorentz (Greek) and space (Latin)
indices is understood.
 Our convention for the matrix Green's functions on the
time-loop Schwinger-Keldysh contour follows Botermans and
Malfliet~\cite{BotMalf90}.

\section{Green's Function Formalism in Coulomb Gauge}
   \label{Sec:GF}

\subsection{Basic Green's Functions}
   \label{Subs:BasicGF}
   We start with some notations and definitions.
  Following the quantum many-particle formulation in terms of
Green's functions  developed by Bezzerides and DuBois~\cite{BezDub72}, we
assume that the system was perturbed from its initial state by some
prescribed $c$-number external four-current
 $J^{(\text{ext})\,\mu}(\vec{r},t)=
 \big(\varrho^{\,(\text{ext})},\vec{J}^{\,(\text{ext})}\big)$,
 where $\varrho^{\,(\text{ext})}(\vec{r},t)$ is the external charge density and
$\vec{J}^{\,(\text{ext})}(\vec{r},t)$ is the external current density. Note,
however, that the introduction of these quantities is only a trick to define
correlation functions and propagators for particles and electromagnetic
fluctuations  in the system. Since we are not interested in the effects of
initial correlations which die out after a few collisions, the initial time
$t^{}_{0}$ will be taken in the remote past, i.e., the limit $t^{}_{0}\to
-\infty$ will be assumed.

To describe statistical properties of the system, we introduce field and
particle Green's functions defined on the time-loop Schwinger-Keldysh
contour $C$ which runs from $-\infty$ to $+\infty$ along the chronological
branch $C^{}_{+}$ and then backwards along the antichronological branch
 $C^{}_{-}$~\cite{BotMalf90}.
From now on, the underlined variables
 $\und{k}=(\und{t}^{}_{\,k},\vec{r}^{}_{k})$ indicate that
 $\und{t}^{}_{\,k}$ lies on the contour $C$, while the notation
 $(k)=(t^{}_{k},\vec{r}^{}_{k})$ is used for space-time variables.
 Integrals along the contour are understood as
   \begin{equation}
 \label{Int:C}
  \int d\und{1}\, F(\und{1})=
 \int^{\infty}_{-\infty}  d{1}\, \left[F(1^{}_{+})
 -  F(1^{}_{-})\right] ,
 \end{equation}
where $F(1^{}_{\pm})$ stands for functions with time arguments  taken on
the branches $C^{}_{\pm}$ of the contour\footnote{Sometimes the
integration rule is defined with the plus sign in Eq.~(\ref{Int:C}). Then
it is necessary to introduce the sign factor $\eta=\pm$ for each argument
on the contour \cite{Bonitz98,BezDub72}.
 We will use the
convention~(\ref{Int:C}) (see also~\cite{BotMalf90})
 which makes formulas more
compact. One only has to keep in mind that the delta function  satisfies
$\delta(\und{t}^{}_{\,1}-\und{t}^{}_{\,2}) =\pm\,\delta({t}^{}_{1}-
{t}^{}_{2})$ on the branches $C^{}_{\pm}$.}.

 For any function
$F(\und{1}\,\und{2})$, we introduce the canonical form~\cite{BotMalf90}
 \begin{equation}
  \label{F:can}
 F(\und{1}\,\und{2})= \left(
 \begin{array}{cc}
 F(1^{}_{+}2^{}_{+}) &\ F(1^{}_{+}2^{}_{-})  \\
 F(1^{}_{-}2^{}_{+}) &\ F(1^{}_{-}2^{}_{-})
 \end{array}
 \right)=
 \left(
 \begin{array}{cc}
 F^{<}(12) + F^{+}(12) &\ F^{<}(12)  \\
 F^{>}(12) &\ F^{<}(12) - F^{-}(12)
 \end{array}
  \right)
 \end{equation}
with the space-time ``correlation functions'' $F^{\mgl}(12)$ and the
 retarded/advanced functions (usually called the ``propagators'')
 \begin{equation}
  \label{F+-}
  F^{\pm}(12)= F^{\delta}(12)\,
  \pm \theta\big(\pm(t^{}_{1} -t^{}_{2})\big)
 \left\{F^{>}(12) -  F^{<}(12)   \right\},
 \end{equation}
 where $\theta(x)$ is the step function, and $F^{\delta}(12)$ is a singular
 part of $F^{\pm}(12)$.
 A possibility of singularities in
 the retarded/advanced  functions
 for equal time-variables on the contour $C$
 was discussed in detail by Danielewicz~\cite{Danielew84}. In
our case, such singular terms appear
 in the propagators for longitudinal field fluctuations [see
 Eq.~(\ref{D(12)})]. Note the useful relation that follows from the above definitions:
 \begin{equation}
  \label{Rel:F}
 F^{>}(12) - F^{<}(12)= F^{+}(12) -F^{-}(12).
 \end{equation}

 It is convenient to treat the external
charge and current on different branches of the contour $C$
as \textit{independent\/} quantities and  formally define
 the ensemble  average $O(\und{1})$ for any operator
$\hat{O}(\und{1})$ in the Heisenberg picture as~\cite{BezDub72}
 \begin{equation}
 \label{ave:C}
 O(\und{1})= \frac{\big\langle
 T^{}_{C}\big\{S\,\hat{O}^{}_{I}(\und{1})\big\}\big\rangle}
 {\langle S\rangle}\, ,
 \end{equation}
where $\hat{O}^{}_{I}$ is the operator in the interaction picture and
$T^{}_{C}$ is the path-ordering operator on the contour $C$. The evolution
operator
 $S$ describes the interaction with the external current. With the
 four-potential operator
$\hat{A}^{\mu}(\vec{r},t)=(\hat{\phi},\hat{\vec{A}}\,)$, the evolution
 operator is written as
 \begin{equation}
  \label{S-C}
 S= T^{}_{C}\exp\left\{
 -i \int
 d\und{1}\,\hat{A}^{\mu}_{I}(\und{1})\,{J}^{\,(\text{ext})}_{\mu}(\und{1})
 \right\}.
  \end{equation}
 At the end of calculations the physical
limit is implied:
$\varrho^{\,(\text{ext})}(1^{}_{+})=\varrho^{\,(\text{ext})}(1^{}_{-})$ and
$\vec{J}^{\,(\text{ext})}(1^{}_{+})=\vec{J}^{\,(\text{ext})}(1^{}_{-})$. In
this limit $\langle S\rangle=1$,
 so that the
quantity~(\ref{ave:C}) coincides with the conventional ensemble average of
a Heisenberg  operator.

 The field Green's functions are defined as
 functional derivatives of the averaged four-potential,
 \beq
 \label{A:mu-def}
  A^{\mu}(\und{1})=\langle \hat{A}^{\mu}(\und{1})\rangle\equiv
  \left(\phi(\und{1}), \vec{A}(\und{1})\right),
 \eeq
 with respect to the external current:
   \beq
     \label{Dmn:def}
   D^{\mu\nu}(\und{1}\,\und{2})=
    \frac{\delta A^{\mu}(\und{1})}{\delta J^{\,(\text{ext})}_{\nu}(\und{2})}.
   \eeq
 The most important Green's functions are
   \beq
     \label{D:l-t}
  D^{00}(\und{1}\,\und{2})=
 \frac{\delta \phi(\und{1})}{\delta\varrho^{\,(\text{ext})}(\und{2})},
 \qquad
 D^{ij}(\und{1}\,\und{2})=
 \frac{\delta A^{i}(\und{1})}{\delta J^{\,(\text{ext})}_{j}(\und{2})},
 \quad
  i,j=1,2,3.
   \eeq
 They characterize  the longitudinal and transverse fluctuations of the
 electromagnetic field, respectively. The components $D^{0i}$ and
  $D^{i0}$ describe the direct coupling between
 longitudinal
 and transverse modes.
 Recalling Eq.~(\ref{ave:C}) for averages on the contour $C$, it is
 easy to show that in the physical limit
    \begin{equation}
   \label{D-T:phys}
    D^{ij}(\und{1}\,\und{2})=
 - i\big\langle
 T^{}_{C}\,\Delta\hat{A}^{i}(\und{1})\,\Delta\hat{A}^{j}(\und{2})\big\rangle
   \end{equation}
with $\Delta\hat{A}^{i}(\und{1})=\hat{A}^{i}(\und{1})-{A}^{i}(\und{1})$.
 In evaluating $D^{00}(\und{1}\,\und{2})$ with Eq.~(\ref{ave:C}),
 one must take account of the relation which is valid in the Coulomb gauge:
   \beq
      \label{V:rel}
  \hat\phi(\und{1})= \int d\und{2}\, V(\und{1}-\und{2})
 \left(\hat\varrho(\und{2}) + \varrho^{(\text{ext})}(\und{2})\right),
   \eeq
 where $\hat\varrho(\und{1})$ is the induced charge density operator, and
  \begin{equation}
  \label{V}
 V(\und{1}-\und{2})= \frac{1}{4\pi\big|\,\vec{r}^{}_{1} -\vec{r}^{}_{2}\big|}\,
 \delta\big(\und{t}^{}_{\,1} -\und{t}^{}_{\,2}\big).
  \end{equation}
From Eq.~(\ref{V:rel}) follows
 $\delta\hat\phi^{}_{I}(\und{1})/\delta\varrho^{\text{ext}}(\und{2})
 =V(\und{1}-\und{2})$, so that in the physical limit
 the longitudinal field Green's function takes the form
      \begin{equation}
  \label{D(12)}
   D^{00}(\und{1}\,\und{2})=
   -i
 \big\langle T^{}_{C}\,\Delta\hat{\phi}(\und{1})\,
  \Delta\hat{\phi}(\und{2})\big\rangle
 + V(\und{1}-\und{2})
 \end{equation}
with $\Delta\hat{\phi}(\und{1})=\hat{\phi}(\und{1})-\phi(\und{1})$.

We now introduce the path-ordered Green's functions for particles. Since our
main interest is in problems where electrons may be relativistic, the
corresponding Green's function is defined in terms of the Dirac field
operators:
    \begin{equation}
  \label{G-e}
  G(\und{1}\,\und{2})= -i
  \left.\left\langle
 T^{}_{C}[S\,\psi^{}_{I}(\und{1})\bar\psi^{}_{I}(\und{2})]\right\rangle\right/
  \left\langle{S}\right\rangle \, .
  \end{equation}
 Note that each of the contour components of
$G(\und{1}\,\und{2})$ is a $4\times 4$ spinor matrix.
 Finally,  the nonrelativistic Green's functions for heavy particles
 (protons and ions)  are defined as
    \begin{equation}
  \label{GF-ion}
  {\mathcal G}^{}_{B}(\und{1}\,\und{2})=
 -i
  \left.\left\langle
 T^{}_{C}[S\Psi^{}_{BI}(\und{1})
 \Psi^{\dagger}_{BI}(\und{2})]\right\rangle\right/
  \left\langle{S}\right\rangle,
  \end{equation}
where the index ``$B$'' labels the particle species.
The operators $\Psi^{}_{B}(\vec{r},t)$ and
$\Psi^{\dagger}_{B}(\vec{r},t)$ obey Fermi or Bose commutation rules for
equal time arguments.  For definiteness, ions will be treated as fermions.
 The ion subsystem is assumed to be non-degenerate, so that
 the final results will not
depend on this assumption. On the other hand, the Fermi statistics is
natural for protons which may thus be regarded as one of the ion species.

\subsection{Equations of Motion for Green's Functions}
    \label{Subs:FG-eqm}
 A convenient starting point in deriving equations of motion for the field
 Green's functions is the set of
 Maxwell's equations for the averaged four-vector potential
 (\ref{A:mu-def}).
  In the Coulomb gauge and Heaviside's units, these equations read
    \begin{equation}
  \label{Maxw:C}
 -\nabla^{2}_{1} \phi(\und{1}) = \varrho(\und{1}) +
 \varrho^{ (\text{ext})}(\und{1}),
 \qquad
 -\Box^{}_{1}\vec{A}(\und{1})=
 \vec{J}^{\,T}(\und{1}) + \vec{J}^{\,(\text{ext})}(\und{1}),
\end{equation}
where $\Box=\nabla^2-\partial^2/\partial t^2$ is the wave operator, and
$\vec{J}^{\,T}(\und{1})$ is the induced transverse current density. Using the
four-component notation, Eqs.~(\ref{Maxw:C}) are summarized as
  \beq
     \label{Maxw:4}
   - \Delta^{\mu}_{\ \lambda}(1)\,A^{\lambda}(\und{1})=
    J^{\mu}(\und{1}) + J^{\,(\text{ext})\,\mu}(\und{1}),
  \eeq
where
 $J^{\mu}(\und{1})=\big(\varrho(\und{1}), \vec{J}^{\,T}(\und{1})\big)$ and
  \beq
    \label{Diff:mn}
  \Delta^{\mu}_{\ \nu}(1)=
   \begin{pmatrix}
   \nabla^{2}_{1} & 0            & 0 & 0\\
       0          & \Box^{}_{1}  & 0 & 0\\
       0          & 0            & \Box^{}_{1} & 0\\
       0          & 0            & 0 & \Box^{}_{1}
  \end{pmatrix}
  \, .
  \eeq

 Equations of motion for the matrix Green's function
(\ref{Dmn:def}) can now be obtained by taking the functional derivatives
of Eq.~(\ref{Maxw:4}) with respect to $J^{\,(\text{ext})}_{\nu}$. Here one
 point needs to be made. Since
   in the Coulomb gauge we have $\vec{\nabla}\cdot \hat{\vec{A}}=0$, only the
   transverse part of the external current enters
   the evolution operator (\ref{S-C}). Because of the condition
   $ \vec{\nabla}\cdot \vec{J}^{\,(\text{ext})}(\und{1})=0$,
the components ${J}^{\,(\text{ext})}_{i}(\und{1})$ cannot be treated as
independent variables. It is therefore convenient to define the
functional derivative with respect to any transverse field
$\vec{V}^{\,T}(\und{1})$ as
         \begin{equation}
 \label{T-deriv}
 \frac{\delta}{\delta V^{\,T}_{i}(\und{1})}\ \longrightarrow\
 \int d\und{1}'\, \delta^{T}_{ij}(\und{1} - \und{1}')\,
 \frac{\delta}{\delta V^{\,T}_{j}(\und{1}')}\, ,
 \end{equation}
 where
   \begin{equation}
    \label{delta-T:C}
  \delta^{T}_{ij}(\und{1} -\und{2})=
  \delta(\und{t}^{}_{1} -\und{t}^{}_{2})\,
  \delta^{T}_{ij}(\vec{r}^{}_{1} -\vec{r}^{}_{2}),
  \end{equation}
 and
   \begin{equation}
    \label{delta-T}
   \delta^{T}_{ij}(\vec{r})=
  \int \frac{d^3 k}{(2\pi)^3}\,
  {\rm e}^{i\vec{k}\cdot\vec{r}}
  \left(
  \delta^{}_{ij} - \frac{k^{}_{i} k^{}_{j}}{|\vec{k}|^2}
  \right)
   \end{equation}
is the transverse delta function.
 Performing now functional differentiation $\delta/\delta V^{T}_{j}(\und{1}')$
in Eq.~(\ref{T-deriv}), all components $V^{T}_{j}(\und{1}')$ can
 be regarded  as independent variables.

Let us return to Eq.~(\ref{Maxw:4}) and take the functional derivative of
both sides with respect to $J^{\,(\text{ext})}_{\nu}(\und{2})$. Noting
that
  $\delta J^{\,(\text{ext})\,\mu}(\und{1})/
  \delta J^{\,(\text{ext})}_{\nu}(\und{2})=
  {\delta}^{\mu\nu}(\und{1}-\und{2})$, where
   \beq
      \label{delta:mn}
  {\delta}^{\mu\nu}(\und{1}-\und{2})=
  {\delta}^{}_{\mu\nu}(\und{1}-\und{2})=
    \begin{pmatrix}
      \delta(\und{1}-\und{2}) & 0\\
      0 & - \delta^{T}_{ij}(\und{1}-\und{2})
    \end{pmatrix} \, ,
   \eeq
and using the chain rule\footnote{From now on, except as otherwise noted,
integration over
 ``primed'' variables is understood.}
 \beq
   \label{chain:mn}
  \frac{\delta J^{\mu}(\und{1})}{\delta J^{\,(\text{ext})}_{\nu}(\und{2})}
  = \frac{\delta J^{\mu}(\und{1})}{\delta A^{\lambda}(\und{1}')}\,
    \frac{\delta A^{\lambda}(\und{1}')}{\delta J^{\,(\text{ext})}_{\nu}(\und{2})},
 \eeq
we obtain the equation of motion for the field Green's function
 \beq
   \label{Dmn:eq}
  - \Delta^{\mu}_{\ \lambda}(1)\,D^{\lambda\nu}(\und{1}\,\und{2})=
   {\delta}^{\mu\nu}(\und{1}-\und{2})
   + \Pi^{\mu}_{\ \lambda}(\und{1}\,\und{1}')\,
 D^{\lambda\nu}(\und{1}'\,\und{2})
 \eeq
with the \textit{polarization matrix}
  \beq
   \label{Pol:mn}
   \Pi^{}_{\mu\nu}(\und{1}\,\und{2})=
  \frac{\delta J^{}_{\mu}(\und{1})}{\delta A^{\nu}_{}(\und{2})}.
  \eeq
Physically, this matrix characterizes the system response to variations of
the total electromagnetic field. The calculation of $\Pi^{}_{\mu\nu}$ for a
many-component relativistic plasma will be detailed later.

 The equation of motion for the electron Green's function follows directly from
the Dirac equation for  $\psi(\und{1})$ on the contour $C$
  \beq
    \label{Dirac:eq}
 \left(i\mynot{\partial}^{}_{1}-
  e \,\,\hat{\mmynot{A}}(\und{1}) -m \right)\psi(\und{1})=0.
  \eeq
 Here and in what follows, we use the conventional abbreviation
  $\mynot{a}=\gamma^{\mu} a^{}_{\mu}$ for any four-vector $a^{\mu}$.
 Recalling the definition (\ref{G-e}), we can write
  \beq
    \label{G:eq}
  \left(i\mynot{\partial}^{}_{1} -m \right)
 G(\und{1}\,\und{2})
 + ie\gamma^{}_{\mu}
  \left.\left\langle T^{}_{C}
  \left[
   S \hat{A}^{\mu}_{I}(\und{1})
   \psi^{}_{I}(\und{1})\bar\psi^{}_{I}(\und{2})
  \right]
  \right\rangle\right/\langle S\rangle
 = \delta(\und{1} -\und{2}).
  \eeq
 Strictly speaking,
the delta function on the right-hand side
 should be multiplied by the unit spinor matrix $I$. For brevity,
this matrix will   be usually (but not always) omitted.
The second term in Eq.~(\ref{G:eq}) can
 be further transformed  by noting that
  \beq
   \label{var:G}
    \frac{\delta G(\und{1}\,\und{2})}{\delta J^{\,(\text{ext})}_{\mu}(\und{1})}
    =
    -  \left.\left\langle T^{}_{C}
  \left[
   S \hat{A}^{\mu}_{I}(\und{1})
   \psi^{}_{I}(\und{1})\bar\psi^{}_{I}(\und{2})
  \right]
  \right\rangle\right/\langle S\rangle
   +i A^{\mu}(\und{1})G(\und{1}\,\und{2})
  \eeq
which is a consequence of Eq.~(\ref{S-C}). Using the matrix identity
   \begin{equation}
  \label{F-Der}
 \delta F(\und{1}\,\und{2})= - \,
  F(\und{1}\,\und{1}')\,
 \delta F^{-1}(\und{1}'\,\und{2}')\, F(\und{2}'\,\und{2})\, ,
 \end{equation}
 Eq.~(\ref{G:eq}) is manipulated to Dyson's equation
  \begin{equation}
      \label{Dyson-e}
  \left(i\mynot{\partial}^{}_{1} - e\mmynot{A}(\und{1}) -m \right)
 G(\und{1}\,\und{2}) - \Sigma^{}_{}(\und{1}\,\und{1}')\,G(\und{1}'\,\und{2})
 = \delta(\und{1} -\und{2})
    \end{equation}
 with the matrix self-energy
   \beq
    \label{Sig-e:cov}
  \Sigma(\und{1}\,\und{2})= -ie\,\gamma^{}_{\mu}
   G(\und{1}\,\und{1}')\,
   \frac{\delta G^{-1}(\und{1}'\,\und{2})}{\delta J^{(\text{ext})}_{\mu}(\und{1})}.
   \eeq
Let us also write down the adjoint of Eq.~(\ref{Dyson-e}), which will be
needed in the following:
    \beq
   \label{Dyson-e:adj}
   G(\und{1}\,\und{2})
 \left(-i\! \stackrel{\leftarrow}{\mynot{\partial}}^{}_{2} -
 e\mmynot{A}(\und{2}) -m \right)
  = \delta(\und{1} -\und{2})
  +
  G(\und{1}\,\und{1}') \Sigma^{}_{}(\und{1}'\,\und{2})\, .
 \eeq

 We would like to mention that the very existence of the inverse Green's function
 $G^{-1}(\und{1}\,\und{2})$ on the time-loop contour $C$ and, consequently,
 the existence of Dyson's
 equation is not a trivial fact. It can be justified only if the
 initial density operator admits Wick's decomposition of correlation
 functions. In the context of the real-time Green's function
 formalism, this implies a factorization of higher-order Green's
 functions into  products of one-particle Green's functions
 in the limit $t^{}_{0}\to -\infty$ \cite{BotMalf90,Kremp85}.
  This mathematical limit should be interpreted in the coarse-grained
 sense, i.e., $|t^{}_{0}|$ should be a time long compared to some microscopic
 ``interaction time'' but short compared to a macroscopic ``relaxation
 time scale''. The above boundary condition thus implies that there are no  bound
 states and other long-lived many-particle correlations in the system.
 Otherwise, in order to have a Dyson equation for Green's functions,
 it is necessary to modify the form of the contour $C$ \cite{ZMR97,MorRoep99}.
 In the case of a weakly coupled relativistic plasma considered in the present
  paper, the boundary condition of weakening of initial correlations
 is well founded, so that one may use expression (\ref{Sig-e:cov}) for the
 electron self-energy.

Nonrelativistic equations of motion for the ion field operators
$\Psi^{}_{B}(1)$ in the Heisenberg picture have exactly the form of the
one-particle Schr\"odinger equation where the four-potential $A^{\mu}(1)$ is
replaced by the corresponding operator $\hat{A}^{\mu}(1)$ \cite{Dub67}. Since
in all practical applications the ion transverse current is very small
compared to the electron transverse current, we shall neglect the direct
interaction between ions and the transverse field $\vec{A}$. Then we have
instead of Eq.~(\ref{G:eq}) the equation
  \beq
    \label{Ga:eq}
  \left(i\,\frac{\partial}{\partial t^{}_{1}} +
 \frac{\nabla^{2}_{1}}{2m^{}_{B}}\right) {\mathcal G}^{}_{B}(\und{1}\,\und{2})
 + ie^{}_{B}
  \left.\left\langle T^{}_{C}
  \left[
   S \hat{\phi}^{}_{I}(\und{1})
   \Psi^{}_{BI}(\und{1})\Psi^{\dagger}_{BI}(\und{2})
  \right]
  \right\rangle\right/\langle S\rangle
 = \delta(\und{1} -\und{2}),
  \eeq
where $m^{}_{B}$ and $e^{}_{B}$ are the ion mass and charge. The second term
in this equation can be transformed in the same way as the second term
 in Eq.~(\ref{G:eq}) to derive Dyson's equation for the heavy particles
 (no summation over $B$)
   \beq
      \label{Dyson-a}
     \left(i\,\frac{\partial}{\partial t^{}_{1}} +
 \frac{\nabla^{2}_{1}}{2m^{}_{B}}\,
  -e^{}_{B} \phi(\und{1}) \right) {\mathcal G}^{}_{B}(\und{1}\,\und{2})
     -  \Sigma^{}_{B}(\und{1}\,\und{1}')\,
 {\mathcal G}^{}_{B}(\und{1}'\,\und{2}) =
 \delta(\und{1} -\und{2}),
   \eeq
where
   \begin{equation}
  \label{Sig-a}
 \Sigma^{}_{B}(\und{1}\,\und{2})=
  -ie^{}_{B}
  {\mathcal G}^{}_{B}(\und{1}\,\und{1}')\,
   \frac{\delta {\mathcal G}^{-1}_{B}(\und{1}'\und{2})}
 {\delta \varrho^{(\text{ext})}(\und{1})}
 \end{equation}
are the ion self-energies.

\subsection{Vertex Functions}
  \label{Subs:Vert}

A convenient way to analyze the polarization matrix and the particle
self-energies is to  express them in terms of vertex functions. The
electron four-vertex is defined as
 \beq
   \label{Gamma-e:mu}
 \Gamma^{}_{\mu}(\und{1}\,\und{2}\,;\und{3})=
  - \frac{\delta G^{-1}(\und{1}\,\und{2})}{\delta A^{\mu}_{}(\und{3})}.
 \eeq
It is evident  that each component of
    $\Gamma^{}_{\mu}$ is a $4\times 4$ spinor matrix.
   Using the chain rule
    $$
 \frac{\delta G^{-1}(\und{1}\,\und{2})}{\delta J^{\,(\text{ext})}_{\mu}(\und{3})}=
  \frac{\delta G^{-1}(\und{1}\,\und{2})}{\delta A^{\nu}(\und{1}')}\,
  \frac{\delta A^{\nu}(\und{1}')}{\delta J^{\,(\text{ext})}_{\mu}(\und{3})}=
   - \Gamma^{}_{\nu}(\und{1}\,\und{2}\,;\und{1}')\,
   D^{\nu\mu}(\und{1}'\,\und{3}),
    $$
the electron self-energy (\ref{Sig-e:cov}) can be rewritten in the form
   \beq
      \label{Sig-e:Gamm}
   \Sigma(\und{1}\,\und{2})=
  i\, \Gamma^{(0)}_{\mu}(\und{1}\,\und{1}';\und{4}')\,
   G(\und{1}'\,\und{2}')\,
  \Gamma^{}_{\nu}(\und{2}'\,\und{2};\und{3}')\,
   D^{\nu\mu}(\und{3}'\,\und{4}'),
   \eeq
where
  \beq
     \label{Gam-e:bare}
 \Gamma^{(0)}_{\mu}(\und{1}\,\und{2}\, ;\und{3})=
  e\,\delta(\und{1}-\und{2})\,{\delta}^{}_{\mu\nu}(\und{1}-\und{3})\,
  \gamma^{\nu}\, .
  \eeq
 Equation (\ref{Sig-e:Gamm}) relates the electron self-energy
to the four-vertex $\Gamma^{}_{\mu}$. Another relation between these
quantities follows from Dyson's equation (\ref{Dyson-e}). Writing
  $$
  G^{-1}(\und{1}\,\und{2})=
  \left(i\mynot{\partial}^{}_{1} - e\mmynot{A}(\und{1})
  -m \right)\delta(\und{1} -\und{2})
  -\Sigma^{}_{}(\und{1}\, \und{2})
  $$
and recalling the definition (\ref{Gamma-e:mu}) of the four-vertex, we find
immediately that
   \beq
     \label{Gamm-e:Sig}
     \Gamma^{}_{\mu}(\und{1}\,\und{2}\,;\und{3})
     =\Gamma^{(0)}_{\mu}(\und{1}\,\und{2}\,;\und{3})
     + \frac{\delta \Sigma(\und{1}\,\und{2})}{\delta A^{\mu}(\und{3})}.
   \eeq
It is thus seen that $\Gamma^{(0)}_{\mu}$ is the \textit{bare
four-vertex\/}, i.e. the vertex in the absence of field fluctuations.

 The ion four-vertices are defined as
  \beq
    \label{Ions-Gamm}
     \Gamma^{}_{B \mu}(\und{1}\,\und{2}\,;\und{3})=
 -\, \frac{\delta {\mathcal G}^{-1}_{B}(\und{1}\,\und{2})}{\delta A^{\mu}(\und{3})}.
  \eeq
 The arguments then go along the same way as for the electron self-energy
and Eq.~(\ref{Sig-a}) is transformed into
  \beq
     \label{Sig-i:Gamm}
  \Sigma^{}_{B}(\und{1}\,\und{2})=
    i\, \Gamma^{(0)}_{B\mu}(\und{1}\,\und{1}';\und{4}')\,
   {\mathcal G}^{}_{B}(\und{1}'\,\und{2}')\,
  \Gamma^{}_{B\nu}(\und{2}'\,\und{2};\und{3}')\,
   D^{\nu\mu}(\und{3}'\,\und{4}')
  \eeq
with the bare vertices
  \beq
     \label{Gam-i:bare}
 \Gamma^{(0)}_{B\mu}(\und{1}\,\und{2}\,;\und{3})=
  e^{}_{B}\,\delta(\und{1}-\und{2})\,
   {\delta}^{}_{\mu 0}(\und{1}-\und{3}).
  \eeq
Using the ion Dyson's equation (\ref{Dyson-a}), it is a simple matter to
see that the analogue of Eq.~(\ref{Gamm-e:Sig}) reads
      \beq
     \label{Gamm-i:Sig}
     \Gamma^{}_{B\mu}(\und{1}\,\und{2}\,;\und{3})
     =\Gamma^{(0)}_{B\mu}(\und{1}\,\und{2}\,;\und{3})
     + \frac{\delta \Sigma^{}_{B}(\und{1}\,\und{2})}{\delta A^{\mu}(\und{3})}.
   \eeq
 To write the polarization matrix
(\ref{Pol:mn}) in terms of the Green's functions and vertices, we need an
expression for the induced four-current
 $J^{}_{\mu}(\und{1})=\langle \hat{J}^{}_{\mu}(\und{1})\rangle$.
The component $J^{}_{0}(\und{1})$ is the induced charge density. It can be
written as the sum of the electron and ion contributions:
   \beq
     \label{current:0}
   J^{}_{0}(\und{1})=
    e \langle \bar{\psi}(\und{1})\,\gamma^{0} \psi(\und{1})\rangle
  + \sum_{B} e^{}_{B}
  {\rm tr}^{}_{S}
  \langle\Psi^{\dagger}_{B}(\und{1})\Psi^{}_{B}(\und{1})\rangle
   \eeq
with ${\rm tr}^{}_{S}$ denoting the trace over spin indices.
 Since the transverse ion current is neglected, we have
   \beq
     \label{current:i}
    J^{T}_{i}(\und{1})= e\,\delta^{T}_{ij}(\und{1}-\und{1}')
     \langle \bar{\psi}(\und{1}')\,\gamma^{j}\psi(\und{1}')\rangle.
   \eeq
Thus, in terms of Green's functions (\ref{G-e}) and (\ref{GF-ion}),
   \beq
      \label{J:mu}
   J^{}_{\mu}(\und{1})=
   -ie\, {\delta}^{}_{\mu\lambda}(\und{1}-\und{1}')\,
   {\rm tr}^{}_{D} \left[\gamma^{\lambda}G(\und{1}'\,\und{1}^{\prime +})\right]
    -i\,{\delta}^{}_{\mu 0}
     \sum_{B} e^{}_{B}\,{\rm tr}^{}_{S}\,
     {\mathcal G}^{}_{B}(\und{1}\,\und{1}^{+}),
   \eeq
where ${\rm tr}^{}_{D}$ stands for the trace over the Dirac spinor
indices, and the notation   $\und{1}^{\prime +}$ shows that the time
 $\und{t}^{\prime +}_{\,1}$ is taken infinitesimally later on the contour $C$ than
$\und{t}^{\prime}_{\,1}$. The polarization matrix (\ref{Pol:mn}) can now
be obtained by differentiating Eq.~(\ref{J:mu}) with respect to
$A^{\nu}(\und{2})$ and then using the identity (\ref{F-Der}) to write
 $\delta G/\delta A^{\nu}$ and
  $\delta {\mathcal G}^{}_{B}/\delta A^{\nu}$ in terms of the vertices.
 A little algebra gives
    \begin{eqnarray}
       \label{Pi-mn:gen}
  \Pi^{}_{\mu\nu}(\und{1}\,\und{2})=
   & - & i\, {\rm tr}^{}_{D}
  \left[
  \Gamma^{(0)}_{\mu}(\und{1}'\,\und{2}';\und{1})
  G(\und{2}'\,\und{3}')
 \Gamma^{}_{\nu}(\und{3}'\,\und{4}';\und{2})
 G(\und{4}'\,\und{1}')
  \right]
   \nonumber\\[5pt]
  & - & i \sum_{B} {\rm tr}^{}_{S}
  \left[
  \Gamma^{(0)}_{B\mu}(\und{1}'\,\und{2}';\und{1})
  {\mathcal G}^{}_{B}(\und{2}'\,\und{3}')
 \Gamma^{}_{B\nu}(\und{3}'\,\und{4}';\und{2})
 {\mathcal G}^{}_{B}(\und{4}'\,\und{1}')
  \right].
    \end{eqnarray}
The above formalism  provides a basis for studying various processes
 in a many-component plasma
 with relativistic electrons.
 It is remarkable that the polarization matrix (\ref{Pi-mn:gen})
  involves the same vertex functions as the
particle self-energies (\ref{Sig-e:Gamm}) and (\ref{Sig-i:Gamm}). Thus,
 any approximation for the vertex functions leads to the corresponding
self-consistent approximation for the polarization functions and the
self-energies in terms of Green's functions.

The method we use for calculating the vertex functions is based on
 Eqs.~(\ref{Gamm-e:Sig}) and (\ref{Gamm-i:Sig}).  Neglecting the last term in
 Eq.~(\ref{Gamm-e:Sig}), i.e. replacing everywhere $\Gamma^{}_{\mu}$
 by the
 bare vertex  $\Gamma^{(0)}_{\mu}$, we recover the Bezzerides-DuBois
 approximation~\cite{BezDub72} in kinetic theory of
  electron-positron  relativistic plasmas. However, this simplest
  approximation is unable to describe radiative processes, so that
   the last terms in Eqs.~(\ref{Gamm-e:Sig}) and (\ref{Gamm-i:Sig})
   must be taken into consideration. Recalling expressions (\ref{Sig-e:Gamm}) and
   (\ref{Sig-i:Gamm})
   for the self-energies, we get the following equations for the vertices:
    \begin{subequations}
     \label{Gamm-eq:e-i}
    \begin{eqnarray}
      & &
   \label{Gamm-eq:e}
 \hspace*{-20pt}
      \Gamma^{}_{\mu}(\und{1}\,\und{2}\,;\und{3})
      =\Gamma^{(0)}_{\mu}(\und{1}\,\und{2}\,;\und{3})
      \nonumber\\[3pt]
       & &
 \hspace*{-10pt}
       {}+ i\,\Gamma^{(0)}_{\lambda}(\und{1}\,\und{1}'\,;\und{4}')
        G(\und{1}'\,\und{1}'')
        \Gamma^{}_{\mu}(\und{1}''\,\und{2}'';\und{3})
        G(\und{2}''\, \und{2}')
         \Gamma^{}_{\lambda'}(\und{2}'\,\und{2}\,;\und{3}')
         D^{\lambda'\lambda}(\und{3}'\,\und{4}')
         \nonumber\\[3pt]
         & &
  \hspace*{-10pt}
         {}+
         i\,\Gamma^{(0)}_{\lambda}(\und{1}\,\und{1}'\,;\und{4}')
          G(\und{1}'\,\und{2}')\,
          \frac{\delta}{\delta A^{\mu}(\und{3})}
          \left[
         \Gamma^{}_{\lambda'}(\und{2}'\,\und{2}\,;\und{3}')\,
      D^{\lambda'\lambda}(\und{3}'\,\und{4}')
          \right],
          \\[8pt]
        & &
   \label{Gamm-eq:i}
 \hspace*{-20pt}
      \Gamma^{}_{B\mu}(\und{1}\,\und{2}\,;\und{3})
      =\Gamma^{(0)}_{B\mu}(\und{1}\,\und{2}\,;\und{3})
      \nonumber\\[3pt]
       & &
 \hspace*{-10pt}
       {}+ i\,\Gamma^{(0)}_{B\lambda}(\und{1}\,\und{1}'\,;\und{4}')
        {\mathcal G}^{}_{B}(\und{1}'\,\und{1}'')
        \Gamma^{}_{B\mu}(\und{1}''\,\und{2}'';\und{3})
        {\mathcal G}^{}_{B}(\und{2}''\, \und{2}')
         \Gamma^{}_{B\lambda'}(\und{2}'\,\und{2}\,;\und{3}')
         D^{\lambda'\lambda}(\und{3}'\,\und{4}')
         \nonumber\\[3pt]
         & &
 \hspace*{-10pt}
         {}+
         i\,\Gamma^{(0)}_{B\lambda}(\und{1}\,\und{1}'\,;\und{4}')
          {\mathcal G}^{}_{B}(\und{1}'\,\und{2}')\,
          \frac{\delta}{\delta A^{\mu}(\und{3})}
          \left[
         \Gamma^{}_{B\lambda'}(\und{2}'\,\und{2}\,;\und{3}')\,
      D^{\lambda'\lambda}(\und{3}'\,\und{4}')
          \right].
    \end{eqnarray}
    \end{subequations}
Although these equations are too complicated to solve them in a general
form, approximate solutions can be found for \textit{weakly coupled
plasmas\/} where collisional interaction is taken into account to lowest
orders.
 It should be noted that a simple asymptotic expansion of
 vacuum electrodynamics in powers of the fine structure constant
 $\alpha=e^2/\hbar c$ is not appropriate even for weakly coupled plasmas
 due to collective
  effects (polarization and screening).
  It is therefore natural to work with the \textit{full\/} Green's functions
$G$, ${\mathcal G}^{}_{B}$, and $D^{\mu\nu}$, which contain polarization
 and screening to all orders in $\alpha$.  As discussed by Bezzerides and
DuBois~\cite{BezDub72}, the weak-coupling approximation for a plasma should
be regarded as an expansion in terms of the field Green's functions $D^{\mu\nu}$
which play the role of intensity measures for fluctuations of the
electromagnetic field, rather than being an expansion in $\alpha$.
This scheme applies
 if several
 conditions are fulfilled. The first condition reads
 $\lambda^{}_{\text{pl}}\ll 1$, where
 $\lambda^{}_{\text{pl}}=1/(n^{}_{e}r^{3}_{\text{sc}})$
 is the plasma parameter. Here, $n^{}_{e}$ is the electron number density
 and $r^{}_{\text{sc}}$ is the screening (Debye) length.
 The second condition, $e^{2}/(\hbar v )\ll 1$, where $v$ a characteristic
 particle velocity, ensures the validity of the Born approximation for
 scattering processes. The above conditions are usually met for relativistic
 plasmas. Finally, it is assumed that
 the plasma modes are not excited considerably above
 their local equilibrium level. This  condition is not satisfied for a
 strongly turbulent regime. In what follows  we shall assume that the system
 can be described within the weak-coupling approximation.

\begin{figure}[h]
 \centerline{
   $\Gamma^{}_{\mu}(\und{1}\,\und{2}\,;\und{3})=$
   \raisebox{-8pt}{\includegraphics[scale=0.36]{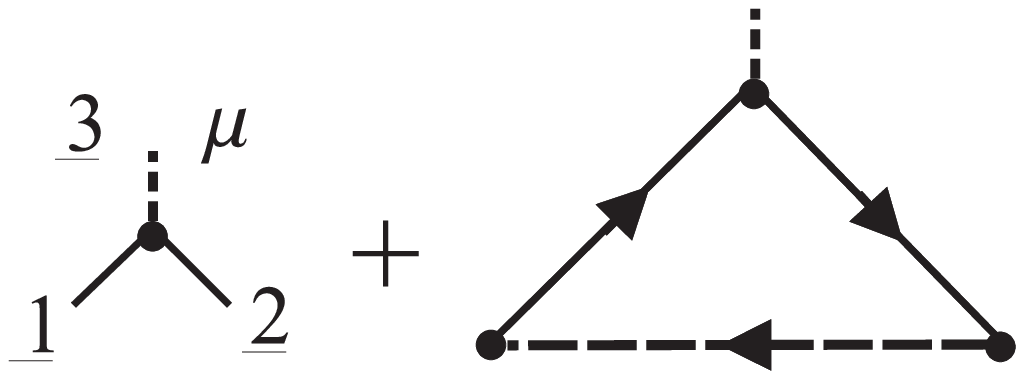}}
 \qquad
   $\Gamma^{}_{B\mu}(\und{1}\,\und{2}\,;\und{3})=$
  \raisebox{-8pt}{\includegraphics[scale=0.36]{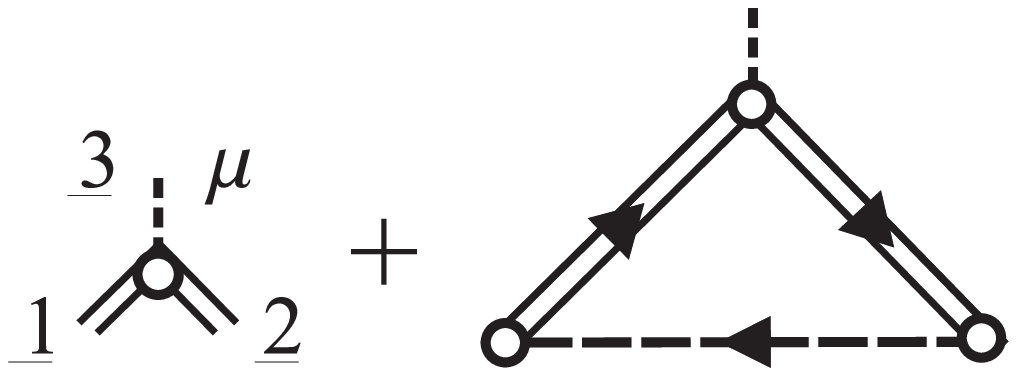}}
   }
 \caption{\label{fig:Vert-e-i}Lowest order diagrams for
the vertices. The first terms are the bare vertices,
  straight and doubled lines denote respectively $G$ and
 ${\mathcal G}^{}_{B}$. Dashed  lines denote
 $iD^{\lambda\sigma}$.
 }
\end{figure}

As can be seen from Eqs.~(\ref{Gamm-eq:e-i}), the
derivatives
 $\delta\Gamma/\delta A^\mu$ and
 $\delta\Gamma^{}_{B}/\delta A^\mu$ are at least of first order
 in the field Green's functions $D$.
 On the other hand, using the identity (\ref{F-Der}) and Eq.~(\ref{Dmn:eq}),
 we may write symbolically
   $$
    \frac{\delta D}{\delta A^\mu} \sim
   D\,\frac{\delta D^{-1}}{\delta A^\mu}\,D\sim
  D\,\frac{\delta \Pi}{\delta A^\mu}\,D.
   $$
Thus, for a weakly coupled plasma, the last terms in
Eqs.~(\ref{Gamm-eq:e-i}) are small compared to other terms on the
right-hand sides. This suggests a self-consistent approximation in which
the last terms in Eqs.~(\ref{Gamm-eq:e-i}) are neglected. The resulting
equations for the vertices are still rather complicated but
 more tractable for specific problems\footnote{Analogous equations for
nonrelativistic plasmas were discussed, e.g., by DuBois~\cite{Dub67}.}.
 Another way is to solve Eqs.~(\ref{Gamm-eq:e-i})
 by iteration. In this paper  we restrict our analysis
 to the first iteration of Eqs.~(\ref{Gamm-eq:e-i}), which
 means that the vertices in the second terms on
 the right-hand sides  are replaced by
 the bare vertices and
 the last terms are neglected.
 This approximation can be
  represented graphically by the Feynman  diagrams shown in
  Fig.~\ref{fig:Vert-e-i}.

\begin{figure}[h]
 \centerline{
   $\Sigma(\und{1}\,\und{2})=$\
   \raisebox{-6pt}{\includegraphics[scale=0.4]{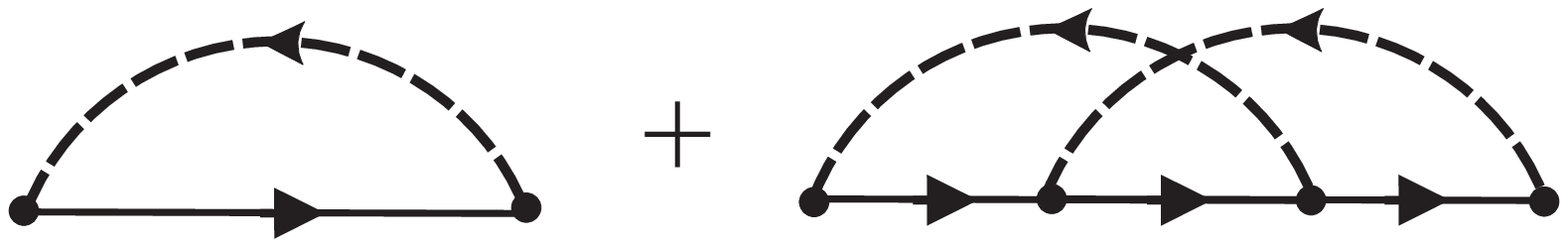}}
    }
   \vspace*{12pt}
  \centerline{
   $\Sigma^{}_{B}(\und{1}\,\und{2})=$\
  \raisebox{-6pt}{\includegraphics[scale=0.4]{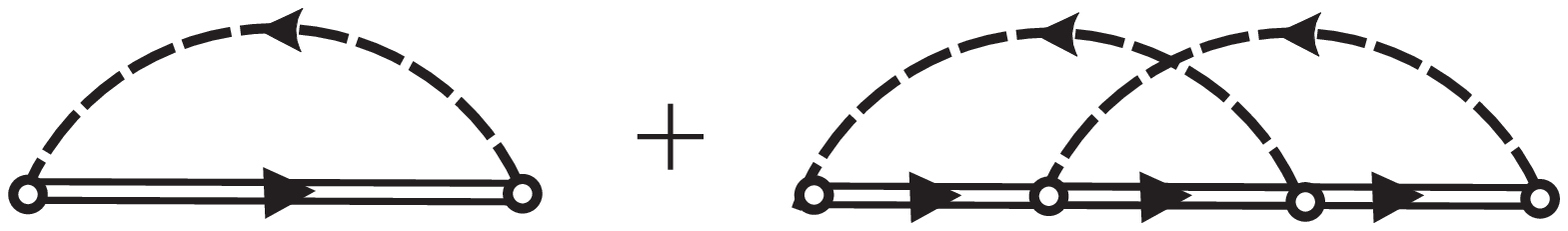}}
   }
  \vspace*{15pt}
  \centerline{
   $i\,\Pi^{}_{\mu\nu}(\und{1}\,\und{2})=$\hspace*{-10pt}
  \raisebox{-50pt}{\includegraphics[scale=0.4]{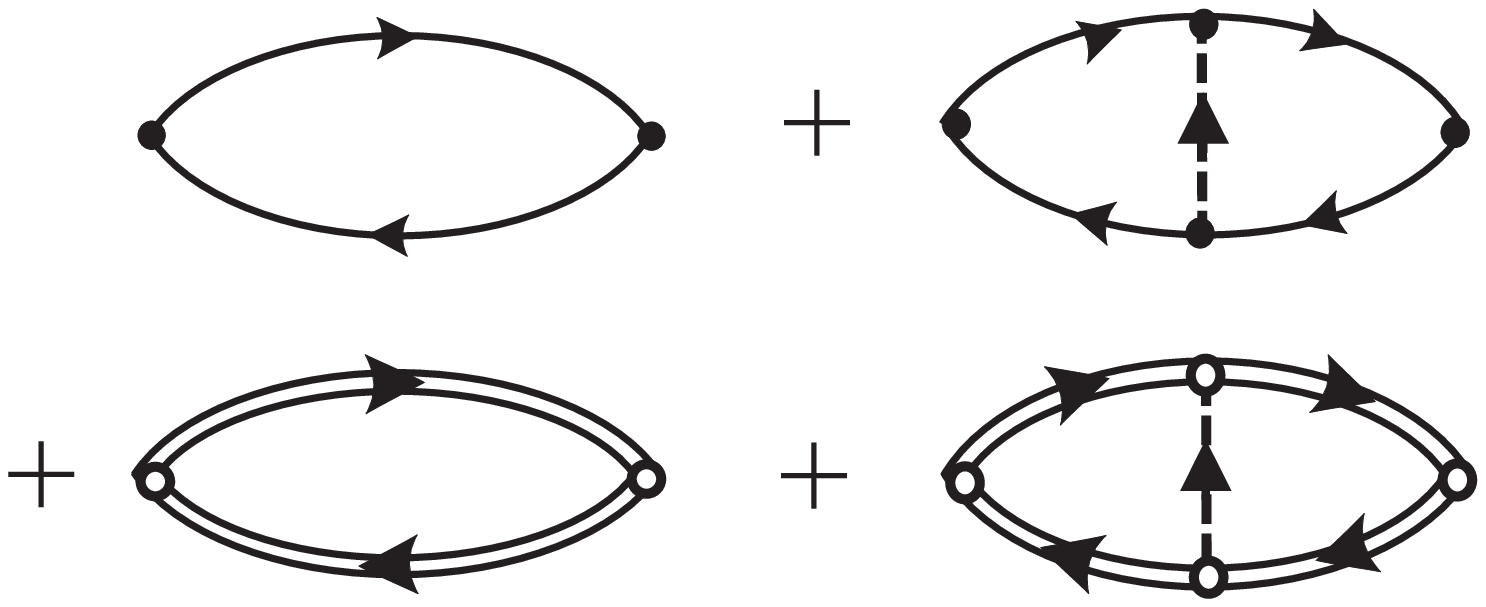}}
   }
 \caption{\label{fig:SelfEn;Pol}Lowest order diagrams for
the electron self-energy (first line), the ion self-energies (second line), and
the polarization matrix. The ion contribution to $\Pi^{}_{\mu\nu}$ is
obtained as a sum of the last two diagrams over the species index $B$.
 }
\end{figure}

This approximation for the vertices
 generates the corresponding
 self-energies (\ref{Sig-e:Gamm}), (\ref{Sig-i:Gamm}), and the
 polarization matrix (\ref{Pi-mn:gen}). They are summarized in
Fig.~\ref{fig:SelfEn;Pol}.

Once expressions for the self-energies and the polarization matrix are
specified, Eqs.~(\ref{Dmn:eq}), (\ref{Dyson-e}), and (\ref{Ga:eq}),
together with
 Maxwell's equations for the mean electromagnetic field
 $A^{\mu}(1)$,  provide a closed and self-consistent description of the
 particle dynamics and the field fluctuations in many-component weakly coupled
plasmas with relativistic electrons (positrons).
 It is, of course, clear that these equations
are prohibitively difficult to solve, so that
 one has to introduce further reasonable approximations,
  depending on the character of the problem. As already noted, our
 prime interest is with photon kinetics in a nonequilibrium relativistic
 plasma. The kinetic description of radiative processes is adequate if
 the medium is approximately isotropic on the scale of the characteristic photon
 wavelength and therefore the
 transverse and longitudinal electromagnetic
 fluctuations do not mix\footnote{The direct coupling between
  transverse and longitudinal field fluctuations described by
  the components $D^{}_{0i}$ and $\Pi^{}_{0i}$ may be important for
 long-wavelength and low-frequency modes in the presence of
 electron beams and their return plasma currents~\cite{Bret04,Tautz07}.}.
In our subsequent discussion we shall assume this condition to be
satisfied, so that the field Green's function $D^{}_{\mu\nu}$ and the
polarization matrix $\Pi^{}_{\mu\nu}$ will be taken in block form:
  \beq
     \label{Block:Isotr}
   D^{}_{\mu\nu}(\und{1}\,\und{2})=
    \begin{pmatrix}
      D(\und{1}\,\und{2}) & 0\\
    0  &  D^{}_{ij}(\und{1}\,\und{2})
   \end{pmatrix}\, ,
   \qquad
   \Pi^{}_{\mu\nu}(\und{1}\,\und{2})=
    \begin{pmatrix}
     \Pi(\und{1}\,\und{2}) & 0\\
    0  &  \Pi^{}_{ij}(\und{1}\,\und{2})
   \end{pmatrix}\, .
  \eeq
Here $D\equiv D^{}_{00}$ and
 $\Pi\equiv \Pi^{}_{00}$ are the longitudinal (plasmon) components,
  whereas $D^{}_{ij}$ and $\Pi^{}_{ij}$ are the
transverse (photon) components.

 \setcounter{equation}{0}

\section{Kinetic Equation for Photons in Plasmas}
 \label{Sec:KinEq}

\subsection{Transport and Mass-shell Equations}
  \label{Subs:TransMass}
 With the assumption of the block structure (\ref{Block:Isotr})
for $D^{}_{\mu\nu}$ and
  $\Pi^{}_{\mu\nu}$, the equation of motion for the photon Green's
functions $D^{}_{ij}(\und{1}\,\und{2})$ is obtained from
Eq.~(\ref{Dmn:eq}) in the form
    \begin{equation}
  \label{T-Dyson}
  \Box^{}_{1} D^{}_{ij}=
 \delta^{T}_{ij}(\und{1}-\und{2})
 + \Pi^{}_{ik}(\und{1}\,\und{1}')\,
 D^{}_{kj}(\und{1}'\,\und{2}).
 \end{equation}
 We shall also
  need the adjoint of this equation which reads
      \begin{equation}
  \label{T-Dyson-adj}
  \Box^{}_{2} D^{}_{ij}(\und{1}\,\und{2})=
 \delta^{T}_{ij}(\und{1}-\und{2})
 +
 D^{}_{ik}(\und{1}\,\und{1}')\,\Pi^{}_{kj}(\und{1}'\,\und{2}).
 \end{equation}

The first step toward a photon kinetic equation is to rewrite the above
equations in terms of the canonical components (\ref{F:can}) of
 $D^{}_{ij}(\und{1}\,\und{2})$,
which will be denoted as $d^{\mgl}_{ij}(12)$ and $d^{\pm}_{ij}(12)$. Without
entering into details of algebraic manipulations described, e.g., in
 the paper by Botermans and Malfliet~\cite{BotMalf90}, let us write down
  the so-called  \textit{Kadanoff-Baym (KB) equations\/} for the transverse
  correlation functions
       \begin{subequations}
  \label{d<>:eq}
  \begin{eqnarray}
 \label{d<>:eq-a}
 & &
   \Box^{}_{1}\, d^{\mgl}_{ij}(1 2)=
  \pi^{+}_{ik}(1 1')\,d^{\mgl}_{kj}(1' 2)
  + \pi^{\mgl}_{ik}(1 1')\,d^{-}_{kj}(1' 2),
  \\[5pt]
 \label{d<>:eq-b}
 & &
   \Box^{}_{2}\, d^{\mgl}_{ij}(1 2)=
  d^{\mgl}_{ik}(1 1')\, \pi^{-}_{kj}(1' 2)
  + d^{+}_{ik}(1 1')\pi^{\mgl}_{kj}(1' 2),
 \end{eqnarray}
 \end{subequations}
and the equations for the propagators
 \begin{subequations}
 \label{d+-:eq}
 \begin{eqnarray}
  \label{d+-:eq-a}
 & &
  \Box^{}_{1}\,d^{\pm}_{ij}(1 2)= \delta^{T}_{ij}(1-2)
 +
 \pi^{\pm}_{ik}(1 1')\,d^{\pm}_{kj}(1' 2),
 \\[5pt]
  \label{d+-:eq-b}
 & &
  \Box^{}_{2}\,d^{\pm}_{ij}(1 2)= \delta^{T}_{ij}(1-2)
 +
 d^{\pm}_{ik}(1 1')\, \pi^{\pm}_{kj}(1' 2),
 \end{eqnarray}
 \end{subequations}
 where $\pi^{\mgl}_{ij}$ and $\pi^{\pm}_{ij}$ are the canonical components
  of $\Pi^{}_{ij}$, and  the primed space-time variables are integrated
  according to the rule
   $$
  \int d1'\,\ldots =
 \int^{\infty}_{-\infty} dt^{\prime}_{1}\int d^3 \vec{r}^{\,\prime}_{1}\,\ldots\, .
  $$
Note that the space-time correlation functions $d^{\mgl}_{ij}(12)$ and the
propagators $d^{\pm}_{ij}(12)$ enjoy symmetry properties which are useful
in working with Eqs.~(\ref{d<>:eq}) and (\ref{d+-:eq}):
      \begin{subequations}
 \label{symm-d<>;d/pm}
  \begin{eqnarray}
  \label{symm-d<>}
  & &
  d^{<}_{ij}(12)= d^{>}_{ji}(21),
 \qquad
   \left[  d^{<}_{ij}(12)  \right]^*= - d^{<}_{ji}(21)\, ,
 \\[3pt]
  \label{symm-d/pm}
  & &
  d^{+}_{ij}(12)= d^{-}_{ji}(21),
 \qquad
   \left[  d^{+}_{ij}(12)  \right]^*= d^{+}_{ij}(12)\, .
 \end{eqnarray}
 \end{subequations}
These properties follows directly from the fact that
 $\hat{A}^{}_{i}$ are Hermitian operators\footnote{Using Eqs.~(\ref{d<>:eq})
 and (\ref{d+-:eq}), it can be verified that the polarization matrices
  $\pi^{\mgl}_{ij}(12)$ and $\pi^{\pm}_{ij}(12)$ have the same
  symmetry properties  as $d^{\mgl}_{ij}(12)$ and $d^{\pm}_{ij}(12)$. }.

The phase space description of photon dynamics is achieved by  using
 the Wigner representation
 which is defined for any
$F(12)\equiv F(x^{}_{1},x^{}_{2})$ as
 \begin{equation}
  \label{W-tr}
 F(X,k)= \int d^4 x\, {\rm e}^{ik\cdot x}\,F\left(X+x/2,X-x/2\right),
 \end{equation}
 where $k\cdot x=k^{\mu} x^{}_{\mu}=k^{0}t - \vec{k}\cdot \vec{r}$.
  Within the framework of kinetic theory, the variations of  the field
Green's functions and the polarization matrix
 in the space-time variable
 $X^{\mu}=(T,\vec{R})$ are assumed to be slow on the scales
 of $\lambda$ and $1/\omega$, where $\lambda$ and
$\omega$ are respectively some characteristic radiation wavelength and
frequency. Therefore, going over to the Wigner representation in
 Eqs.~(\ref{d<>:eq}) and (\ref{d+-:eq}), we shall keep terms only
 to first order in $X$-gradients\footnote{This gradient expansion
 scheme is a usual way for deriving kinetic equations in the Green's function
  formalism~\cite{KadBaym62,Danielew84,BotMalf90,BezDub72}.}.
Formal manipulations are simplified by using the first-order
transformation rule~\cite{BotMalf90}
 \beq
   \label{Transf:W}
   F^{}_{1}(11')F^{}_{2}(1'2)\ \rightarrow\
 F^{}_{1}(X,k)\,F^{}_{2}(X,k)
 -\frac{i}{2}\left\{F^{}_{1}(X,k),F^{}_{2}(X,k)\right\},
 \eeq
 where
 \begin{equation}
  \label{Poiss}
 \left\{F^{}_{1}(X,k),F^{}_{2}(X,k)\right\}=
 \frac{\partial F^{}_{1}}{\partial X^{\mu}_{}}
 \frac{\partial F^{}_{2}}{\partial k^{}_{\mu}}
 -
 \frac{\partial F^{}_{1}}{\partial k^{\mu}}
 \frac{\partial F^{}_{2}}{\partial X^{}_{\mu}}
 \end{equation}
is the four-dimensional Poisson bracket. After some algebra,
 Eqs.~(\ref{d<>:eq}) and (\ref{d+-:eq}) reduce to the equations of motion
 for $d^{\mgl}_{ij}(X,k)$ and $d^{\pm}_{ij}(X,k)$
 (the arguments
$X$ and $k$ are omitted to save writing):
  \begin{eqnarray}
  \label{d<>:W1}
  & &
 \hspace*{-25pt}
  \left(k^2 + ik^{\mu} \,
 \frac{\partial}{\partial X^{\mu}}\right) d^{\mgl}_{ij}=
  \pi^{+}_{in}\,d^{\mgl}_{nj}
  + \pi^{\mgl}_{in}\,d^{-}_{nj}
  -
 \, \frac{i}{2} \left(
 \left\{\pi^{+}_{in}\, ,d^{\mgl}_{nj}\right\}
 + \left\{ \pi^{\mgl}_{in}\, ,d^{-}_{nj}\right\}
 \right),
 \\[10pt]
   \label{d<>:W2}
  & &
  \hspace*{-25pt}
  \left(k^2 - ik^{\mu} \,
 \frac{\partial}{\partial X^{\mu}}\right) d^{\mgl}_{ij}=
 d^{\mgl}_{in}\, \pi^{-}_{nj}
  +d^{+}_{in}\, \pi^{\mgl}_{nj}
  -
 \, \frac{i}{2} \left(
  \left\{d^{\mgl}_{in}\, , \pi^{-}_{nj}\right\}
  + \left\{d^{+}_{in}\, ,\pi^{\mgl}_{nj}\right\}
 \right),
 \\[10pt]
  & &
  \hspace*{-25pt}
  \label{d+-:W1}
    \left(k^2 + ik^{\mu} \,
 \frac{\partial}{\partial X^{\mu}}\right) d^{\pm}_{ij}=
 \delta^{}_{ij} - \frac{k^{}_{i} k^{}_{j}}{|\vec{k}|^2}
 +  \pi^{\pm}_{in}\,d^{\pm}_{nj}
   -\, \frac{i}{2} \left\{ \pi^{\pm}_{in}\, ,d^{\pm}_{nj}\right\},
   \\[10pt]
  & &
  \label{d+-:W2}
  \hspace*{-25pt}
    \left(k^2 - ik^{\mu} \,
 \frac{\partial}{\partial X^{\mu}}\right) d^{\pm}_{ij}=
 \delta^{}_{ij} - \frac{k^{}_{i} k^{}_{j}}{|\vec{k}|^2}
 +  d^{\pm}_{in}\,\pi^{\pm}_{nj}
   -\, \frac{i}{2} \left\{ d^{\pm}_{in}\, , \pi^{\pm}_{nj}\right\},
  \end{eqnarray}
where $k^2= k^{\mu}k^{}_{\mu}= k^{2}_{0} -|\vec{k}|^2_{}$.

In the context of kinetic theory, it is natural to interpret
 $k^\mu=(k^0,\vec{k})$ as the photon
four-momentum at the space-time point $X$. However, here one is faced with
a difficulty which is a consequence of the Wigner
transformation~(\ref{W-tr}). To explain this point, let us consider some
transverse tensor $T^{}_{ij}(12)$ satisfying
     \begin{equation}
  \label{Transv}
  \nabla^{}_{1i}\, T^{}_{ij}(12)=0,
 \qquad
    \nabla^{}_{2j}\, T^{}_{ij}(12)=0.
  \end{equation}
Note that the components of $D^{}_{ij}(\und{1}\, \und{2})$ and
$\Pi^{}_{ij}(\und{1}\,\und{2})$ satisfy
 these relations due to the gauge constraint  $\vec{\nabla}\cdot
\hat{\vec{A}}=0$. In terms of the Wigner transforms, Eqs.~(\ref{Transv}) read
 \begin{equation}
  \label{Transv:W}
 \left(\frac{1}{2}\, \frac{\partial}{\partial R^{i}} +
 ik^{i} \right) T^{}_{ij}(X,k)=0,
 \qquad
  \left(\frac{1}{2}\, \frac{\partial}{\partial R^{j}} -
 ik^{j}\right) T^{}_{ij}(X,k)=0.
 \end{equation}
Thus,  in an inhomogeneous medium, $k^{i}T^{}_{ij}(X,k)\not=0$
 and $k^{j} T^{}_{ij}(X,k)\not=0$.  In particular, we conclude that
$d^{\mgl}_{ij}(X,k)$ and $d^{\pm}_{ij}(X,k)$ in
Eqs. (\ref{d<>:W1})\,--\,(\ref{d+-:W2}) contain \textit{longitudinal
parts\/} with respect to $\vec{k}$.
 However, in Appendix~\ref{App:Flux}
it is shown that, up to first-order $X$-gradients, the energy flux of the
radiation field is completely determined by the \textit{transverse part\/}
 of $d^{<}_{ij}(X,k)$, which, for any
tensor $T^{}_{ij}(X,k)$, is defined as
    \begin{equation}
  \label{Transv:k}
 T^{\perp}_{ij}(X,k)=\Delta^{\perp}_{im}(\vec{k})\,
 T^{}_{mn}(X,k)
  \Delta^{\perp}_{nj}(\vec{k}),
  \end{equation}
where
  \begin{equation}
 \label{projec}
 \Delta^{\perp}_{ij}(\vec{k})=\delta^{}_{ij} -
 {k^{i} k^{j}}/{|\vec{k}|^2}
  \end{equation}
is the transverse projector. Since the longitudinal parts of the field
correlation functions do not contribute to the observable energy flux, it is
reasonable to eliminate them  in Eqs.~(\ref{d<>:W1})\,--\,(\ref{d+-:W2}).
This can be done in the following way.

Let us first show that  any tensor $T^{}_{ij}(X,k)$ satisfying
Eqs.~(\ref{Transv:W}) can be expressed in terms of its transverse
part~(\ref{Transv:k}). We start from the identity (the arguments $X$ and
$k$ are omitted for brevity)
  $$
 T^{}_{ij}=T^{\perp}_{ij} +
  \frac{k^{i} k^{m}}{|\vec{k}|^2}\,T^{}_{mn}\,
 \Delta^{\perp}_{nj} +
 \Delta^{\perp}_{im}\,T^{}_{mn}\,
 \frac{k^{n} k^{j}}{|\vec{k}|^2}+
  \frac{k^{i} k^{j}}{|\vec{k}|^4}\,
 k^{m} T^{}_{mn}k^{n},
  $$
 which follows directly from the obvious relation
 $\delta^{}_{ij}=\Delta^{\perp}_{ij} + k^{i}k^{j}/|\vec{k}|^2$.
The terms $k^{m}T^{}_{mn}$ and $k^{n}T^{}_{mn}$ can then be eliminated
with the help  of Eqs.~(\ref{Transv:W}). After a simple algebra we obtain
 $$
 T^{}_{ij}= T^{\perp}_{ij}
 + \frac{i}{2|\vec{k}|^2}\,\frac{\partial}{\partial R^{m}}
 \left(
  k^{i} T^{}_{mj} -k^{j} T^{}_{im}
 \right)
 - \frac{k^{i} k^{j}}{4|\vec{k}|^4}\,
\frac{\partial^2 T^{}_{mn}}{\partial R^{m}\,\partial R^{n}}.
 $$
 Solving this equation by iteration, one can find $T$ as a
 series in derivatives of $T^{\perp}$. Keeping only  first-order
gradients yields
  \begin{equation}
  \label{Tens-Transv}
 T^{}_{ij}(X,k) = T^{\perp}_{ij}(X,k)
 + \frac{i}{2\,|\vec{k}|^2}\left(
 k^{i}\, \frac{\partial T^{\perp}_{mj}}{\partial R^{m}}
 -
  k^{j}\, \frac{\partial T^{\perp}_{im}}{\partial R^{m}}
 \right).
 \end{equation}
 We next consider the
contraction of two tensors, $T^{}_{ij}(X,k)$ and $Q^{}_{ij}(X,k)$, each of
which satisfies (\ref{Transv:W}). Using Eq.~(\ref{Tens-Transv}), we
 find, up to first-order corrections,
 \begin{eqnarray}
 \label{Contr:W}
  & &
 T^{}_{il}(X,k)Q^{}_{lj}(X,k)=
  T^{\perp}_{il}(X,k)Q^{\perp}_{lj}(X,k)
 \nonumber\\[5pt]
  & &
  \hspace*{100pt}
 {}+\,\frac{i}{2\,|\vec{k}|^2}
 \left(
 k^{i}\,\frac{\partial T^{\perp}_{ml}}{\partial R^{m}}\,
 Q^{\perp}_{lj}
 -k^{j}\, T^{\perp}_{il}
 \,\frac{\partial Q^{\perp}_{lm}}{\partial R^{m}}
  \right).
 \end{eqnarray}
Formulas~(\ref{Tens-Transv}) and (\ref{Contr:W}) allow  one to rewrite
Eqs.~(\ref{d<>:W1})\,--\,(\ref{d+-:W2}) in terms of the transverse parts of
the correlation functions, photon propagators, and polarization matrices.
Since the Poisson brackets  are already first order in $X$-gradients, the
corresponding matrices may be replaced by their transverse parts. Then,
acting on both sides of Eqs.~(\ref{d<>:W1})\,--\,(\ref{d+-:W2}) on the left
and right by the projector~(\ref{projec}), the desired equations can easily
 be obtained. Note that in this way the peculiar gradient terms, such as the
last term in Eq.~(\ref{Contr:W}), are eliminated\footnote{From now on all
$d$- and $\pi$-matrices in the Wigner representation are understood to be
transverse with respect to $\vec{k}$,  although this will not always be
written explicitly.}.

We will not write down the resulting equations
because it is more convenient to deal with the equivalent
\textit{transport\/} and \textit{mass-shell\/}  equations
 for the transverse parts of $d^{\mgl}_{ij}(X,k)$ and
$d^{\pm}_{ij}(X,k)$.
 The transport equations are obtained by taking differences
of Eqs.~(\ref{d<>:W1})\,--\,(\ref{d+-:W2}) and then eliminating the
longitudinal parts of all matrices
 as described above. In condensed matrix notation, the
resulting transport equations read
 \begin{eqnarray}
 \label{d<>:TrEq}
 & &
 \hspace*{-25pt}
  \left\{k^2, d^{\mgl}\right\} -\,
 \frac{1}{2}
 \left( \left\{\pi^{+}_{},d^{\mgl}_{}\right\}^{\perp}
 + \left\{ \pi^{\mgl}_{},d^{-}_{}\right\}^{\perp}
 -   \left\{d^{\mgl}_{}, \pi^{-}_{}\right\}^{\perp}
 -  \left\{d^{+}_{},\pi^{\mgl}_{}\right\}^{\perp}
   \right)
 \nonumber\\[5pt]
 & &
  \hspace*{-20pt}
 {}= \frac{i}{2}
 \left(
 \left[ (\pi^{+} +\pi^{-}), d^{\mgl}\right]^{}_{-}
  -  \left[ (d^{+} + d^{-}), \pi^{\mgl}\right]^{}_{-}
 +
  \left[\pi^{>}, d^{<}\right]^{}_{+} -
  \left[\pi^{<}, d^{>}\right]^{}_{+}
 \right),
 \\[10pt]
 \label{d+-:TrEq}
 & &
 \hspace*{-25pt}
 \left\{k^2, d^{\pm}\right\}
 -\,\frac{1}{2}
 \left(
 \left\{ \pi^{\pm}_{},d^{\pm}_{}\right\}^{\perp}
- \left\{ d^{\pm}_{}, \pi^{\pm}_{}\right\}^{\perp}
 \right) = i  \left[ \pi^{\pm}_{},d^{\pm}_{}\right]_{-},
 \end{eqnarray}
where $[A,B]^{}_{\pm}=AB\pm BA$ are the anticommutators/commutators of
matrices, and the superscript $\perp$ denotes the transverse
 part of a tensor.

Taking  sums of Eqs.~(\ref{d<>:W1})\,--\,(\ref{d+-:W2})
 instead of differences, one obtains  the mass-shell equations that do
 not contain the dominant drift terms:
   \begin{eqnarray}
 \label{d<>:Mass}
 & &
 \hspace*{-25pt}
  k^2\, d^{\mgl} +
 \frac{i}{4}
 \left( \left\{\pi^{+}_{},d^{\mgl}_{}\right\}^{\perp}
 + \left\{ \pi^{\mgl}_{},d^{-}_{}\right\}^{\perp}
 +   \left\{d^{\mgl}_{}, \pi^{-}_{}\right\}^{\perp}
 +  \left\{d^{+}_{},\pi^{\mgl}_{}\right\}^{\perp}
   \right)
 \nonumber\\[5pt]
 & &
  \hspace*{-20pt}
 {}= \frac{1}{4}
 \left(
 \left[ (\pi^{+} +\pi^{-}), d^{\mgl}\right]^{}_{+}
  -  \left[ (d^{+} + d^{-}), \pi^{\mgl}\right]^{}_{+}
 +
  \left[\pi^{>}, d^{<}\right]^{}_{-} -
  \left[\pi^{<}, d^{>}\right]^{}_{-}
 \right),
 \\[10pt]
 \label{d+-:Mass}
 & &
 \hspace*{-25pt}
 k^2\, d^{\pm}
 +\,\frac{i}{4}
 \left(
 \left\{ \pi^{\pm}_{},d^{\pm}_{}\right\}^{\perp}
 + \left\{ d^{\pm}_{}, \pi^{\pm}_{}\right\}^{\perp}
 \right) = \Delta^{\perp}(\vec{k})
 +\,\frac{1}{2}  \left[ \pi^{\pm}_{},d^{\pm}_{}\right]_{+}.
 \end{eqnarray}
In vacuum, Eq.~(\ref{d<>:Mass}) reduces to the mass-shell condition
$d^{\mgl}_{ij}(k)\propto \delta(k^2)$.

By analogy with vacuum QED it is convenient to go over in
Eqs.~(\ref{d<>:TrEq})\,--\,(\ref{d+-:Mass}) to a representation using
photon polarization states.  It seems natural to introduce these states
through a set of  eigenvectors of some Hermitian transverse tensor
 $T^{\perp}_{ij}(X,k)$. The properties of photon modes are closely related to
symmetry properties of the polarization tensor $\pi^{+}_{ij}(12)$
which determine the medium response to the transverse electromagnetic
field~\cite{Dub67}. This tensor can be chosen to define the photon
polarization states.

In the Wigner representation, the tensor $\pi^{+}_{ij}(X,k)$ can
always be written as  a sum
 \begin{equation}
 \label{pi+:dec}
 \pi^{+}_{ij}(X,k)= {\rm Re}\,\pi^{+}_{ij}(X,k)
 + i\,{\rm Im}\,\pi^{+}_{ij}(X,k),
 \end{equation}
where  $ {\rm Re}\,\pi^{+}_{ij}$ and ${\rm Im}\,\pi^{+}_{ij}$ are Hermitian.
Using relation
  $
 \left[\pi^{+}_{ij}(12)\right]^{*}= \pi^{-}_{ji}(21),
 $
we find that
 \begin{equation}
 \label{ReIm:+}
  \begin{array}{l}
 \displaystyle
  {\rm Re}\,\pi^{+}_{ij}(X,k)= \frac{1}{2}
 \left( \pi^{+}_{ij}(X,k) +  \pi^{-}_{ij}(X,k)\right),
 \\[8pt]
  \displaystyle
   {\rm Im}\,\pi^{+}_{ij}(X,k)= \frac{1}{2i}
 \left( \pi^{+}_{ij}(X,k) -  \pi^{-}_{ij}(X,k)\right).
 \end{array}
 \end{equation}
Let $\vec{\epsilon}^{}_{s}(X,k)$ be  the eigenvectors of
 ${\rm Re}\,\pi^{+}_{ij}(X,k)$. Since in general these eigenvectors are complex,
 we may write
 \begin{equation}
  \label{pi+-:gen}
  {\rm Re}\,\pi^{+}_{ij}(X,k)=
 \sum_{s} \epsilon^{}_{si}(X,k) \,\pi^{}_{s}(X,k)\,
 \epsilon^{*}_{sj}(X,k),
 \end{equation}
where $\pi^{}_{s}(X,k)$, ($s=1,2$), are \textit{real\/} eigenvalues. This
does not mean, however, that the \textit{full\/} tensors
$\pi^{\pm}_{ij}(X,k)$, as well as $\pi^{\mgl}_{ij}(X,k)$,
 $d^{\mgl}_{ij}(X,k)$, and  $d^{\pm}_{ij}(X,k)$, are diagonalized by
 the same set of eigenvectors. In an anisotropic medium,
 the nonzero off-diagonal components
of these tensors describe coupling between different polarization states. The
 tensor ${\rm Re}\,\pi^{+}_{ij}(X,k)$ for a weakly coupled relativistic plasma
is considered in Appendix~\ref{App:Dispers}. It is shown that effects of
anisotropy on the photon polarization states are
 inessential for $|k^{}_{0}|\gg\omega^{}_{e}$,
where $\omega^{}_{e}$ is the electron plasma frequency
 [see Eq.~(\ref{omega-e})]. In
what follows this condition is assumed to be fulfilled.
 Note also that, for a weakly coupled plasma,
 the photon polarization states can  be defined through the
 eigenvectors of the tensor ${\rm Re}\,\pi^{+}_{ij}(X,k)$
in which only real parts of the elements  are retained. The corresponding
eigenvectors  are real and satisfy
 \begin{equation}
 \label{ComplRel}
 \vec{\epsilon}^{}_{s}(X,k)\cdot\vec{\epsilon}^{}_{s'}(X,k)=
  \delta^{}_{ss'},
 \qquad
 \sum_{s} \epsilon^{}_{si}(X,k)\epsilon^{}_{sj}(X,k)=
 \delta^{}_{ij} - \frac{k^{}_{i}k^{}_{j}}{|\vec{k}|^2}.
 \end{equation}
In this \textit{principal-axis representation\/}, off-diagonal
 components of the tensors $b^{}_{ij}(X,k)=
\left\{d^{\mgl}_{ij},d^{\pm}_{ij},{\pi}^{\mgl}_{ij},{\pi}^{\pm}_{ij}\right\}$
 are assumed to be small for $|k^{}_{0}|\gg \omega^{}_{e}$, so that all these tensors
 can be  expressed in terms of their diagonal components, $b^{}_{s}(X,k)$, as
\begin{equation}
   \label{expan-pol}
  b^{}_{ij}(X,k)= \sum_{s}\epsilon^{}_{si}(X,k)
  b^{}_{s}(X,k) \epsilon^{}_{sj}(X,k).
  \end{equation}

Calculating the diagonal projections  of matrix equations (\ref{d<>:TrEq})
and (\ref{d<>:Mass}) on the polarization vectors  $\vec{\epsilon}^{}_{s}$,
one obtains the transport and mass-shell equations for the field correlation
functions in the principal-axis representation:
 \begin{subequations}
 \label{Kin-MS}
 \begin{eqnarray}
  \label{Kin}
 & &
  \left\{k^2 - {\rm Re}\,\pi^{+}_{s}, d^{\mgl}_{s}\right\}
 + \left\{{\rm Re}\,d^{+}_{s},\pi^{\mgl}_{s}\right\}
  = i\left(\pi^{>}_{s}\, d^{<}_{s} -
  \pi^{<}_{s}\, d^{>}_{s}\right),
 \\[5pt]
 & &
 \label{MS}
   \left\{ {\rm Im}\,\pi^{+}_{s},d^{\mgl}_{s}\right\} +
 \left\{{\rm Im}\, d^{+}_{s},\pi^{\mgl}_{s}\right\}
 = 2 \left(k^2 - {\rm Re}\,\pi^{+}_{s}\right)\left(
  d^{\mgl}_{s} -
   \left|\, d^{+}_{s}\right|^2 \pi^{\mgl}_{s}
  \right).
 \end{eqnarray}
 \end{subequations}
The same procedure applied to Eqs.~(\ref{d+-:TrEq}) and~(\ref{d+-:Mass})
yields
 \begin{equation}
 \label{d+-:K-MS}
   \left\{k^{2} - \pi{}^{\pm}_{s},d^{\pm}_{s}\right\}=0,
 \qquad
 \left(k^{2} - \pi{}^{\pm}_{s}\right)d^{\pm}_{s}=1.
  \end{equation}
Note that in the mass-shell equation for the photon propagators the
gradient corrections cancel each other.

A few remarks should be made about Eqs.~(\ref{Kin-MS}).
Equation~(\ref{Kin}) may be regarded as a particular case of the
gauge-invariant transport equation derived by Bezzerides and
DuBois~\cite{BezDub72} if  the polarization states are chosen in
accordance with the Coulomb gauge. The mass-shell equation~(\ref{MS}) was
ignored in Ref.~\cite{BezDub72}.
 There is  no reason, however, to do this because the field
correlation functions must satisfy \textit{both\/} equations. The mass-shell
equation is in a sense a constraint for approximations in the transport
equation.

\subsection{Resonant and Virtual Photons in Plasmas}
 \label{Subs:ResVirtPhot}

Let us turn first to Eqs.~(\ref{d+-:K-MS}). From the second (mass-shell)
equation one readily finds the ``explicit'' expression for the photon
propagators:
      \begin{equation}
  \label{d+-:expl}
 d^{\pm}_{s}(X,k)= \frac{1}{k^{2} -
 {\rm Re}\,\pi^{+}_{s}(X,k) \pm i\,k^{}_{0}\Gamma^{}_{s}(X,k)},
 \end{equation}
where
 \begin{equation}
  \label{Damping}
 \Gamma^{}_{s}(X,k)= -\, k^{-1}_{0}\, {\rm Im}\,\pi^{+}_{s}(X,k)
  \end{equation}
is the $k$-dependent damping width for the photon mode. It is clear that the first of
Eqs.~(\ref{d+-:K-MS}) is automatically satisfied due to the identity
$\{A,f(A)\}=0$.

We now consider  Eqs.~(\ref{Kin-MS}).
 From the point of view of kinetic theory
the physical meaning of these equations remains to be seen
 because the correlation
 functions $d^{\mgl}_{s}(X,k)$ involve contributions from the resonant
 (propagating) and virtual photons. Since a kinetic equation describes
only the resonant photons, there is a need to pick out the corresponding
 terms from the field correlation functions.
 Our analysis of this problem parallels
 the approach discussed previously by
\v{S}pi\v{c}ka and Lipavsk\'{y} in the context of nonrelativistic
 solid state physics~\cite{SpLip94,SpLip95}.

Recalling the definition (\ref{Poiss}) of the four-dimensional Poisson
bracket, it is easy to see that
 $\{k^2 - {\rm Re}\,\pi^{+}_{s}, d^{\mgl}_{s}\}$ has the structure
of the drift term  in a kinetic equation for quasiparticles with energies
given by the solution of the dispersion equation
 $k^2 - {\rm Re}\,\pi^{+}_{s}=0$. Analogous observations
 serve as a starting point  for  derivation of Boltzmann-type quasiparticle
 kinetic equations from transport equations for correlation
 functions. In most derivations, the
 peculiar terms like the second term on the left-hand side
 of Eq.~(\ref{Kin}) are ignored. But, as was first noted by Botermans and
 Malfliet~\cite{BotMalf90} (see also~\cite{SpLip94,SpLip95}),
 such terms contribute to the drift. To show this in our case, we
 introduce the notation
$\Delta^{\pm}= k^2 - {\rm Re}\,\pi^{+}_{s} \pm ik^{}_{0}\Gamma^{}_{s}$ and
use Eq.~(\ref{d+-:expl}) to write
    \begin{eqnarray*}
 \left\{{\rm Re}\,d^{+}_{s},\pi^{\mgl}_{s}\right\}&=&
  \frac{1}{2}\left\{ \left(\frac{1}{\Delta^{+}}+
    \frac{1}{\Delta^{-}}\right),\pi^{\mgl}_{s}\right\}
   \\[8pt]
{}&=& -\frac{1}{2}\left\{
  \Delta^{+}, \frac{\pi^{\mgl}_{s}}{\left(\Delta^{+}\right)^2}\right\}
 -\frac{1}{2}\left\{
  \Delta^{-}, \frac{\pi^{\mgl}_{s}}{\left(\Delta^{-}\right)^2}\right\},
  \end{eqnarray*}
or
 \begin{equation}
  \label{pi:trfm}
  \hspace*{-6pt}
  \left\{{\rm Re}\,d^{+}_{s},\pi^{\mgl}_{s}\right\}=\! -\!
 \left\{
  \left( k^{2} - {\rm Re}\,\pi^{+}_{s}\right),
  \pi^{\mgl}_{s}\,{\rm Re}\left(d^{+}_{s}\right)^2
  \right\}
 \!+\!\left\{
  k^{}_{0}\Gamma^{}_{s}, \pi^{\mgl}_{s}\,{\rm Im}\left(d^{+}_{s}\right)^2
  \right\}\! ,
 \end{equation}
where the first term dominates in the case of small damping. Let us now
define new functions $\widetilde{d}^{\,\mgl}_{s}(X,k)$ through the relation
  \begin{equation}
    \label{Res-d:def}
  d^{\mgl}_{s}(X,k)=\widetilde{d}^{\,\mgl}_{s}(X,k) +
  \pi^{\mgl}_{s}(X,k)\, {\rm Re}\! \left[\left( d^{+}_{s}(X,k)\right)^2\right].
  \end{equation}
Inserting expressions (\ref{pi:trfm}) and (\ref{Res-d:def}) into
 Eqs.~(\ref{Kin-MS}), we obtain the transport and mass-shell
equations for $\widetilde{d}^{\,\mgl}_{s}$:
   \begin{subequations}
 \label{Kin-MS2}
 \begin{eqnarray}
  \label{Kin2}
 & &
  \hspace*{-20pt}
 \left\{k^2 - {\rm Re}\,\pi^{+}_{s},\widetilde{d}^{\,\mgl}_{s}
 \right\} + \left\{ k^{}_{0}\Gamma^{}_{s},
   \pi^{\mgl}_{s}\,{\rm Im}\left( d^{+}_{s}\right)^2 \right\}
   = i \left( \pi^{>}_{s}\, \widetilde{d}^{\, <}_{s}
   - \pi^{<}_{s}\, \widetilde{d}^{\, >}_{s}  \right),
 \\[5pt]
 \label{MS2}
 & &
 \hspace*{-20pt}
 \left\{k^{}_{0}\Gamma^{}_{s}, \widetilde{d}^{\,\mgl}_{s}\right\}
  + \left\{k^2 - {\rm Re}\,\pi^{+}_{s},
 \pi^{\mgl}_{s}\,{\rm Im}\left( d^{+}_{s}\right)^2   \right\}
 + 2\left\{k^{}_{0}\Gamma^{}_{s},
 \pi^{\mgl}_{s}\,{\rm Re} \left( d^{+}_{s}\right)^2   \right\}
  \nonumber\\[10pt]
   & &
  \hspace*{60pt}
 {}  =-2 \left(k^2 - {\rm Re}\,\pi^{+}_{s}   \right)
  \left( \widetilde{d}^{\,\mgl}_{s}
  - 2 \left|\,d^{+}_{s}\right|^4\left( k^{}_{0}\Gamma^{}_{s}\right)^2
  \pi^{\mgl}_{s}   \right).
 \end{eqnarray}
\end{subequations}
 We wish to emphasize that the contributions from the
last term of Eq.~(\ref{Res-d:def}) to the
right-hand side of Eq.~(\ref{Kin2}) cancel identically.

Before going any further, it is instructive to consider the spectral
properties of
 $\widetilde{d}^{\,\mgl}_{s}(X,k)$. First we recall the conventional
\textit{full spectral function\/} which is defined in terms of correlation
functions or propagators
 (see, e.g.,~\cite{BotMalf90}). For the photon modes,
 the full spectral function is given by
     \begin{equation}
  \label{SpF-def}
 a^{}_{s}(X,k)= i\left(d^{>}_{s} - d^{<}_{s}\right)
 =i\left(d^{+}_{s} - d^{-}_{s}   \right).
  \end{equation}
In the first gradient approximation, $a^{}_{s}(X,k)$ is obtained from
Eq.~(\ref{d+-:expl}):
      \begin{equation}
  \label{SpF-exp}
 a^{}_{s}(X,k)=
 \frac{2k^{}_{0}\Gamma^{}_{s}}
 {\left( k^2 - {\rm Re}\,\pi^{+}_{s}\right)^2 +
 \left( k^{}_{0}\Gamma^{}_{s}\right)^2}.
 \end{equation}
We now introduce the quantity
  \begin{equation}
   \label{SpF-def2}
 \widetilde{a}^{}_{s}(X,k)= i\left(\widetilde{d}^{\,>}_{s} -
 \widetilde{d}^{\,<}_{s}\right)
  \end{equation}
which will be referred to as the \textit{resonant spectral function}.
 Using Eqs.~(\ref{Res-d:def}), (\ref{SpF-exp}), and relations
   \begin{equation}
  \label{pi<>:+-}
 \pi^{>}_{s}- \pi^{<}_{s}=
 \pi^{+}_{s} -\pi^{-}_{s}= -2ik^{}_{0}\Gamma^{}_{s},
 \end{equation}
we find
      \begin{equation}
  \label{SpF-res}
 \widetilde{a}^{}_{s}(X,k)=
 \frac{4 \left( k^{}_{0}\Gamma^{}_{s}\right)^3}{\left[\left(
  k^2 - {\rm Re}\,\pi^{+}_{s}\right)^2 +
 \left( k^{}_{0}\Gamma^{}_{s}\right)^2
 \right]^2}.
  \end{equation}
When this expression is compared with (\ref{SpF-exp}), two important
observations can be made. First, it is easy to check that both spectral
functions take the same form in the zero damping limit:
   \begin{equation}
 \label{aa:lim}
 \lim_{\Gamma^{}_{s}\to 0} a^{}_{s}=
\lim_{\Gamma^{}_{s}\to 0} \widetilde{a}^{}_{s}=
 2\pi \eta(k^{}_{0})\,\delta\!\left(k^2 - {\rm Re}\,\pi^{+}_{s}\right),
 \end{equation}
where $\eta(k^{}_{0})= k^{}_{0}/|k^{}_{0}|$. Second, for a finite
$\Gamma^{}_{s}$, the resonant spectral function (\ref{SpF-res}) falls off
faster than the full spectral function (\ref{SpF-exp}). Thus, for weakly
damped field excitations, the first term in Eq.~(\ref{Res-d:def})
dominates in the vicinity of the photon mass-shell ($k^2\approx {\rm
Re}\,\pi^{+}_{s}$), while the second term
  dominates in the off-shell region where it falls as
 $(k^2 - {\rm Re}\,\pi^{+}_{s})^{-2}$. In other words,
  $\widetilde{a}^{}_{s}$ has a stronger peak and smaller wings than
  ${a}^{}_{s}$. We see that physically the first term in
  Eq.~(\ref{Res-d:def}) may be regarded as
 the ``resonant'' part of the field correlation functions.
 The second term represents the ``off-shell'' part which  may be
identified as the contribution of \textit{virtual\/} photons\footnote{We
would like to emphasize that this interpretation
 makes sense only in the case of small damping.
 Formally, Eq.~(\ref{Res-d:def}) in itself is not related to any
interpretation.}.

It is interesting to note that the spectral
 function~(\ref{SpF-res}) also occurs in the self-consistent
 calculation of
 thermodynamic quantities of \textit{equilibrium\/} QED plasmas.
  As shown by Vanderheyden and Baym~\cite{VandBaym98},
  the resonant spectral function rather than the
  Lorentzian-like full spectral function
   (\ref{SpF-exp}) determines the contribution of photons to
   the equilibrium entropy. That is why in Ref.~\cite{VandBaym98}
   $\widetilde{a}^{}_{s}$ was called ``the entropy spectral function''.
   We have seen, however, that this function has a deeper physical
 meaning; it characterizes the spectral properties of resonant
(propagating) photons in a medium. Therefore we prefer to call
   $\widetilde{a}^{}_{s}$  the resonant spectral function.

\subsection{Kinetic Equation for Resonant Photons}
 \label{Subs:KinEqPhot}
 On the basis of the above considerations,
 it is reasonable to formulate
 a kinetic description of resonant photons in terms of
two functions $N^{\mgl}(X,k)$ defined through the relations
        \begin{equation}
   \label{SpF}
  \hspace*{-10pt}
 \widetilde{d}^{\,<}_{s}(X,k)=
   -i\, \widetilde{a}^{}_{s}(X,k) N^{<}_{s}(X,k),
   \quad
 \widetilde{d}^{\,>}_{s}(X,k)=
   -i\, \widetilde{a}^{}_{s}(X,k)N^{>}_{s}(X,k),
   \end{equation}
where
  \begin{equation}
   \label{N<>}
  N^{>}_{s}(X,k) - N^{<}_{s}(X,k)=1.
  \end{equation}
It is also convenient to use the representation
 \begin{eqnarray}
   \label{N(p)}
   & &
   N^{<}_{s}(X,k)=\theta(k^{}_{0})\,N^{}_{s}(X,k) -
   \theta(-k^{}_{0})\,\left[1+N^{}_{s}(X,-k)\right],
   \nonumber\\[5pt]
   & &
  N^{>}_{s}(X,k)=\theta(k^{}_{0}\left[1+N^{}_{s}(X,k)\right]
  - \theta(-k^{}_{0})\,N^{}_{s}(X,-k),
 \end{eqnarray}
 which serve as the definition of the local photon distribution function
  $N^{}_{s}(X,k)$ in four-dimensional phase space.

We will now convert Eq.~(\ref{Kin2})  into a kinetic equation for
 $N^{\mgl}_{s}(X,k)$.
It suffices to consider only the equation for $\widetilde{d}^{\,<}_{s}$
since the transport equation for $\widetilde{d}^{\,>}_{s}$ leads to the
same kinetic equation. Substituting $\widetilde{d}^{\,<}_{s}= -i\,
\widetilde{a}^{}_{s} N^{<}_{s}$
 into Eq.~(\ref{Kin2}) for $\widetilde{d}^{\,<}_{s}$ and calculating the first
  Poisson bracket on the left-hand side, we obtain a
 peculiar drift term
 $N^{<}_{s}\!\left\{k^2 - {\rm
     Re}\,\pi^{+}_{s},\widetilde{a}^{}_{s}\right\}$. Here we
shall follow
 Botermans and Malfliet~\cite{BotMalf90} who  have shown how
this term can be
 eliminated by another
peculiar term coming from the second Poisson
 bracket\footnote{The original arguments presented by Botermans and Malfliet
 refer to non-relativistic systems, but they are in essence applicable
 to any kinetic equation
 derived from the transport equations.}.
 Since Eq.~(\ref{Kin2}) itself  is correct to
first-order $X$-gradients, in calculating Poisson brackets the polarization
functions $\pi^{\mgl}_{s}$ may be expressed in terms of the correlation
functions from the local balance relation
 \begin{equation}
  \label{Balance}
  \pi^{>}_{s}\, \widetilde{d}^{\, <}_{s}
   - \pi^{<}_{s}\, \widetilde{d}^{\, >}_{s} = 0.
  \end{equation}
With the help of Eqs.~(\ref{SpF-def2}), (\ref{pi<>:+-}), and (\ref{SpF})
 we find that
 \begin{equation}
  \label{Balance1}
 \pi^{\mgl}_{s}= -2i k^{}_{0}\Gamma^{}_{s} N^{\mgl}_{s}.
  \end{equation}
This expression for $\pi^{<}_{s}$ can now be used
to calculate the second Poisson bracket
in Eq.~(\ref{Kin2}). A straightforward algebra then leads to
    \begin{equation}
   \label{Kin1}
 \widetilde{a}^{}_{s}
 \left[
 \left\{k^2 - {\rm Re}\,\pi^{+}_{s},N^{<}_{s}\right\}
  - \frac{k^2 - {\rm Re}\,\pi^{+}_{s}}{k^{}_{0}\Gamma^{}_{s}}
  \left\{ k^{}_{0}\Gamma^{}_{s},N^{<}_{s} \right\} -
  i\left(\pi^{>}_{s} N^{<}_{s} -
  \pi^{<}_{s} N^{>}_{s}\right)
 \right]=0.
 \end{equation}
The desired kinetic equation is obtained  by setting the expression in
square brackets equal to zero. We now apply the same procedure to the
mass-shell equation~(\ref{MS2}). As a result of simple manipulations we get
  \begin{equation}
  \label{}
 \left(k^2 - {\rm Re}\,\pi^{+}_{s}\right) \widetilde{a}^{}_{s}
 \left[ \ldots\right]=0
  \end{equation}
with the same expression in square brackets as in Eq.~(\ref{Kin1}). We see
that in the present approach the mass-shell equation is consistent with the
kinetic equation.

 For weakly damped photons,
 $\widetilde{a}^{}_{s}$ is a sharply peaked function of $k^{}_{0}$
near the effective photon frequencies, $\omega^{}_{s}(X,\vec{k})$, which are
solutions of the dispersion equation
  \begin{equation}
  \label{DispEq}
 k^2 -  {\rm Re}\,\pi^{+}_{s}(X,k)=0.
 \end{equation}
  For definiteness, it will be assumed that the photon
frequencies are  \textit{positive\/} solutions. If
$k^{}_{0}=\omega^{}_{s}(X,\vec{k})$ is such a solution, then, using the
property
 $\left[\pi^{+}_{s}(k)\right]^*= \pi^{+}_{s}(-k)$,  it is easy
 to see that the corresponding negative solution is
$k^{}_{0}=-\omega^{}_{s}(X,-\vec{k})$. A more
 detailed discussion of Eq.~(\ref{DispEq})
 is given in Appendix~\ref{App:Dispers}.
In the small damping limit, the resonant
spectral function may be approximated as [see Eq.~(\ref{aa:lim})]
   \begin{equation}
 \label{a:free}
 \widetilde{a}^{}_{s}(X,k)=
 2\pi\, \eta(k^{}_{0})\,
 \delta\!\left(k^2 - {\rm Re}\,\pi^{+}_{s}\right).
 \end{equation}
Then, integrating Eq.~(\ref{Kin1})  over $k^{}_{0}>0$, we arrive at the
 kinetic equation
   \begin{equation}
  \label{Kin:MS}
 \left(\frac{\partial}{\partial T} +
  \frac{\partial \omega^{}_{s}}{\partial \vec{k}}\cdot
 \frac{\partial}{\partial \vec{R}}
 -\frac{\partial\omega^{}_{s} }{\partial \vec{R}}
\cdot \frac{\partial}{\partial\vec{k}}\right)n^{}_{s}(X,\vec{k})=
 I^{\text{(emiss)}}_{s}(X,\vec{k}) -
 I^{\text{(abs)}}_{s}(X,\vec{k})
 \end{equation}
with the \textit{on-shell\/} photon distribution function
  \begin{equation}
 \label{OccN:def}
  \left. n^{}_{s}(X,\vec{k})= {N}^{}_{s}(X,k)
  \right|^{}_{k^{}_{0}=\omega^{}_{s}(X,\vec{k})}.
  \end{equation}
The first and the second terms on the right-hand side
 of Eq.~(\ref{Kin:MS})
 are respectively the photon emission and absorption rates:
 \begin{subequations}
 \label{prod-abs}
  \begin{eqnarray}
 \label{prod}
  & &
  I^{\text{(emiss)}}_{s}(X,\vec{k})=
 iZ^{}_{s}(X,\vec{k})\,\pi^{<}_{s}(X,\vec{k})
 \left[ 1+ n^{}_{s}(X,\vec{k}) \right],
 \\[5pt]
  \label{abs}
  & &
  I^{\text{(abs)}}_{s}(X,\vec{k})=
  iZ^{}_{s}(X,\vec{k})\,\pi^{>}_{s}(X,\vec{k})\,
  n^{}_{s}(X,\vec{k}),
 \end{eqnarray}
  \end{subequations}
where
 \begin{equation}
 \label{pi<>:k}
 \pi^{\mgl}_{s}(X,\vec{k}) = \left.
 \pi^{\mgl}_{s}(X,k)\right|^{}_{k^{}_{0}=\omega^{}_{s}(X,\vec{k})},
 \end{equation}
and $Z^{}_{s}$ is given by
 \begin{equation}
  \label{Z}
 Z^{-1}_{s}(X,\vec{k})= \left.
\frac{\partial}{\partial k^{}_{0}} \left(k^2 - {\rm
Re}\,\pi^{+}_{s}(X,k)\right)
 \right|^{}_{k^{}_{0}=\omega^{}_{s}(X,\vec{k})}.
 \end{equation}
The photon kinetic equation (\ref{Kin:MS}) with emission and absorption rates
 (\ref{prod-abs}) was derived many years ago by DuBois~\cite{Dub67}
for nonrelativistic plasmas. The generalization to relativistic plasmas is
obvious until the polarization functions $\pi^{\mgl}_{s}(X,k)$ are
specified. There was a reason, however, to discuss here the derivation of
the photon kinetic equation. As we have seen, the
 kinetic equation is only related to the \textit{resonant parts\/} of the
 transverse field correlation functions while their
 \textit{off-shell parts\/} are represented by the last term
  in Eq.~(\ref{Res-d:def}). We shall see shortly that these off-shell parts
 must be taken into account in calculating the emission and absorption
 rates.  Note in this connection that
 the photon distribution function is usually
 introduced through the relations
 $d^{\mgl}_{s}=-ia^{}_{s} N^{\mgl}_{s}$ with the
\textit{full\/} spectral function $a^{}_{s}(X,k)$ which is then taken in the
  singular form (\ref{a:free}) (see, e.g.,~\cite{Dub67}).
 In doing so, the off-shell
corrections  to $d^{\mgl}_{s}$ are missing.

 \setcounter{equation}{0}

\section{Electron Propagators and Correlation Functions}
    \label{Sec:Electrons}
  As seen from Eqs.~(\ref{prod-abs}),
the transverse polarization functions $\pi^{\mgl}_{s}(X,k)$ play a key role
in computing radiation effects. Before going into a detailed analysis of
these quantities, we will discuss some features of the electron Green's
function $G(\und{1}\,\und{2})$ and its components, the propagators
 $G^{\pm}(12)$ and the correlation functions $G^{\mgl}(12)$, which
are the necessary ingredients to calculate the polarization
 matrix~(\ref{Pi-mn:gen}).

 \subsection{Electron Propagators}
 Using Dyson's equations (\ref{Dyson-e}) and
 (\ref{Dyson-e:adj}) together with
  the canonical representation~(\ref{F:can}) of the electron Green's function
  $G(\und{1}\,\und{2})$ and the self-energy  $\Sigma(\und{1}\,\und{2})$,
  one can derive the equations of motion for the
  retarded/advanced electron propagators:
     \begin{subequations}
\label{Prop-eq12}
  \begin{eqnarray}
  \label{Prop-eq1}
   & &
  \left(i\!\mynot{\partial}^{}_{1} - e\mmynot{A}(1) -m \right)
 G^{\pm}(12) = \delta(1 -2)
  +
   \Sigma^{\pm}_{}(1 1')\,G^{\pm}(1'2),
    \\[5pt]
 \label{Prop-eq2}
   & &
   G^{\pm}(12)
   \left(-i\!
   \stackrel{\leftarrow}{\mynot{\partial}}^{}_{2} - e\mmynot{A}(2) -m \right)
  = \delta(1 - 2)
  +  G^{\pm}(1 1')\,\Sigma^{\pm}_{}(1' 2).
 \end{eqnarray}
  \end{subequations}
 The derivation follows  a standard way (see, e.g.,
Ref.~\cite{BotMalf90}), so that we do not repeat the algebra here.
  Going over to the Wigner
representation and keeping only linear terms in the gradient expansion, we
obtain
    \begin{equation}
 \label{G+-:W}
   (g^{\pm})^{-1} G^{\pm}=1 +
 \frac{i}{2}\left\{(g^{\pm})^{-1},G^{\pm}\right\},
 \qquad
   G^{\pm}(g^{\pm})^{-1}=1 +
 \frac{i}{2}\left\{G^{\pm},(g^{\pm})^{-1}\right\},
   \end{equation}
where $G^{\pm}=G^{\pm}(X,p)$, and we have introduced the \textit{local\/}
 propagators
    \begin{equation}
  \label{g+-}
 g^{\pm}(X,p)= \frac{1}{\mmynot{\mPi}(X,p) -m
 -\Sigma^{\pm}(X,p)}
  \end{equation}
 with the kinematic momentum $\mPi^{\mu}(X,p)=p^{\mu} -e A^{\mu}(X)$. Up to
 first-order terms in $X$-gradients,  we find from Eqs.~(\ref{G+-:W})
  two formally different ``explicit'' expressions for the propagators:
      \begin{equation}
   \label{G+-}
 G^{\pm}(X,p)= g^{\pm} +
 \frac{i}{2}\, g^{\pm} \left\{(g^{\pm})^{-1}, g^{\pm}\right\},
   \quad
  G^{\pm}(X,p)= g^{\pm} +
 \frac{i}{2}\left\{g^{\pm},(g^{\pm})^{-1}\right\}g^{\pm}.
  \end{equation}
It is easy to verify, however,  that these expressions are equivalent to each
other due to the identity
  \begin{equation}
  \label{ident:1}
 A\left\{A^{-1}, A \right\}= \left\{A,A^{-1}\right\} A \, ,
  \end{equation}
which holds for any matrix $A(X,p)$ and follows directly from the definition
(\ref{Poiss}) of the Poisson bracket. The first-order gradient terms in
 Eqs.~(\ref{G+-}) arise due to the matrix structure of the relativistic
  propagators (\ref{g+-}). These terms are zero for \textit{scalar\/}
 propagators since, in the latter case,
  $\{(g^{\pm})^{-1}, g^{\pm}\}= \{g^{\pm},(g^{\pm})^{-1}\}=0$.
Note, however, that the gradient corrections to the electron propagators may
be neglected when evaluating
 \textit{local\/} quantities, say, the polarization functions
  $\pi^{\mgl}_{s}(X,k)$ that enter the photon emission and absorption rates.

The local propagators (\ref{g+-}) have in general a rather complicated spinor
 structure due to the presence of the matrix self-energies $\Sigma^{\pm}(X,p)$.
Any spinor-dependent quantity ${\mathcal Q}$ can be expanded in a complete
basis in spinor space as~\cite{ItzZuber80}
  \beq
    \label{Spinor:dec}
 {\mathcal Q}=
  I Q^{}_{(S)}+
   \gamma^{}_{\mu} Q^{\mu}_{(V)}+
   \gamma^{}_{5} Q^{}_{(P)} +
   \gamma^{}_{5}\gamma^{}_{\mu} Q^{\mu}_{(A)}+
   \frac{1}{2}\,\sigma^{}_{\mu\nu} Q^{\mu\nu}_{(T)},
  \eeq
where $I$ is the unit matrix, and $Q^{}_{(S)}$, $Q^{\mu}_{(V)}$,
 $Q^{}_{(P)}$, $Q^{\mu}_{(A)}$, $Q^{\mu\nu}_{(T)}$ are the scalar, vector,
 pseudo-scalar, axial-vector, and tensor components. The decomposition
  (\ref{Spinor:dec}) can in principle be obtained for the local propagators
  (\ref{g+-}) in terms of the components of the self-energies. It is clear,
  however,
  that deriving the propagators along this way is generally rather complicated.
The situation becomes simpler for equal probabilities of the spin
polarization, when a reasonable approximation is to retain only the first
two terms in the decomposition~(\ref{Spinor:dec}) of the
 self-energies for relativistic fermions\footnote{For nuclear matter an analogous
 approximation was discussed, e.g., in~\cite{BotMalf90,MrowHeinz94}.}.
 The local propagators (\ref{g+-}) then become
  \beq
    \label{g+-:appr}
    g^{\pm}(X,p)=
     \frac{\mynot{P}^{\pm}+M^{\pm}}{(P^{\pm})^2-(M^{\pm})^2}
  \eeq
with
 \beq
    \label{PM}
 P^{\pm\,\mu}=\mPi^{\mu} -\Sigma^{\pm\,\mu}_{(V)},
 \qquad
 M^{\pm}= m +\Sigma^{\pm}_{(S)}.
 \eeq
A very simplified  version of this approximation,
 which is used sometimes in thermal field theory, is
the ansatz~\cite{QuackHenning96,Aurenche97,Aurenche00}
  \beq
    \label{Sig+-:Gamma}
 \Sigma^{\pm}(X,p)= \mp
 i(\Gamma^{}_{p}/2)\gamma^0,
  \eeq
 where $\Gamma^{}_{p}$ is an adjustable ``spectral width parameter''.

\subsection{Quasiparticle and Off-Shell Parts of Correlation Functions}
 \label{Subsec:QPCorr}
 We now consider the electron (positron) correlation functions $G^{\mgl}$.
 Following the standard scheme~\cite{BotMalf90}, we obtain
 from Eqs.~(\ref{Dyson-e}) and (\ref{Dyson-e:adj}) the KB equations
 for the correlation functions:
   \begin{subequations}
\label{KB-12}
  \begin{eqnarray}
  \label{KB-1}
  & &
   \hspace*{-20pt}
    \big(i\!\mynot{\partial}^{}_{1} - e\mmynot{A}(1) -m \big)G^{\mgl}(1 2)
  = \Sigma^{+}(1 1')\,G^{\mgl}(1' 2) + \Sigma^{\mgl}(1 1')\,G^{-}(1'2),
  \\[5pt]
 \label{KB-2}
   & &
   \hspace*{-20pt}
  G^{\mgl}(1 2)
   \big(\!-
   i\!\stackrel{\leftarrow}{\mynot{\partial}}^{}_{\!2}-e\mmynot{A}(2) -m \big)
   = G^{\mgl}(11')\, \Sigma^{-}(1'2) + G^{+}(11')\,\Sigma^{\mgl}(1'2).
 \end{eqnarray}
 \end{subequations}
As in the case of transverse photons, it is natural to expect that the
electron correlation functions $G^{\mgl}(X,p)$ contain sharply peaked
``quasiparticle'' parts and off-shell parts. To separate these
contributions, we have to analyze drift terms in the transport equations
which follow from the KB equations~(\ref{KB-12}).
 This is detailed in
Appendix~\ref{App:Electrons} (see also Ref.~\cite{MorRoep06}). In the local
form, the decomposition of the electron correlation functions reads
(cf.~Eq.~(\ref{Res-d:def}) for photons)
    \begin{eqnarray}
   \label{G<>:dec}
   & &
   \hspace*{-15pt}
  G^{\mgl}(X,p)=\widetilde{G}^{\mgl}(X,p)
  \nonumber\\
  & &
  {}+
  \frac{1}{2}\left[g^{+}(X,p)\,\Sigma^{\mgl}(X,p)\,g^{+}(X,p) +
   g^{-}(X,p)\,\Sigma^{\mgl}(X,p)\,g^{-}(X,p)\right].
 \end{eqnarray}
 To show that $\widetilde{G}^{\mgl}(X,p)$ may be interpreted
 as the quasiparticle parts of the electron correlation functions,
 it is instructive to consider their spectral
 properties.
 The \textit{full spectral function\/} in spinor space is defined
as~\cite{BotMalf90,MrowHeinz94}
 \begin{equation}
 \label{SF-1}
 {\mathcal A}(X,p)=  i \left(G^{>}(X,p)- G^{<}(X,p)\right)=
  i \left(G^{+}(X,p)- G^{-}(X,p)\right).
 \end{equation}
In the local approximation $G^{\pm}(X,p)=g^{\pm}(X,p)$. Then,
recalling Eq.~(\ref{g+-}), we find that
   \begin{equation}
 \label{SF}
 {\mathcal A}(X,p)=
 i\, g^{+}\Delta\Sigma\,g^{-},
 \end{equation}
where $ \Delta\Sigma(X,p)= \Sigma^{+}(X,p) - \Sigma^{-}(X,p)$. We now
introduce the \textit{quasiparticle spectral function\/} associated with
$\widetilde{G}^{\mgl}(X,p)$:
  \begin{equation}
  \label{QSF-1}
 \widetilde{\mathcal A}(X,p)=
  i \left(\widetilde{G}^{>}(X,p)- \widetilde{G}^{<}(X,p)\right).
  \end{equation}
Using Eqs.~(\ref{G<>:dec}), (\ref{SF}), and the relation $\Sigma^{>} -
\Sigma^{<}= \Sigma^{+}-\Sigma^{-}$, a simple algebra gives
    \begin{equation}
  \label{QSF}
 \widetilde{\mathcal A}(X,p)=
 - \frac{i}{2}\,g^{+}\Delta\Sigma\,g^{+}
 \Delta\Sigma\,g^{-} \Delta\Sigma\,g^{-}.
 \end{equation}
 To gain some insight into an important
 difference between the spectral functions
${\mathcal A}(X,p)$ and $ \widetilde{\mathcal A}(X,p)$,
 let us consider the zero
 damping limit where the propagators (\ref{g+-}) reduce to
     \begin{equation}
   \label{g+-:free}
  g^{\pm}_{0}(X,p)=
  \frac{1}{\mmynot{\mPi} -m \pm i\! \mynot{\varepsilon}}
  \end{equation}
  with $\varepsilon^{\mu}=(\varepsilon,0,0,0)$, $\varepsilon\to +0$. The
corresponding retarded/advanced self-energies are $\Sigma^{\pm}=\mp i
 \varepsilon \gamma^0$, so that $\Delta\Sigma=-i\Gamma\gamma^0$, where
 $\Gamma=2\varepsilon$ is the infinitesimally small spectral width.
  Then some spinor algebra in Eqs.~(\ref{SF}) and (\ref{QSF})
   leads to the following prelimit expressions:
    \begin{eqnarray}
     \label{A:free}
     & &
 \hspace*{-15pt}
 {\mathcal A}(X,p)= \frac{2 \mPi^{}_{0}\Gamma}{
  \left(\mPi^{2}_{0} - {\mathcal E}^{2}_{p}\right)^2 +
 \left(\mPi^{}_{0}\Gamma\right)^2}
 \left[\,\mmynot{\mPi} +m + \frac{{\mathcal E}^{2}_{p}
 -\mPi^{2}_{0}}{2\mPi^{}_{0}}\,\gamma^{0}\right],
  \\[10pt]
    \label{QSF-free}
   & &
   \hspace*{-15pt}
 \widetilde{\mathcal A}(X,p)=
   \frac{4\left(\mPi^{}_0\Gamma\right)^3}
        {\left[ \left(\mPi_{0}^2 -{\mathcal E}^{2}_{p}\right)^2
          +\left(\mPi^{}_0\Gamma\right)^2 \right]^2}
 \left[
  \frac{{\mathcal E}^{2}_{p} + \mPi^{2}_{0}}{2\mPi^{2}_{0}}
   \left(\,\mmynot{\mPi} +m \right)
  \right.
   \nonumber\\[5pt]
   & &
   \hspace*{160pt}
   {}+
   \left.
   \frac{{\mathcal E}^{2}_{p} - \mPi^{2}_{0}}{2\mPi^{}_{0}}
   \left(1 + \frac{{\mathcal E}^{2}_{p} - \mPi^{2}_{0}}
                  {4\mPi^{2}_{0}}
   \right)\gamma^0
  \right],
  \end{eqnarray}
where ${\mathcal E}^{}_{p}(X)=\sqrt{\vec{\mPi}^2 + m^2}$. In the limit
 $\Gamma\to 0$ both spectral functions take the same form:
   \begin{equation}
  \label{A:lim}
 \lim_{\Gamma\to 0} \widetilde{\mathcal A}=
  \lim_{\Gamma\to 0} {\mathcal A}
 =
  2\pi\,\eta(\mPi^{}_{0})\,\delta(\mPi^2 -m^2)
 \left(\,\mmynot{\mPi} +m \right),
  \end{equation}
where $\eta(\mPi^{}_{0})=\mPi^{}_{0}/|\mPi^{}_{0}|\,$. Note, however, that
for finite  but small $\Gamma$, the quasiparticle spectral
function~(\ref{QSF-free})
 falls off faster than the full spectral function~(\ref{A:free}).
 In other words, the ``collisional broadening'' of the quasiparticle spectral
 function is considerably smaller than that of the full spectral function.
 This has much in common with the properties of the resonant and the full
 spectral functions for the transverse
  field fluctuations, Eqs.~(\ref{SpF-exp}) and (\ref{SpF-res}).

\subsection{Distribution Functions of Electrons and Positrons}
  \label{Subsec:Distrib}
In analogy with the kinetic description of resonant photons in
 Subsection~\ref{Subs:ResVirtPhot}, it is reasonable to introduce the
 electron (positron) distribution functions through relations
 between $\widetilde{G}^{\mgl}(X,p)$ and the quasiparticle spectral function
  (\ref{QSF}). Before we go into a discussion of this point, we will briefly
  touch upon the hermicity properties of $\widetilde{G}^{\mgl}(X,p)$ and
 $\widetilde{\mathcal A}(X,p)$.
   Note that the electron correlation, propagators, and self-energies,
 as spinor matrices, satisfy the relations
 \begin{subequations}
 \label{Herm}
 \begin{eqnarray}
\label{Herm:GSig<>} \hspace*{-20pt}
 &
 \left[G^{\mgl}(X,p)\right]^{\dagger}=
 -\gamma^{0} G^{\mgl}(X,p)\gamma^0,
 &
  \quad
 \left[\Sigma^{\mgl}(X,p)\right]^{\dagger}=
 -\gamma^{0} \Sigma^{\mgl}(X,p)\gamma^0,\\[5pt]
 \label{Herm:GSig+-}
  \hspace*{-20pt}
  &
  \hspace*{-6.5pt}
 \left[G^{\pm}(X,p)\right]^{\dagger}=
 \gamma^{0} G^{\mp}(X,p)\gamma^0,
 &
 \quad
 \left[\Sigma^{\pm}(X,p)\right]^{\dagger}=
 \gamma^{0} \Sigma^{\mp}(X,p)\gamma^0,
 \end{eqnarray}
 \end{subequations}
which follow directly from the definition of these quantities.
 Recalling (\ref{g+-}), it is easy to see that
the local propagators $g^{\pm}(X,p)$ satisfy the same relations as
$G^{\pm}(X,p)$.
 Then one derives from Eq.~(\ref{G<>:dec})
 \begin{equation}
 \label{Herm:QPG}
  \left[\widetilde{G}^{\mgl}(X,p)\right]^{\dagger}=
 -\gamma^{0} \widetilde{G}^{\mgl}(X,p)\gamma^0.
 \end{equation}
This immediately leads to the following property of the quasiparticle
spectral function (\ref{QSF-1}):
 \begin{equation}
 \label{Herm:QPSF}
   \left[\widetilde{\mathcal  A}(X,p)\right]^{\dagger}=
 \gamma^{0}\widetilde{\mathcal  A} (X,p)\gamma^0.
 \end{equation}
We now introduce the distribution functions in spinor space,
 ${\mathcal F}^{\mgl}(X,p)$, by the relation
    \begin{equation}
 \label{F<>}
 \widetilde{G}^{\mgl}(X,p)= \mp\, \frac{i}{2}
 \left(\widetilde{\mathcal  A}(X,p)\,{\mathcal F}^{\mgl}(X,p)
 + {\mathcal F}^{\mgl}(X,p)\,\widetilde{\mathcal A}(X,p)\right).
 \end{equation}
Assuming
  \begin{equation}
  \label{F:cond}
{\mathcal F}^{>}_{}(X,p) + {\mathcal F}^{<}_{}(X,p)=I,
 \end{equation}
  it is easy to verify that Eq.~(\ref{QSF-1}) is satisfied,
  whereas Eqs.~(\ref{Herm:QPG}) and (\ref{Herm:QPSF}) require that
      \begin{equation}
  \label{Herm:F<>}
  \left[{\mathcal F}^{\mgl}(X,p)\right]^{\dagger}=
 \gamma^{0} {\mathcal F}^{\mgl}(X,p)\gamma^{0}.
  \end{equation}
This property ensures that all components of  ${\mathcal F}^{\mgl}(X,p)$ in
the expansion (\ref{Spinor:dec}) are real.

The spinor structure of ${\mathcal F}^{\mgl}(X,p)$ can be specified
completely in the zero damping limit where
 $G^{\mgl}(X,p)=\widetilde{G}^{\mgl}(X,p)$, and the quasiparticle spectral
 function is taken in the limiting form (\ref{A:lim}). Then, as shown by
Bezzerides and DuBois~\cite{BezDub72}, in the case of equal probabilities of
the spin polarization\footnote{In what follows we restrict our consideration
to this case. More general distribution functions of relativistic fermions
are discussed, e.g., in Refs.~\cite{BotMalf90,BezDub72,MrowHeinz94}.}, the
matrices ${\mathcal F}^{\mgl}(X,p)$ are diagonal:
    \begin{equation}
   \label{F<>:free1}
\mathcal{F}^{\mgl}(X,p)= I F^{\mgl}(X,p),
 \qquad
   F^{>}(X,p)+F^{<}(X,p)=1,
  \end{equation}
 so that one has
   \begin{equation}
  \label{G<>:free}
   \widetilde{G}^{\mgl}(X,p)=
 \mp\, 2\pi i\,\eta({\mPi}^{0})\,\delta({\mPi}^2-m^2)
 \left(\,\mmynot{\mPi} +m \right)F^{\mgl}(X,p).
  \end{equation}
We would like to make one remark here
 about the kinematic four-momentum
$\mPi^{\mu}(X)=p^{\mu} - eA^{\mu}(X)$ that enters  the above expression and
other formulas. The appearance of the vector potential is due to the fact
that the electron Green's function
 (\ref{G-e}) is not invariant under gauge
 transformation of the mean electromagnetic field. In principle,
  one could work from the beginning with the gauge invariant Green's function
     \begin{equation}
    \label{G:inv}
   {\mathcal G}(\und{1}\,\und{2})=
   G(\und{1}\,\und{2})\,\exp\left[ie\int^{\und{1}}_{\und{2}}
   A^{}_{\mu}(\und{x})\,d\und{x}^{\,\mu}_{} \right],
   \end{equation}
where $\und{x}^{\mu}=(\und{t},\vec{r})$  and integration over $\vec{r}$ is
performed along a straight line connecting the points $\vec{r}^{}_{2}$ and
$\vec{r}^{}_{1}$. In the leading gradient approximation, there are simple
relations between the corresponding propagators and correlation functions:
   \begin{equation}
   \label{inv:non}
 {\mathcal G}^{\pm}(X,p)= g^{\pm}(X,p+ eA(X)),
  \quad
 {\mathcal G}^{\mgl}(X,p)= G^{\mgl}(X,p+ eA(X)).
  \end{equation}
In particular, it follows from Eq.~(\ref{G<>:free}) that the
 gauge invariant
quasiparticle correlation functions in the collisionless approximation are given by
   \begin{equation}
  \label{G<>:free:inv}
   \widetilde{\mathcal G}^{\,\mgl}(X,p)=
 \mp\, 2\pi i\,\eta({p}^{0})\,\delta({p}^2-m^2)
 \left(\mynot{p} +m \right)f^{\mgl}(X,p),
  \end{equation}
where
  \begin{equation}
   \label{f-Ginv}
  f^{\mgl}(X,p)= F^{\mgl}(X,p+ eA(X))
  \end{equation}
are gauge invariant functions. As shown in Ref.~\cite{BezDub72}, they are
related to the on-shell
 distribution functions of electrons
 $(f^{}_{e^-})$ and positrons $(f^{}_{e^+})$:
 \begin{equation}
    \label{f:e+-}
    \hspace*{-10pt}
    \begin{array}{ll}
  \left. f^{<}(X,p)\right|^{}_{p^{0}=E^{}_{p}}= f^{}_{e^-}(X,\vec{p}),
 \quad
  &
 \left. f^{<}(X,p)\right|^{}_{p^{0}=-E^{}_{p}}=
 1-f^{}_{e^+}(X,-\vec{p}),
 \\[3pt]
 \left. f^{>}(X,p)\right|^{}_{p^{0}=E^{}_{p}}=
 1-f^{}_{e^-}(X,\vec{p}),
 \quad
  &
 \left. f^{>}(X,p)\right|^{}_{p^{0}=-E^{}_{p}}=
 f^{}_{e^+}(X,-\vec{p}),
 \end{array}
 \end{equation}
where $E^{}_{p}=\sqrt{\displaystyle \vec{p}^{\,\,2} +m^2}$.
 In working with the components of the non-invariant Green's function
 $ G(\und{1}\,\und{2})$, one must be careful to distinguish between
 the kinematic ($\mPi$) and the canonical ($p$) momenta in evaluating the
  drift terms in a kinetic equation for electrons, because the Poisson
  brackets contain derivatives of $A(X)$. On the other hand,
  for
 \textit{local\/} quantities, such as the emission and absorption rates,
 the vector potential $A(X)$ drops out from the
 final expressions as it must be due to gauge invariance. Formally,
  this can be achieved by the change of variables
   $p^{}_{i}\to p^{}_{i}+eA(X)$ in integrals over the electron four-momenta,
or equivalently by setting $A(X)=0$
   and replacing $F^{\mgl}(X,p)\to f^{\mgl}(X,p)$ in the quasiparticle
   correlation functions.

Since for a weakly coupled relativistic plasma the spectral function
$\widetilde{\mathcal A}(X,p)$ is sharply peaked near the mass-shell, it is
reasonable to assume that the spinor matrices
  ${\mathcal F}^{\mgl}(X,p)$ are related to the distribution functions
   of electrons (positrons) in just the same way as in the
  zero damping limit. Then, in the case of nonpolarized particles,
  Eq.~(\ref{F<>}) reduces to
   \beq
     \label{F<>:appr}
  \widetilde{G}^{\mgl}(X,p)= \mp\, i\,
  \widetilde{\mathcal A}(X,p)F^{\mgl}(X,p).
   \eeq
Where appropriate, this ansatz allows one to recover the singular form
 (\ref{G<>:free}) for the quasiparticle correlation functions.

\subsection{Extended Quasiparticle Approximation for Relativistic Electrons}
\label{Subsec:ExtQPA}

Neglecting the second (off-shell) term in Eq.~(\ref{G<>:dec}), i.e.
identifying $\widetilde{G}^{\mgl}(X,p)$ with the \textit{full\/}
correlation functions, we recover, in essence, the relativistic
quasiparticle approximation used by Bezzerides and DuBois~\cite{BezDub72}.
This approximation is sufficient to derive a particle kinetic equation in
which the collision term involves electron-electron
 (electron-positron) scattering and
 Cherenkov emission/absorption of longitudinal plasma waves,
 but, as will be shown below, is inadequate to
describe radiative processes (e.g., Compton scattering and
bremsstrahlung).
 Therefore, the off-shell term in Eq.~(\ref{G<>:dec}) has to be taken into
 account, at least to leading order in the field fluctuations.

Let us first of all consider the self-energies $\Sigma^{\mgl}(X,p)$ appearing
 in Eq.~(\ref{G<>:dec}). They can be calculated
 from the matrix self-energy $\Sigma(\und{1}\,\und{2})$ shown in
 Fig.~\ref{fig:SelfEn;Pol}. To leading order in the field
 Green's functions, only the first diagram is retained. This gives
   \beq
     \label{Sigma<>}
  \Sigma(\und{1}\,\und{2})= ie^2
   \delta^{}_{\mu\sigma}(\und{1}-\und{3}')\gamma^{\sigma}
   G(\und{1}\,\und{2})\,
   \delta^{}_{\nu\sigma'}(\und{2}-\und{4}')\gamma^{\sigma'}
   D^{\nu\mu}(\und{4}'\,\und{3}').
   \eeq
The self-energies $\Sigma^{\mgl}(12)$ are the components
 $\Sigma^{\mgl}(12)=\Sigma(1^{}_{\mp}2^{}_{\pm})$.
In the local Wigner representation we obtain
   \beq
      \label{Sigma<>:W}
  \Sigma^{\mgl}(X,p)=
   ie^2\int \frac{d^4 k}{(2\pi)^4}\,
   \hat{\gamma}^{\mu}(k)\,G^{\mgl}(X,p+k)\,\hat{\gamma}^{\nu}(k)\,
   D^{\mlg}_{\nu\mu}(X,k),
   \eeq
where we have introduced the notation
      \beq
    \label{delta-k:mu}
 \hat{\gamma}^{\mu}(k)=
    \left(\gamma^0,\Delta^{\perp}_{ij}(\vec{k})\,\gamma^{j}\right).
  \eeq
 It is clear that the relation (\ref{G<>:dec}) is in fact an
 \textit{integral equation\/} for $G^{\mgl}(X,p)$ because
  the self-energies
 $\Sigma^{\mgl}(X,p)$ are functionals of the
 electron correlation functions.
  We recall, however, that the expression~(\ref{Sigma<>:W})
 itself is valid to first
  order in the field correlation functions. Therefore, within the same
  approximation, we may replace $G^{\mgl}(X,p)$ in
  Eq.~(\ref{Sigma<>:W}) by $\widetilde{G}^{\mgl}(X,p)$.
  Then Eq.~(\ref{G<>:dec}) reduces to
     \begin{eqnarray}
       \label{G<>:ansatz}
       & &
       \hspace*{-15pt}
   G^{\mgl}(X,p)= \widetilde{G}^{\mgl}(X,p)
    \nonumber\\[8pt]
       & &
       %\hspace*{15pt}
       {}+i\,\frac{e^2}{2}\int\frac{d^4 k}{(2\pi)^4}\,
       D^{\mlg}_{\nu\mu}(k)
       \left[
     g^{+}(p)\hat{\gamma}^{\mu}(k)
     \widetilde{G}^{\mgl}(p+k)\hat{\gamma}^{\nu}(k) g^{+}(p)
     \right.
     \nonumber\\[5pt]
       & &
       \hspace*{130pt}
        {}+ \left.
     g^{-}(p)\hat{\gamma}^{\mu}(k)
     \widetilde{G}^{\mgl}(p+k)\hat{\gamma}^{\nu}(k) g^{-}(p)
     \right].
     \end{eqnarray}
Note that  in non-relativistic kinetic theory an analogous decomposition
of correlation functions into the quasiparticle and off-shell parts is
called the ``extended quasiparticle
approximation''~\cite{KoelMalf93,SpLip94,SpLip95}. Expression
(\ref{G<>:ansatz}) may thus be regarded as the \textit{extended
quasiparticle approximation for relativistic electrons\/}.
 We  shall see later that the second term on the right-hand side
 of Eq.~(\ref{G<>:ansatz})  plays an important role in calculating
 the photon emission and absorbtion rates.

\setcounter{equation}{0}

\section{Transverse Polarization Functions}
  \label{Sec:TransPolFunct}
We are now in a position to perform calculations of the transverse
polarization functions $\pi^{\mgl}_{s}(X,k)$ determining the photon emission
and absorption rates.  The starting point is the
 polarization matrix
 $\Pi^{}_{\mu\nu}(\und{1}\,\und{2})$ on the time-loop contour $C$, which is
represented by the diagrams shown in Fig.~\ref{fig:SelfEn;Pol}.
       \begin{figure}[h]
 \centerline{\includegraphics[width=0.7\linewidth]{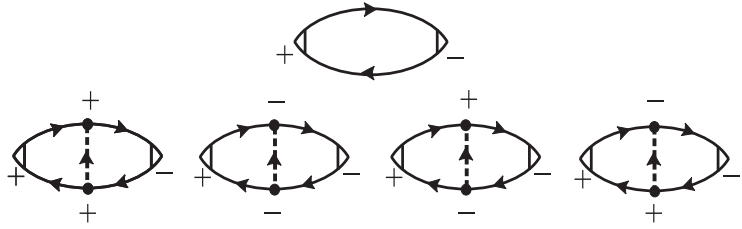}}
 \caption{\label{fig:PolMatr_Trans_pi}Space-time diagrams for
 $i\pi^{<}_{ij}(12)$. The $\pm$ signs indicate the branches of the contour $C$.}
\end{figure}
 Recalling the definition (\ref{Gam-i:bare}) of the bare ion vertex,
 it is easy to verify that the last two diagrams do not
contribute to $\Pi^{}_{ij}(\und{1}\,\und{2})$, so that the component
 $\pi^{<}_{ij}(12)=\Pi^{}_{ij}(1^{}_{+} 2^{}_{-})$ is
  given by the diagrams shown in
 Fig.~\ref{fig:PolMatr_Trans_pi}, where we have introduced
  the transverse bare vertex
    \beq
     \label{Gamma:trans}
     \Gamma^{}_{i}(12;3)=
     \raisebox{-8pt}{\includegraphics[scale=0.4]{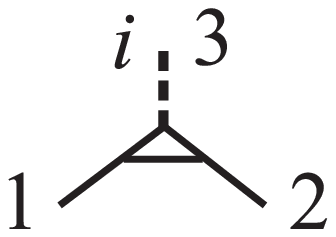}}\
     \equiv -e\,\delta(1-2)\,\delta^{T}_{ij}(1-3)\,\gamma^{j}.
     \eeq
The analogous  representation for
 $\pi^{>}_{ij}(12)=\Pi^{}_{ij}(1^{}_{-} 2^{}_{+})$ is obvious.

The next step is to find the contributions of the diagrams in
Fig.~\ref{fig:PolMatr_Trans_pi} to $\pi^{<}_{s}(X,k)$ by Wigner
 transforming all functions and then using the
 diagonal principal-axis representation
 (\ref{expan-pol}) for $\pi^{<}_{ij}(X,k)$. Since the polarization functions
 will be used for the calculation of the local radiating power,
  the $X$-gradient corrections to the Green's functions can be omitted.
  Then, the contribution of the diagram in the first line of
  Fig.~\ref{fig:PolMatr_Trans_pi} to
 $\pi^{<}_{s}(X,k)$ is found to be
         \begin{equation}
  \label{pi-1}
    \pi^{<\,(1)}_{s}(X,k)=
 -ie^{2} \int \frac{d^4 p}{(2\pi)^4}\,
 {\rm tr}^{}_{D}\left\{\,
 \mynot{\epsilon}^{}_{s}(k)\,
  G^{<}_{}(X,p)\mynot{\epsilon}^{}_{s}(k)\,G^{>}_{}(X,p-k)
 \right\}
  \end{equation}
with the polarization four-vectors defined as
 $\epsilon^{\mu}_{s}(X,k)=(0,\vec{\epsilon}^{}_{s}(X,k))$.

\begin{figure}[h]
 \centerline{
 \includegraphics[width=0.8\linewidth]{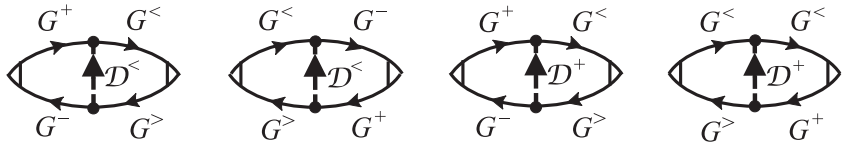}}
 \caption{\label{fig:One_loop_correct}
 Corrections to the one-loop diagram for $i\pi^{<}_{ij}(12)$.
 The symbol ${\mathcal D}$ denotes longitudinal or
 transverse  components of $D^{}_{\mu\nu}(12)$.}
 \end{figure}

 The computation of the diagrams in the second line of
 Fig.~\ref{fig:PolMatr_Trans_pi} requires
 a more elaborate analysis. To understand the physical meaning of different
 processes represented by these diagrams, it is convenient to use the
 canonical
 form (\ref{F:can}) for the electron and field Green's functions.
 Within the approximation where the direct coupling
 between transverse and
 longitudinal field fluctuations is neglected,
 each canonical component of
   $D^{}_{\mu\nu}(\und{1}\,\und{2})$ is a block matrix:
    \beq
       \label{Dmn:comp}
    D^{\mgl}_{\mu\nu}(12)=
     \begin{pmatrix}
       D^{\mgl}(12) & 0\\
       0 & d^{\mgl}_{ij}(12)
     \end{pmatrix}\, ,
     \qquad
 D^{\pm}_{\mu\nu}(12)=
     \begin{pmatrix}
       D^{\pm}(12) & 0\\
       0 & d^{\pm}_{ij}(12)
     \end{pmatrix}\, ,
    \eeq
where the correlation functions $d^{\mgl}_{ij}(12)$ and $D^{\mgl}(12)$
characterize the degree of excitation of the field fluctuations,
$D^{\pm}(12)$ describes the screened Coulomb interaction, and
 $d^{\pm}_{ij}(12)$ are the photon propagators. In terms of the canonical
 components of Green's functions, the second line of
  Fig.~\ref{fig:PolMatr_Trans_pi} generates ten space-time diagrams
  which differ from each other  by the positioning of the
electron correlation functions $G^{\mgl}$. In eight diagrams both
correlation functions appear at the left or right of the intermediate
 vertices.
 Examples  are given in Fig.~\ref{fig:One_loop_correct}.
Writing down explicit expressions, it can easily be verified that such
diagrams give small corrections to the one-loop diagram in the first line of
 Fig.~\ref{fig:PolMatr_Trans_pi}. For more details see
 Appendix~\ref{App:TwoLoop}. In the following the contributions due to these
 diagrams will be neglected.
 \begin{figure}[h]
 \centerline{\includegraphics[width=0.4\linewidth]{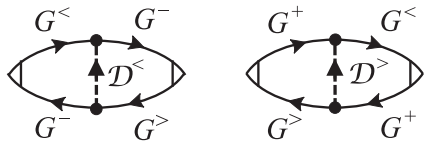}}
 \caption{\label{fig:pi_higher_order} Diagrams for $ i \pi^{<}_{ij}(12)$
 contributing to higher-order radiative processes. The symbol ${\mathcal D}$
  has the same meaning as in Fig.~\ref{fig:One_loop_correct}.
  }
 \end{figure}

 The two remaining
 diagrams shown in  Fig. \ref{fig:pi_higher_order} have an entirely different
 structure. They describe  interactions between quantum states
of incoming and outgoing particles. As we shall see later, these diagrams
contribute to the rates of higher-order radiative processes
 (e.g., bremsstrahlung and Compton scattering). In the local
 Wigner form, the contribution  of
 the diagrams in Fig.\,\ref{fig:pi_higher_order} to $\pi^{<}_{s}(X,k)$
  is given by (the fixed argument $X$ is omitted)
     \begin{eqnarray}
     \label{pi-2}
     & &
     \hspace*{-23pt}
 {\pi}^{<(2)}_{s}(k)=\frac{e^4}{(2\pi)^8}
  \int  d^4 k'\,
  \prod^{4}_{i=1} d^4 p^{}_{i}\,
   \nonumber\\[5pt]
     & &
      \hspace*{-23pt}
   {}\times \delta^4(k- p^{}_{2}+ p^{}_{3})\,
 \delta^4(k'- p^{}_{3}+ p^{}_{4})\,
  \delta^4(p^{}_{1}+p^{}_{3} -p^{}_{2} - p^{}_{4})\,
   \nonumber\\[-5pt]
    & &\quad\\[-5pt]
    & &
    \hspace*{-23pt}
    {}\times
    \left[
    D^{<}_{\lambda\lambda'}(k')\,
     {\rm tr}^{}_{D}\!
     \left\{
  \mynot{\epsilon}^{}_{s}(k)
   G^{<}(p^{}_{1})\,
  \hat{\gamma}^{\lambda'}(k')
   g^{-}(p^{}_{2})
  \mynot{\epsilon}^{}_{s}(k)
    G^{>}(p^{}_{3})\,
 \hat{\gamma}^{\lambda}(k')
  g^{-}(p^{}_{4})
     \right\}\right.
     \nonumber\\[5pt]
      & &
    \hspace*{-10pt}
     {}+\left.
   D^{>}_{\lambda\lambda'}(k')\,
     {\rm tr}^{}_{D}\!
     \left\{
  \mynot{\epsilon}^{}_{s}(k)
   g^{+}(p^{}_{1})
  \hat{\gamma}^{\lambda'}(k')
  G^{<}(p^{}_{2})
  \mynot{\epsilon}^{}_{s}(k)
    g^{+}(p^{}_{3})
 \hat{\gamma}^{\lambda}(k')
  G^{>}(p^{}_{4}) \right\}
     \right]\! ,
     \nonumber
  \end{eqnarray}
where we have used the notation (\ref{delta-k:mu}).
 The analysis of diagrams corresponding to
 $\pi^{>}_{ij}(12)=\Pi^{}_{ij}(1^{}_{-} 2^{}_{+})$ follows along exactly
 the same lines and therefore will not be discussed here. The
 contributions to $\pi^{>}_{s}(X,k)$ are the same as
 Eqs.~(\ref{pi-1}) and (\ref{pi-2}) with
 $>$ and $<$ signs interchanged.

 Recalling Eq.~(\ref{G<>:ansatz}), we can now express the relevant
 contributions to $\pi^{<}_{s}(X,k)$ in terms of the quasiparticle parts
 of the electron correlation functions. Since our consideration is restricted
 to the lowest order in the field fluctuations, the full electron
 correlation functions $G^{\mgl}(X,p)$ entering
 Eq.~(\ref{pi-2}) are to be replaced by $\widetilde{G}^{\mgl}(X,p)$.
 Note, however, that in
 Eq.~(\ref{pi-1}) one must keep first-order off-shell
 corrections to the electron correlation functions
 coming from the integral term in Eq.~(\ref{G<>:ansatz}). Collecting all
 contributions to $\pi^{<}_{s}(X,k)$, it is convenient to eliminate
 the field correlation functions $D^{>}_{\mu\nu}(X,k)$ with the help
 of the symmetry relation
      \beq
       \label{Dmn:symm}
     D^{>}_{\mu\nu}(X,k)= D^{<}_{\nu\mu}(X,-k).
      \eeq
Then some algebra gives
    \begin{eqnarray}
       \label{pi<s:final}
    & &
    \hspace*{-10pt}
     \pi^{<}_{s}(X,k)=
      -ie^{2}(2\pi)^4
      \int \frac{d^4 p^{}_{1}}{(2\pi)^4}\,\frac{d^4 p^{}_{2}}{(2\pi)^4}\,
      \delta^4\left(p^{}_{1}-p^{}_{2}-k\right)
      \nonumber\\[5pt]
      & &
   \hspace*{130pt}
      {}\times
  {\rm tr}^{}_{D}\left\{\,\mynot{\epsilon}^{}_{s}(k)\,
  \widetilde{G}^{<}_{}(p^{}_{1})\mynot{\epsilon}^{}_{s}(k)\,
  \widetilde{G}^{>}_{}(p^{}_{2})
   \right\}
   \nonumber\\[8pt]
    & &
    {}+
    e^4\left(2\pi\right)^4 \int
     \frac{d^4 p^{}_{1}}{(2\pi)^4}\,\frac{d^4 p^{}_{2}}{(2\pi)^4}\,
     \frac{d^4 k^{\prime}}{(2\pi)^4}\,
     \delta^4\left(p^{}_{1} +k^{\prime}-p^{}_{2}-k \right)
     \nonumber\\[5pt]
      & &
      \hspace*{180pt}
      {}\times
      K^{\lambda\lambda'}_{s}(p^{}_{1},p^{}_{2};k,k')\,
     D^{<}_{\lambda\lambda'}(k^{\prime}),
    \end{eqnarray}
   where
   \begin{eqnarray}
      \label{K}
   & &
   \hspace*{-22pt}
  K^{\lambda\lambda'}_{s}(p^{}_{1},p^{}_{2};k,k')=
   \frac{1}{2}\,
    {\rm tr}^{}_{D}
    \left\{ \mynot{\epsilon}^{}_{s}(k)\left[
     g^{+}(p^{}_{2}+k)\hat{\gamma}^{\lambda}(k')
     \widetilde{G}^{<}(p^{}_{1})\hat{\gamma}^{\lambda'}(k')
     g^{+}(p^{}_{2}+k)
     \right.\right.
     \nonumber\\[5pt]
      & &
      \hspace*{50pt}
      \left.\left.
      {}+ g^{-}(p^{}_{2}+k)\hat{\gamma}^{\lambda}(k')
     \widetilde{G}^{<}(p^{}_{1})\hat{\gamma}^{\lambda'}(k')
     g^{-}(p^{}_{2}+k)\right]\mynot{\epsilon}^{}_{s}(k)
     \widetilde{G}^{>}(p^{}_{2})\right\}
     \nonumber\\[5pt]
      & &
      \hspace*{-17pt}
     {}+
  \frac{1}{2}\,
    {\rm tr}^{}_{D}
    \left\{\hat{\gamma}^{\lambda}(k')\left[
     g^{+}(p^{}_{1}-k)\mynot{\epsilon}^{}_{s}(k)
     \widetilde{G}^{<}(p^{}_{1})\mynot{\epsilon}^{}_{s}(k)
     g^{+}(p^{}_{1}-k)
     \right.\right.
     \nonumber\\[5pt]
      & &
      \hspace*{10pt}
      \left.\left.
      {}+ g^{-}(p^{}_{1}-k)\mynot{\epsilon}^{}_{s}(k)
     \widetilde{G}^{<}(p^{}_{1}) \mynot{\epsilon}^{}_{s}(k)
     g^{-}(p^{}_{1}-k)\right]\hat{\gamma}^{\lambda'}(k')
     \widetilde{G}^{>}(p^{}_{2})\right\}
     \nonumber\\[5pt]
      & &
      \hspace*{-17pt}
     {}+
     {\rm tr}^{}_{D}\left\{
    \mynot{\epsilon}^{}_{s}(k)g^{+}(p^{}_{2}+k)
    \hat{\gamma}^{\lambda'}(k') \widetilde{G}^{<}(p^{}_{1})
    \mynot{\epsilon}^{}_{s}(k)g^{+}(p^{}_{1}-k)
    \hat{\gamma}^{\lambda}(k')\widetilde{G}^{>}(p^{}_{2})
     \right\}
     \nonumber\\[5pt]
      & &
      \hspace*{-17pt}
    {}+
     {\rm tr}^{}_{D}\left\{\hat{\gamma}^{\lambda}(k')
    g^{-}(p^{}_{1}-k)\mynot{\epsilon}^{}_{s}(k)
     \widetilde{G}^{<}(p^{}_{1})\hat{\gamma}^{\lambda'}(k')
    g^{-}(p^{}_{2}+k)\mynot{\epsilon}^{}_{s}(k)
    \widetilde{G}^{>}(p^{}_{2})
     \right\}.
   \end{eqnarray}
The expression (\ref{pi<s:final})   deserves some comments. Formally, it looks
 like an expansion in powers of $e^2$ in vacuum electrodynamics,
 but, as has already
  been noted above, the situation is more complicated for the plasma case
  since medium effects must be included
   to all orders in $e^2$. These effects
   enter Eq.~(\ref{pi<s:final}) in three types.
 First, the quasiparticle correlation functions $\widetilde{G}^{\mgl}$ are corrected for
 multiple-scattering effects through the spectral function (\ref{QSF}).
  Second, the local propagators $g^{\pm}$ contain the quasiparticle damping
   described by the self-energies in Eq.~(\ref{g+-}). Finally, the
  correlation functions $D^{<}_{\lambda\lambda'}$ contain polarization
  effects and contributions from resonant excitations of the
  electromagnetic field (photons and plasmons). Neglecting the collisional
  broadening of $\widetilde{G}^{\mgl}$ in the first term of
  Eq.~(\ref{pi<s:final}), i.e. taking these functions in the
   approximation (\ref{G<>:free}), we recover the well-known expression for
  the transverse polarization function in a nonequilibrium
   QED plasma~\cite{BezDub72}. The second term in Eq.~(\ref{pi<s:final})
    is one of our main results. It may be worth recalling that the
    structure of the trace factor (\ref{K}) in this term is closely related to the
    decomposition (\ref{G<>:dec}) of the electron correlation functions,
  which leads to the extended quasiparticle ansatz (\ref{G<>:ansatz}).
     In the next section we will use the expression
    (\ref{pi<s:final}) to discuss radiative processes in
     relativistic plasmas.

 \setcounter{equation}{0}

 \section{Photon Production: Interpretation of Elementary Processes}
 \label{Sec:Rates}

\subsection{Local Radiating Power}
 \label{Subsec:RadPower}

The kinetic results (\ref{prod-abs}) for the photon production and
absorption rates imply the singular approximation (\ref{a:free}) for the
resonant spectral function $\widetilde{a}^{}_{s}(X,k)$.
 In Appendix~\ref{App:EnProd} we derive a general
expression for the local energy production associated with emission and
 absorption of resonant photons, in which the finite width of
$\widetilde{a}^{}_{s}(X,k)$ is taken into account. Here we shall quote the
result:
     \begin{eqnarray}
    \label{Emitt-E:App2}
    & &
    \hspace*{-10pt}
  \left(\frac{\partial {\mathcal E}(X)}{\partial T}\right)^{}_{\rm phot}=
 i\sum_{s} \int\frac{d^4 k}{(2\pi)^4}\,\theta(k^{}_{0})
  k^{}_{0}
  \widetilde{a}^{}_{s}(X,k)
   \nonumber\\[5pt]
    & &
    \hspace*{60pt}
  {}\times
  \left\{\pi^{<}_{s}(X,k)
  \left[1+ N^{}_{s}(X,k)\right]
  - \pi^{>}_{s}(X,k) N^{}_{s}(X,k)
  \right\},
  \end{eqnarray}
where $N^{}_{s}(X,k)$ is the local photon distribution
function in four-dimensional phase space defined by Eqs.~(\ref{N(p)}).
 The two integrals in Eq.~(\ref{Emitt-E:App2})
  have a straightforward physical interpretation in terms of the
radiated and absorbed power. In particular, the quantity in which we are
 interested here primarily  is the \textit{differential radiating power\/}
associated with the photon production:
  \beq
    \label{dR:gen}
   \frac{dR(X,k)}{d^{4} k}=\frac{i}{(2\pi)^4}\,
   k^{}_{0}\,
   \sum_{s}
   \widetilde{a}^{}_{s}(X,k)\,\pi^{<}_{s}(X,k) \left[1+ N^{}_{s}(X,k)\right],
  \eeq
where $k^{}_{0}>0$ is implied. We observe that, for a fixed $\vec{k}$,
 the $\widetilde{a}^{}_{s}(X,k)$ is just the weight function that
determines the emission profile.
 This supports the interpretation of  $\widetilde{a}^{}_{s}(X,k)$
 as the spectral function for propagating photons
 in a plasma. In the approximation (\ref{a:free}), the differential
 radiating power can be expressed in terms of
the three-momentum $\vec{k}$ of the emitted photons by integrating
 Eq.~(\ref{dR:gen})  over $\omega=k^{}_{0}$:
  \beq
    \label{dR:k}
   \frac{dR(X,\vec{k})}{d^3 \vec{k}}=\frac{i}{(2\pi)^3}\,
   \sum_{s}
   \omega^{}_{s}(X,\vec{k})\,
   Z^{}_{s}(X,\vec{k})\,\pi^{<}_{s}(X,\vec{k})
 \left[ 1+ n^{}_{s}(X,\vec{k}) \right],
  \eeq
 with the quantities on the right-hand side defined
earlier (see Subsection~\ref{Subs:KinEqPhot}).
 Note that the expression (\ref{dR:k}) is consistent with
the photon production rate (\ref{prod}) derived from the kinetic equation.

 In most situations the effective photon
frequencies of interest satisfy
 the inequality $\omega^{}_{s}(X,\vec{k})\gg
 \omega^{}_{e}$, where
    \begin{equation}
     \label{omega-e}
  \omega^{2}_{e}(X)= 2e^2 \int \frac{d^3\vec{p}}{(2\pi)^3}\,
   \frac{f(X,\vec{p})}{E^{}_{p}}
   \end{equation}
is the relativistically corrected plasma frequency and $f(X,\vec{p})\equiv
f^{}_{e^{-}}(X,\vec{p})$ is the electron distribution function.
 Then, as shown in
Appendix~\ref{App:Dispers}, a good approximation for
 the photon dispersion is
    \begin{equation}
     \label{omega-s:isotr}
   \omega(X,\vec{k})\equiv \omega^{}_{s}(X,\vec{k})= \left(
     \vec{k}^{\,2} + \omega^{2}_{e}(X)
     \right)^{1/2} .
   \end{equation}
Physically, this implies local isotropy for the photon
 polarization states.
 Under the assumption that $\omega(X,\vec{k})\gg\omega^{}_{e}$,
   Eq.~(\ref{Z}) gives $Z^{-1}_{s}=2\omega(X,\vec{k})$, so that
   Eq.~(\ref{dR:k}) reduces to
  \beq
  \label{dR:k2}
   \frac{dR(X,\vec{k})}{d^3 \vec{k}}=\frac{i}{2(2\pi)^3}\,
   \sum_{s}\pi^{<}_{s}(X,\vec{k})
  \left[1+ n^{}_{s}(X,\vec{k})\right].
  \eeq
In the following we shall restrict ourselves to this simple expression for
the radiating power. It is not difficult, however, to generalize the
 results using Eq.~(\ref{dR:gen}). Since the width of the resonant
spectral function is assumed to be small, even in the latter case
$N^{}_{s}(X,k)$ can be identified with the on-shell photon distribution
function $n^{}_{s}(X,\vec{k})$.

 \subsection{Cherenkov Emission in Plasmas}
  \label{SubSec:Cherenkov}
\begin{figure}[h]
 \centerline{\includegraphics[width=0.85\linewidth]{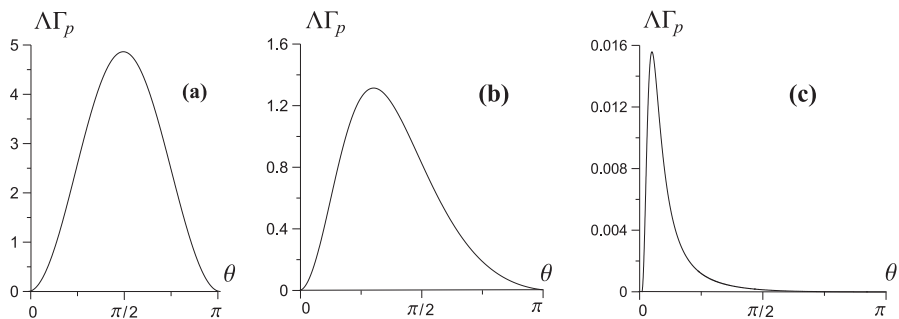}}
 \caption{\label{fig:Angular_total_new}Angular distribution of emitted
 photons in the Cherenkov channel: {\bf (a)} $\omega(\vec{k})=0.1\Gamma^{}_{p}$;
  {\bf (b)} $\omega(\vec{k})=\Gamma^{}_{p}$;
  {\bf (c)} $\omega(\vec{k})=10\,\Gamma^{}_{p}$.
  Here, $\theta$ is the angle between the
 electron momentum $\vec{p}$ and the photon momentum $\vec{k}$.
 The electron kinetic energy is assumed to be $E^{}_{\text{kin}}=10\ \text{MeV}$.
 }
\end{figure}

 We begin with the contribution of the first term of Eq.~(\ref{pi<s:final})
 to the radiating power. Usually the corresponding process is referred to as
  Cherenkov radiation.
  Note, however, that the energy-momentum conserving
  Cherenkov emission of transverse photons in QED plasmas is kinematically
  forbidden~\cite{BezDub72} because, for all expected
   densities, the transverse dispersion
  curve stays above the $\omega=|\vec{k}|$. Strictly speaking, this is only
  true when the quasiparticle spectral function in Eq.~(\ref{F<>}) is
  approximated by the singular mass-shell expression (\ref{A:lim}), i.e., when
   medium effects are ignored. The collisional broadening of
   the spectral function
   $\widetilde{\mathcal A}(p)$ results in a statistical energy uncertainty for
   electrons (positrons), so that the emission of low-energy  photons
   becomes possible\footnote{A similar situation occurs in the theory of a quark-gluon
   plasma~\cite{QuackHenning96} where the thermal broadening of the quark
    spectral function is one of the mechanisms for the soft photon
     production.}. Assuming, for simplicity, equal probabilities for the
     photon polarizations and the particle spin states and using
     Eq.~(\ref{F<>:appr}), we obtain the local radiating
 power in the ``Cherenkov channel''\footnote{From now on Green's
functions, distribution functions, etc. are understood to be
 gauge-invariant, i.e., $A^{\mu}(X)$ is set equal to zero as explained in
 Subsection~\ref{Subsec:Distrib}.}
    \begin{eqnarray}
    \label{R:Cherenkov}
   & &
    \hspace*{-20pt}
   \left(\frac{dR(X,\vec{k})}{d^3\vec{k}}\right)^{}_{\text{Cher}}=
   8\pi e^2\left[1+n(X,\vec{k})\right]
   \nonumber\\[5pt]
   & &
   %\hspace*{5pt}
    {}\times
   \int \frac{d^4 p^{}_{1}}{(2\pi)^4}\,\frac{d^4 p^{}_{2}}{(2\pi)^4}\,
   \delta^4\left(p^{}_{1}-p^{}_{2}-k\right)
   C(p^{}_{1},p^{}_{2};k)\,   f^{<}(X,p^{}_{1})f^{>}(X,p^{}_{2}),
  \end{eqnarray}
where $n(X,\vec{k})\equiv n^{}_{s}(X,\vec{k})$,
$k^{}_{0}=\omega(\vec{k})$, and
    \beq
       \label{Cher:C}
    C(p^{}_{1},p^{}_{2};k)=\frac{1}{8}
     \sum_{s} {\rm tr}^{}_{D}\left\{
   \mynot{\epsilon}^{}_{s}(k) \widetilde{\mathcal A}(p^{}_{1})
   \mynot{\epsilon}^{}_{s}(k) \widetilde{{\mathcal A}}(p^{}_{2})
     \right\}.
    \eeq
 The actual evaluation of expression (\ref{R:Cherenkov}) requires a
knowledge of the quasiparticle spectral function $\widetilde{\mathcal
A}(p)$ and a model for the distribution functions $f^{\mgl}(X,p)$ in the
plasma. Here we choose a simple parametrization of the quasiparticle
spectral function
    \beq
    \label{QA:Appr}
 \widetilde{\mathcal A}(p)= S(p)\left(\,\mynot{p}+m\right)
  \eeq
 with
  \beq
   \label{S(p)}
 S(p)=
 \frac{4\left(p^{}_0\Gamma^{}_{p}\right)^3  }
        {\left[ \left(p^{2}_{} - m^2\right)^2
          +\left(p^{}_0\Gamma^{}_{p}\right)^2 \right]^2}.
  \eeq
This ansatz may be regarded as a generalization of Eq.~(\ref{QSF-free}) to
the case of a finite but small spectral width parameter $\Gamma^{}_{p}\ll
  E^{}_{p}$. Within this approximation, the $\Gamma^{}_{p}$ is assumed
  to be taken at $p^2=m^2$.  Note that
 in the limit $\Gamma^{}_{p}\to 0$ we recover from Eq.~(\ref{QA:Appr})
 the singular spectral function
(\ref{A:lim}) (in the gauge-invariant form). With Eq.~(\ref{QA:Appr}) the
trace in Eq.~(\ref{Cher:C}) is evaluated explicitly and we obtain
  \beq
   \label{Cher:C-Appr}
 C(p^{}_{1},p^{}_{2};k)=
 S(p^{}_{1})S(p^{}_{2})
 \left[
  p^{0}_{1}p^{0}_{2} - m^2
  - \big(\vec{p}^{}_{1}\cdot\hat{\vec{k}}\big)
  \big(\vec{p}^{}_{2}\cdot\hat{\vec{k}}\big)
 \right],
  \eeq
where $\hat{\vec{k}}=\vec{k}/|\vec{k}|$. Under the conditions currently
available in laser-plasma
 experiments, the positron contribution to the radiating power
(\ref{R:Cherenkov}) is negligible. In that case $S(p^{}_{1})$ and
 $S(p^{}_{2})$ are sharply peaked near
  $p^{0}_{1}=E^{}_{p^{}_{1}}$ and $p^{0}_{2}=E^{}_{p^{}_{2}}$, so that
 $f^{<}(p^{}_{1})$ and $f^{>}(p^{}_{2})$ are expressed in terms of the
electron distribution function $f(\vec{p})$, i.e.
 $f^{<}(p^{}_{1})=f(\vec{p}^{}_{1})$,
 $f^{>}(p^{}_{2})=1-f(\vec{p}^{}_{2})$.
 Then, since in real laboratory plasmas the subsystem of relativistic electrons
  is non-degenerate, the blocking factor
 $1-f(\vec{p}^{}_{2})$ can be replaced by 1. Finally, it is clear that the dominant
 contribution to the integrals in Eq.~(\ref{R:Cherenkov}) comes
from $|p^{0}_{1}-E^{}_{p^{}_{1}}|\lesssim \Gamma^{}_{p^{}_{1}}$ and
$|p^{0}_{2}-E^{}_{p^{}_{2}}|\lesssim \Gamma^{}_{p^{}_{2}}$
 because the
quasiparticle spectral function (\ref{QA:Appr}) has very small wings.
  The characteristic frequencies of emitted photons are thus expected to
 satisfy $\omega(\vec{k})\lesssim \Gamma^{}_{p^{}_{1}}\ll E^{}_{p^{}_{1}}$
  and $\omega(\vec{k})\lesssim \Gamma^{}_{p^{}_{2}}\ll E^{}_{p^{}_{2}}$.
 This observation allows one to use for
  the function (\ref{Cher:C-Appr}) the approximate expression
   \beq
   \label{Cher:C-Appr2}
 C(p^{}_{1},p^{}_{2};k)\approx
  \frac{\big[\vec{p}^{\,2}_{1} - \big( \vec{p}^{}_{1}\cdot \hat{\vec{k}}\big)^2\big]
  \Gamma^{3}_{p^{}_{1}}\Gamma^{3}_{p^{}_{2}}}{\big(4E^{}_{p^{}_{1}}\big)^2
  \left[
   \left(p^{0}_{1} - E^{}_{p^{}_{1}}\right)^2 +
   \left(\frac{1}{2}\,\Gamma^{}_{p^{}_{1}}\right)^2
  \right]^2
  \left[
  \left(
   p^{0}_{2} - E^{}_{p^{}_{2}}
  \right)^2 +\left(\frac{1}{2}\,\Gamma^{}_{p^{}_{2}}\right)^2
  \right]^2
  }\, .
  \eeq
 Here, the difference between the spectral width parameters
$\Gamma^{}_{p^{}_{1}}$ and $\Gamma^{}_{p^{}_{2}}$ may be neglected. On the
other hand, the difference between $E^{}_{p^{}_{1}}$ and $E^{}_{p^{}_{2}}$
 in the denominator
 plays a crucial role. Indeed, in the limit $\Gamma^{}_{p}\to 0$ we have
 $C(p^{}_{1},p^{}_{2};k)\propto \delta(p^{0}_{1}-E^{}_{p^{}_{1}})\,
 \delta(p^{0}_{2}-E^{}_{p^{}_{2}})$.
  Then it can easily be seen that the integral
 in Eq.~(\ref{R:Cherenkov}) vanishes due to the presence of the
 four-dimensional delta function and the
 dispersion relation
  (\ref{omega-s:isotr}) for photons. In the case of a finite quasiparticle
  spectral width, the stringent energy-momentum conservation
  is violated, so that, for sufficiently small
  $k$, the function $S(p^{}_{1})$ overlaps with $S(p^{}_{2})=S(p^{}_{1}-k)$, which leads
  to a finite radiating power in the Cherenkov channel.
Based on the above considerations, one can transform Eq.~(\ref{R:Cherenkov}),
after integration over $p^{}_{2}$ and $p^{0}_{1}$, to
  \beq
     \label{R:Cher:fin}
     \left(\frac{dR(X,\vec{k})}{d^3\vec{k}}\right)^{}_{\text{Cher}}=
     \frac{e^2}{(2\pi)^3}\big[1+ n(\vec{k})\big]
     \int \frac{d^3 \vec{p}}{(2\pi)^3}\,
     {\Lambda}(\vec{p},\vec{k})\,f(\vec{p}),
  \eeq
 where
   \beq
      \label{Cher:Lambda}
    \Lambda(\vec{p},\vec{k})=
    \frac{\Gamma^{3}_{p}
    \left[\vec{p}^{\,2} - ( \vec{p}\cdot \hat{\vec{k}})^2\right]
    \left[5\Gamma^{2}_{p}
    +\omega^{2}(\vec{k})
    \left(1- \displaystyle\frac{\vec{p}\cdot\vec{k}}{E^{}_{p}\,\omega(\vec{k})}
    \right)^2\right]}{E^{2}_{p}
     \left[\Gamma^{2}_{p}  +\omega^{2}(\vec{k})
    \left(1- \displaystyle\frac{\vec{p}\cdot\vec{k}}{E^{}_{p}\,\omega(\vec{k})}
    \right)^2  \right]^3 }\, .
   \eeq
Let us briefly  mention some important properties of this function. Since
in a plasma
 $[1-(\vec{p}\cdot\vec{k})/(E^{}_{p}\,\omega(\vec{k}))]\not= 0 $
 for all possible values of $\vec{p}$ and $\vec{k}$, it is clear that
 $\Lambda(\vec{p},\vec{k})\to 0$ in the limit $\Gamma^{}_{p}\to 0$
 (more precisely, $\Gamma^{}_{p}/\omega(\vec{k})\to 0$). The
 frequencies $\omega(\vec{k})\gg \Gamma^{}_{p}$ correspond to the hard
 photon emission. Although, for a finite $\Gamma^{}_{p}$, in this
  region the radiating power is not zero, its contribution is
 difficult
 to detect experimentally. On the other hand, in the low-frequency region
  ($\omega(\vec{k})\ll \Gamma^{}_{p}$) the $\Lambda(\vec{p},\vec{k})$
 depends only weakly on the photon frequency and can be approximated as
  \beq
    \label{Lambda:low}
  \Lambda(\vec{p},\vec{k})\approx \frac{5}{E^{2}_{p}\,\Gamma^{}_{p}}
     \left[\vec{p}^{\,2} - ( \vec{p}\cdot \hat{\vec{k}})^2\right],
     \quad
     (\omega(\vec{k})\ll \Gamma^{}_{p}).
  \eeq
Finally, we note that for fixed $\vec{p}$ and $\omega(\vec{k})$, the
  quantity $\Lambda(\vec{p},\vec{k})\,\Gamma^{}_{p}$ may be
 regarded as
 the angular distribution of emitted photons. The behavior of this
 distribution in the frequency region
 $\omega(\vec{k})\approx \Gamma^{}_{p}$ is illustrated
 in Fig.~\ref{fig:Angular_total_new}.

 Any further details of
 the radiating power predicted by Eq.~(\ref{R:Cher:fin}) require explicit
 expressions for the electron distribution function $f(X,\vec{p})$ and
  an estimate of the quasiparticle
 spectral width $\Gamma^{}_{p}$ for real experimental situations.

\subsection{Interactions of Electrons with Transverse Field Fluctuations}
\label{SubSec:ScattTransv}
 We now turn to the contribution of the second term of
 Eq.~(\ref{pi<s:final}) to the local radiating power.
 According to Eqs.~(\ref{Dmn:comp}),
 the Wigner transformed field correlation functions are given by
        \beq
       \label{Dmn:comp-Xk}
    D^{\mgl}_{\mu\nu}(X,k)=
     \begin{pmatrix}
       D^{\mgl}(X,k) & 0\\
       0 &
       \displaystyle
       \sum_{s}\epsilon^{}_{si}(k)\,d^{\mgl}_{s}(X,k)\,\epsilon^{}_{sj}(k)
     \end{pmatrix}\, .
     \eeq
We therefore have two different contributions to the radiating power
 arising from the interaction of electrons (positrons) with
 longitudinal  and transverse
 fluctuations of the electromagnetic field.

 Let us first consider
 the interaction of electrons with transverse field fluctuations.
As before, we shall assume equal probabilities for the particle
 spin states. Then, using the representation (\ref{F<>:appr}) for the
 quasiparticle correlation functions
 (in the gauge-invariant form) and the definition
  (\ref{delta-k:mu}) of matrices $\hat{\gamma}^\mu(k)$, the corresponding
 contribution  to the
 local radiating power can be written as
   \begin{eqnarray}
    \label{dR-Trans}
     & &
     \hspace*{-25pt}
  \left(\frac{dR(X,\vec{k})}{d^3 \vec{k}}\right)^{}_{\text{transv}}=
  i\pi e^4\sum_{s,s'}\left[1+n^{}_{s}(X,\vec{k})\right]
  \int
     \frac{d^4 p^{}_{1}}{(2\pi)^4}\,\frac{d^4 p^{}_{2}}{(2\pi)^4}\,
     \frac{d^4 k^{\prime}}{(2\pi)^4}\,
     \nonumber\\[10pt]
    & &
     \hspace*{-15pt}
     {}\times
     \delta^4\left(p^{}_{1} +k^{\prime}-p^{}_{2}-k \right)
     T^{}_{ss'}(p^{}_{1},p^{}_{2};k,k')
     f^{<}(X,p^{}_{1}) f ^{>}(X,p^{}_{2})d^{<}_{s'}(X,k'),
  \end{eqnarray}
 where we have introduced the function
   \begin{eqnarray}
      \label{T:gen}
      & &
      \hspace*{-20pt}
 T^{}_{ss'}(p^{}_{1},p^{}_{2};k,k')=
  {\rm Re}\left\{
  {\rm tr}^{}_{D}\left[
  \mynot{\epsilon}^{}_{s}g^{+}(p^{}_{1}+k')\mynot{\epsilon}^{}_{s'}
  \widetilde{\mathcal A}(p^{}_{1})
  \mynot{\epsilon}^{}_{s'}g^{+}(p^{}_{1}+k')\mynot{\epsilon}^{}_{s}
  \widetilde{\mathcal A}(p^{}_{2})
  \right]
  \right.
  \nonumber\\[5pt]
   & &
   \hspace*{40pt}
   {}+
  {\rm tr}^{}_{D}\left[
  \mynot{\epsilon}^{}_{s'}g^{+}(p^{}_{1}-k)\mynot{\epsilon}^{}_{s}
  \widetilde{\mathcal A}(p^{}_{1})
  \mynot{\epsilon}^{}_{s}g^{+}(p^{}_{1}-k)\mynot{\epsilon}^{}_{s'}
  \widetilde{\mathcal A}(p^{}_{2})
  \right]
 \nonumber\\[5pt]
   & &
    \hspace*{40pt}
   \left.
   {}+
  2\,{\rm tr}^{}_{D}\left[
  \mynot{\epsilon}^{}_{s}g^{+}(p^{}_{1}+k')\mynot{\epsilon}^{}_{s'}
  \widetilde{\mathcal A}(p^{}_{1})
  \mynot{\epsilon}^{}_{s}g^{+}(p^{}_{1}-k)\mynot{\epsilon}^{}_{s'}
  \widetilde{\mathcal A}(p^{}_{2})
  \right]
  \right\}.
   \end{eqnarray}
For brevity, we use the notation $\epsilon^{}_{s}=\epsilon^{}_{s}(k)$,
$\epsilon^{}_{s'}=\epsilon^{}_{s'}(k')$.
 Before continuing, it is expedient to dwell briefly on the approximation
  in which all medium effects
are completely ignored, i.e., the gauge-invariant quasiparticle spectral
function
 is taken in the form
  \beq
   \label{QSF0}
   \widetilde{\mathcal A}(p)=2\pi\,\eta(p^{0})\,\delta(p^2 -m^2)
 \left(\,\mynot{p} +m \right),
  \eeq
 and the off-shell propagators are
 \beq
  \label{g:off-sh}
  g^{\pm}(p^{}_{1}+k')=\frac{1}{\mynot{p}^{}_{1}+\mynot{k}'-m},
   \qquad
 g^{\pm}(p^{}_{1}-k)=\frac{1}{\mynot{p}^{}_{1}-\mynot{k}-m}.
 \eeq
Then the function (\ref{T:gen}) reduces to
   \begin{eqnarray}
     \label{T:0}
      & &
      \hspace*{-15pt}
  T^{(0)}_{ss'}(p^{}_{1},p^{}_{2};k,k')=
  (2\pi)^2 \eta(p^{0}_{1})\,\eta(p^{0}_{2})\,\delta(p^{2}_{1}-m^2)\,
  \delta(p^{2}_{2}-m^2)
  \nonumber\\[5pt]
   & &
   {}\times
   {\rm tr}^{}_{D}
       \left\{
  \left[
 \mynot{\epsilon}^{}_{s'}\,
 \frac{1}{\mynot{p}^{}_{1}\,- \mynot{k}-m}\,
 \mynot{\epsilon}^{}_{s}
  +\mynot{\epsilon}^{}_{s}\,
 \frac{1}{\mynot{p}^{}_{1}\,+ \mynot{k}'-m}\,
  \mynot{\epsilon}^{}_{s'}
   \right](\,\mynot{p}^{}_{1}\,+m)
    \right.
    \nonumber\\[5pt]
    & &
    \hspace*{50pt}
    {}\times
    \left.
    \left[
 \mynot{\epsilon}^{}_{s'}\,
  \frac{1}{\mynot{p}^{}_{1}\,+ \mynot{k}' -m}\,
 \mynot{\epsilon}^{}_{s}
  +\mynot{\epsilon}^{}_{s}\,
 \frac{1}{\mynot{p}^{}_{1}\, - \mynot{k}-m}\,
  \mynot{\epsilon}^{}_{s'}
   \right]
 (\,\mynot{p}^{}_{2}\,+m)
    \right\}.
   \end{eqnarray}
 The trace factor in this expression is just the same
 as in the cross sections for Compton scattering (the Klein-Nishina formula)
 and for electron-positron annihilation
 known from vacuum QED \cite{ItzZuber80}.
 It is important to emphasize that the cross sections of QED are
 correctly reproduced in many-particle Green's function theory
 only if the vertex
corrections in the first term of Eq.~(\ref{Pi-mn:gen}) as well as the
off-shell parts of the electron correlation functions (\ref{G<>:dec}) are
taken into account.

 Medium effects enter Eq.~(\ref{T:gen}) through the collisional
broadening of the propagators $g^{\pm}(p)$ and the quasiparticle spectral
function $\widetilde{\mathcal A}(p)$. To analyze the role of these
effects, let us go back to the decomposition (\ref{Res-d:def}) of the
transverse field correlation functions into the resonant (photon) and
off-shell parts. Physically, the contribution of the resonant correlation
function $\widetilde{d}^{<}_{s'}(k')$ to the
 radiating power (\ref{dR-Trans}) can be ascribed to
 electron-positron annihilation and Compton scattering. As discussed above,
  the collisional effects are of importance when photon frequencies are
comparable to the quasiparticle spectral width $\Gamma^{}_{p}$. For pair
annihilation processes ($p^{0}_{2}<0$),  both $k^{}_{0}$ and
$|k^{\prime}_{0}|$ ($k^{\prime}_{0}<0$) in Eq.~(\ref{dR-Trans}) are much
larger than $\Gamma^{}_{p^{}_{1}}$ and $\Gamma^{}_{p^{}_{2}}$ (provided
$\Gamma^{}_{p}\ll E^{}_{p}$), so that one can use approximation
(\ref{T:0}). On the other hand,  the conditions
  $k^{}_{0}\lesssim \Gamma^{}_{p}$ and (or)
  $k^{\prime}_{0}\lesssim \Gamma^{}_{p}$ can be
 satisfied for Compton scattering, and therefore the collisional broadening should
  be taken into account.

  The contribution of Compton scattering to the radiating power
 is determined by the resonant correlation function
 $\widetilde{d}^{<}_{s'}(X,k')$ with
  $k^{\prime}_{0}>0$. Relations (\ref{SpF}) and (\ref{N(p)}) can be used
  to express $\widetilde{d}^{<}_{s'}(X,k')$ in terms of the
  four phase-space photon
  distribution $N^{}_{s'}(X,k')$. Again, assuming equal probabilities for
  the photon polarizations and taking approximation
  (\ref{a:free}) for the resonant spectral function, we have for
   $k^{\prime}_{0}>0$
     \beq
       \label{d:res-posit}
  \widetilde{d}^{<}_{s'}(X,k')= - \frac{i\,\pi}{\omega(X,\vec{k}')}\,
  \delta\left(k^{\prime}_{0}-\omega(X,\vec{k}^{\prime})\right) n(X,\vec{k}^{\prime}).
     \eeq
Then, neglecting as before the positron contribution to the radiating
power and replacing the blocking factor $1-f(\vec{p}^{}_{2})$ by unity, we
obtain from Eq.~(\ref{dR-Trans})
     \beq
    \label{dR-Compton}
  \left(\frac{dR(X,\vec{k})}{d^3 \vec{k}}\right)^{}_{\text{Comp}}=
  \frac{e^4}{4(2\pi)^3}
  \left[1+n(\vec{k})\right]
  \int
     \frac{d^4 p}{(2\pi)^4}\,
     \frac{d^3 \vec{k}^{\prime}}{(2\pi)^3}\,
     T(p;k,k')f(\vec{p})\,n(\vec{k}^{\prime})\, ,
 \eeq
 where $k^{}_{0}=\omega(\vec{k})$, $k^{\prime}_{0}=\omega(\vec{k}')$, and
  \beq
    \label{T:Compton}
 T(p;k,k')=\frac{1}{\omega(\vec{k}')}
  \sum_{s,s'} T^{}_{ss'}(p,p+k'-k;k,k')\, .
  \eeq
 For high intensity laser-plasma experiments,
 an estimate of the effective temperature of the bulk plasma is
  $T^{}_{\text{pl}}\approx 100\ {\rm eV}$ \cite{Tatarakis03}.
  Since the energies of the laser-generated electrons are in the MeV region,
 their collisions with ambient photons
 should be more correctly classified as the so-called \textit{inverse
Compton scattering\/} in which the energies of photons are increased by a
factor of the order $\gamma^2$, where $\gamma$ is the Lorentz factor of
the electrons. The inverse Compton scattering was studied by many authors
with a view toward astrophysical problems (see,
e.g.,~\cite{Jones68,Blumenthal70,Pacholcz70,Yeung73,Altshul07,Profumo08})
and the thermal QCD~\cite{Braun07}, but without those collisional effects
which are relevant if the characteristic energy of ambient photons is
 comparable to or smaller than the quasiparticle spectral width.
  The importance of the collisional broadening
  in this case is evident from the fact that,
  in the limit $k'\to 0$,
  the function (\ref{T:0}) contains divergent terms.

Note that there is another contribution to the
  radiating power associated with
 the second (off-shell) part of the field correlation
  function ${d}^{<}_{s'}(k')$ given by  Eq.~(\ref{Res-d:def}).
 Since the transverse polarization effects are caused by current
 fluctuations, this contribution may be interpreted physically as
 coming from the scattering of electrons by the current fluctuations in
 the plasma.
 It should be remembered that the formula (\ref{dR-Trans}), as
 such, is valid to first order terms in the field correlation functions.
 Therefore, in calculating the off-shell part of ${d}^{<}_{s'}(k')$,
 we have to retain only the first term in the
transverse  polarization function~(\ref{pi<s:final}). As before,
 we neglect the positron contribution and assume equal
probabilities for the electron spin states. Then, for a non-degenerate
electron subsystem,
    \beq
     \label{pi<:lowest}
  \pi^{<}_{s}(X,k)\approx
   -ie^2 \int \frac{d^4 p}{(2\pi)^4}\,
    {\rm tr}^{}_{D}\left\{
   \mynot{\epsilon}^{}_{s}(k) \widetilde{\mathcal A}(p)
   \mynot{\epsilon}^{}_{s}(k) \widetilde{{\mathcal A}}(p-k)
     \right\} f(X,\vec{p}).
    \eeq
Using this expression to calculate the off-shell term in
Eq.~(\ref{Res-d:def}), we obtain the following result for the
 local radiating power associated with the electron scattering by the
 current fluctuations:
   \beq
     \label{dR-Curr}
    % & &
 \left(\frac{dR(X,\vec{k})}{d^3 \vec{k}}\right)^{}_{\text{curr.\ fl.}}=
  \frac{e^6}{2\left(2\pi\right)^3}
   \left[1+ n(\vec{k})\right]
   %\nonumber\\[10pt]
   % & &
   %\hspace*{110pt}
   %{}\times
   \int \frac{d^4 p}{(2\pi)^4}\,\frac{d^4 p'}{(2\pi)^4}\,
   {\mathcal T}(p,p';k) f(\vec{p})f(\vec{p}^{\,\prime}),
   \eeq
 where
    \begin{eqnarray}
      \label{T:Curr}
     & &
     \hspace*{-10pt}
     {\mathcal T}(p,p';k)=
     \sum_{s,s'}
     \int \frac{d^4 k'}{(2\pi)^4}\,
     T^{}_{ss'}(p,p+k'-k;k,k')\,
     {\rm Re}\left(d^{+}_{s'}(k')\right)^2
     \nonumber\\[8pt]
      & &
      \hspace*{130pt}
     {}\times
        {\rm tr}^{}_{D}\left\{
   \mynot{\epsilon}^{}_{s'}(k') \widetilde{\mathcal A}(p')
   \mynot{\epsilon}^{}_{s'}(k') \widetilde{{\mathcal A}}(p'-k')
     \right\}.
   \end{eqnarray}
To calculate the functions $T(p;k,k')$ and ${\mathcal T}(p,p';k) $ which
determine the scattering probabilities in Eqs.~(\ref{dR-Compton}) and
(\ref{dR-Curr}), one has to specify the quasiparticle spectral
function $\widetilde{\mathcal A}(p)$. For instance, such calculations can
be performed using the parametrization (\ref{QA:Appr}) or some improved
expressions based on a more detailed analysis of the relativistic electron
self-energies $\Sigma^{\pm}(p)$ in plasmas. We leave, however, this
special problem, as well as the discussion of Eqs.~(\ref{dR-Compton}) and
(\ref{dR-Curr}) for realistic electron distributions, to future work.

\subsection{Interactions of Electrons with Longitudinal Field Fluctuations}
\label{SubSec:R-Long}
 The analysis of the
 contribution to the radiating power coming from the interaction of electrons
(positrons) with longitudinal field fluctuations proceeds exactly in
parallel with that in the previous subsection. In the case of equal
probabilities for the particle
 spin states we now have instead of Eq.~(\ref{dR-Trans})
    \begin{eqnarray}
    \label{dR-Long}
     & &
     \hspace*{-25pt}
  \left(\frac{dR(X,\vec{k})}{d^3 \vec{k}}\right)^{}_{\text{longit}}=
  i\pi e^4\sum_{s}\left[1+n^{}_{s}(X,\vec{k})\right]
  \int
     \frac{d^4 p^{}_{1}}{(2\pi)^4}\,\frac{d^4 p^{}_{2}}{(2\pi)^4}\,
     \frac{d^4 k^{\prime}}{(2\pi)^4}\,
     \nonumber\\[10pt]
    & &
     \hspace*{-15pt}
     {}\times
     \delta^4\left(p^{}_{1} +k^{\prime}-p^{}_{2}-k \right)
     L^{}_{s}(p^{}_{1},p^{}_{2};k,k')
     f^{<}(X,p^{}_{1}) f ^{>}(X,p^{}_{2})D^{<}(X,k'),
  \end{eqnarray}
where
     \begin{eqnarray}
      \label{L:gen}
      & &
      \hspace*{-20pt}
 L^{}_{s}(p^{}_{1},p^{}_{2};k,k')=
  {\rm Re}\left\{
  {\rm tr}^{}_{D}\left[
  \mynot{\epsilon}^{}_{s}g^{+}(p^{}_{1}+k')\gamma^{0}
  \widetilde{\mathcal A}(p^{}_{1})
   \gamma^{0} g^{+}(p^{}_{1}+k')\mynot{\epsilon}^{}_{s}
  \widetilde{\mathcal A}(p^{}_{2})
  \right]
  \right.
  \nonumber\\[5pt]
   & &
   \hspace*{40pt}
   {}+
  {\rm tr}^{}_{D}\left[
  \gamma^{0} g^{+}(p^{}_{1}-k)\mynot{\epsilon}^{}_{s}
  \widetilde{\mathcal A}(p^{}_{1})
  \mynot{\epsilon}^{}_{s}g^{+}(p^{}_{1}-k)\gamma^{0}
  \widetilde{\mathcal A}(p^{}_{2})
  \right]
 \nonumber\\[5pt]
   & &
    \hspace*{40pt}
   \left.
   {}+
  2\,{\rm tr}^{}_{D}\left[
  \mynot{\epsilon}^{}_{s}g^{+}(p^{}_{1}+k')\gamma^{0}
  \widetilde{\mathcal A}(p^{}_{1})
  \mynot{\epsilon}^{}_{s}g^{+}(p^{}_{1}-k)\gamma^{0}
  \widetilde{\mathcal A}(p^{}_{2})
  \right]
  \right\}.
   \end{eqnarray}
In the simplest approximation where all medium effects are omitted, this
function reduces to
   \begin{eqnarray}
     \label{L:0}
      & &
      \hspace*{-15pt}
  L^{(0)}_{s}(p^{}_{1},p^{}_{2};k,k')=
  (2\pi)^2 \eta(p^{0}_{1})\,\eta(p^{0}_{2})\,\delta(p^{2}_{1}-m^2)\,
  \delta(p^{2}_{2}-m^2)
  \nonumber\\[5pt]
   & &
   {}\times
   {\rm tr}^{}_{D}
       \left\{
  \left[
 \gamma^{0}\,
 \frac{1}{\mynot{p}^{}_{1}\,- \mynot{k}-m}\,
 \mynot{\epsilon}^{}_{s}
  +\mynot{\epsilon}^{}_{s}\,
 \frac{1}{\mynot{p}^{}_{1}\,+ \mynot{k}'-m}\,
  \gamma^{0}
   \right](\,\mynot{p}^{}_{1}\,+m)
    \right.
    \nonumber\\[5pt]
    & &
    \hspace*{50pt}
    {}\times
    \left.
    \left[
 \gamma^{0}\,
  \frac{1}{\mynot{p}^{}_{1}\,+ \mynot{k}' -m}\,
 \mynot{\epsilon}^{}_{s}
  +\mynot{\epsilon}^{}_{s}\,
 \frac{1}{\mynot{p}^{}_{1}\, - \mynot{k}-m}\,
 \gamma^{0}
   \right]
 (\,\mynot{p}^{}_{2}\,+m)
    \right\}.
   \end{eqnarray}
 Here, the trace factor
 is identical to the one in vacuum QED cross sections for
  bremsstrahlung emission of photons (the Bethe-Heitler formula)
  and for the one-photon
   electron-positron annihilation \cite{ItzZuber80}.
   Once again we see that the results of vacuum QED are
 correctly reproduced within the extended quasiparticle approximation
   formulated in Subsection~\ref{Subsec:ExtQPA}.

 Neglecting the processes involving positrons and assuming equal
 probabilities for the photon polarizations, one obtains from
  Eq.~(\ref{dR-Long}) the expression where only the contribution of
   electron scattering by the
  longitudinal field fluctuations is taken into account:
         \beq
    \label{dR-Long:Sc}
  \left(\frac{dR(X,\vec{k})}{d^3 \vec{k}}\right)^{}_{\text{longit}}=
  \frac{i\,e^4}{2(2\pi)^3} \left[1+n(\vec{k})\right]
  \int
     \frac{d^4 p}{(2\pi)^4}\,
     \frac{d^4 k^{\prime}}{(2\pi)^4}\,
     L(p;k,k')
     f(\vec{p})D^{<}(k')\, .
 \eeq
Here we have defined
 \beq
   \label{L:Scatt}
 L(p;k,k')=\sum_{s} L^{}_{s}(p,p+k'-k;k,k')\, .
 \eeq
Note that the field correlation function $D^{<}(X,k)$ is
closely related to charge fluctuations in a plasma. To show this,
 we use relation (\ref{V:rel})
 which, in the local form, yields
   \beq
    \label{Phi-rho}
    \nabla^{2}_{1}\,\Delta\hat{\phi}(\und{1})= - \Delta \hat{\varrho}(\und{1}),
   \eeq
 where $\Delta{\hat\varrho}(\und{1})=\hat\varrho(\und{1})-
 \langle \hat{\varrho}(\und{1})\rangle$ is the operator for
 fluctuations of the induced charge density. Now recalling definition
(\ref{D(12)}) of the longitudinal field Green's function leads to
   \beq
    \label{D<:rho}
   \nabla^{2}_{1}\nabla^{2}_{2}\,D^{<}(12)=
   -i\,\langle \Delta{\hat\varrho}(2)\Delta{\hat\varrho}(1)\rangle.
   \eeq
Finally, performing the Wigner transformation on both sides of this equation and
neglecting the $X$-gradients, we find that
  \beq
    \label{D<:S}
      D^{<}(X,k)= -i \,\frac{S(X,k)}{|\vec{k}|^4},
  \eeq
where
  \beq
   \label{S}
 S(X,k)=\int d^4 x\, {\rm e}^{ik\cdot x}\,
  \big\langle
  \Delta\hat{\varrho}(X-x/2)\, \Delta\hat{\varrho}(X+x/2)
  \big\rangle
 \eeq
is the local dynamic structure factor.

There are several scattering processes that can contribute to
Eq.~(\ref{dR-Long:Sc}) depending on the structure of the
 longitudinal field correlation function $D^{<}(X,k)$ in different
 regions of the $k$ space. Before discussing this point
 we will consider some important properties of this function which
  can be derived from the plasmon transport equation (\ref{L:Transp}) given in
Appendix~\ref{App:LFieldFunct}.

If the space-time variations of the drift terms on the left-hand side of
 Eq.~(\ref{L:Transp})
 are so slow that these terms are small compared
 with the emission/absorption terms on the right-hand side,
 a good approximation for the field correlation functions is obtained
by neglecting the drift terms. We arrive at the local balance
 equation
   \beq
      \label{L:balance}
   \Pi^{>}(X,k)D^{<}(X,k)=\Pi^{<}(X,k)D^{>}(X,k),
   \eeq
whence follows the well-known local equilibrium
 or the \textit{adiabatic
approximation\/} for the longitudinal field fluctuations~\cite{Dub67}:
   \beq
    \label{L:Adiab}
 D^{\mgl}_{\text{ad}}(X,k)= \Pi^{\mgl}(X,k)\big|D^{+}(X,k)\big|^2 .
   \eeq
 The adiabatic approximation is inapplicable
 in the presence of unstable plasma waves where the resonance approximation
 seems to be appropriate~\cite{Tsytovich77}. To include
 the resonance portion of the spectrum of plasma modes, Bezzerides and
DuBois~\cite{BezDub72} proposed the following ansatz for the field
 correlation functions:
    \beq
    \label{LCF:adi-res}
 D^{\mgl}(X,k)=\Delta D^{\mgl}(X,k) +
 \Pi^{\mgl}(X,k) \big|D^{+}(X,k)\big|^2,
  \eeq
where  $\Delta D^{\mgl}(X,k)$ is the contribution to $D^{\mgl}(X,k)$ in
excess of the local equilibrium value. In general, there is a number of
 resonances with frequencies $k^{}_{0}=\omega^{}_{\alpha}(X,\vec{k})$
which are to be found as solutions of the equation
  \beq
   \label{L:disp-eq}
 {\vec{k}}^{\,2}- {\rm Re}\,\Pi^{+}(X,k)=0,
  \eeq
where $\Pi^{+}(X,k)$ is the retarded component of the longitudinal
polarization function. Assuming sharp resonances, it seems reasonable to
approximate $\Delta D^{\mgl}(X,k)$ by a sum of the delta functions
corresponding to different unstable modes~\cite{BezDub72}. It should be
 noted, however, that the physical meaning of the term
$\Delta D(X,k)$ is not quite clear. In particular,
 expression (\ref{LCF:adi-res}) leads to the kinetic equation for the occupation
numbers of unstable modes which differs from the familiar plasmon
kinetic equation in plasma turbulence theory~\cite{Dub67,Kadomtsev65}.
 Our above analysis of the photon kinetics and the formal similarity between
 the photon transport equation (\ref{Kin}) and the plasmon transport equation
 Eq.~(\ref{L:Transp}) suggest another
representation for the correlation functions $D^{\mgl}(X,k)$.
 The same line of reasoning as in Section~\ref{Sec:KinEq} leads to the
 following decomposition of $D^{\mgl}(X,k)$ into
``resonant'' and ``off-shell'' parts:
    \beq
      \label{LCF:decomp}
 D^{\mgl}(X,k)=\widetilde{D}^{\mgl}(X,k) +
   \Pi^{\mgl}(X,k)\, {\rm Re}\! \left[\left( D^{+}(X,k)\right)^2\right],
   \eeq
where the first term represents the resonant contribution from unstable
plasma modes.
 Formally,
Eqs.~(\ref{LCF:decomp}) and (\ref{LCF:adi-res}) are
 of course equivalent as long as no approximations are invoked.
 The exact relation between $\Delta D^{\mgl}$ and
$\widetilde{D}^{\mgl}$ can easily be found:
   \beq
   \label{L:excess-Res}
 \Delta D^{\mgl}=\widetilde{D}^{\mgl}
 - \frac{2\Pi^{\mgl}\left( {\rm Im}\,\Pi^{+}\right)^2}
   {\left[\left({\vec{k}}^{\,2}- {\rm Re}\,\Pi^{+}\right)^2 +
 \left({\rm Im}\,\Pi^{+}\right)^2\right]^2},
   \eeq
where we have used the expression (\ref{D+-}) for $D^{+}(X,k)$.
 It is seen that $\Delta D^{\mgl}$ and
$\widetilde{D}^{\mgl}$ differ essentially from each other
 just in the vicinity of the strong resonances determined by
 Eq.~(\ref{L:disp-eq}). This means that
  Eqs.~(\ref{LCF:decomp}) and (\ref{LCF:adi-res}) give
  \textit{different prescriptions\/}
for picking out the contribution of the resonant plasma
 modes, i.e., they lead to somewhat different
 pictures of collisions involving
 plasmons. Physically, the use of the decomposition
(\ref{LCF:decomp}) rather than (\ref{LCF:adi-res})
 is preferable since the former is closely related to the
 structure of the drift terms in the plasmon transport equation.
 In the off-resonant region we have $\widetilde{G}^{\mgl}\approx 0$
 and $\Delta G^{\mgl}\approx 0$, so that both decompositions reduce
  to the adiabatic expression (\ref{L:Adiab}).

 By analogy with transverse photons it is convenient to introduce the
 spectral function for the resonant plasma modes,
  \beq
   \label{SF:plasmon}
  \widetilde{a}^{}_{L}(X,k)=i\left(\widetilde{D}^{>}(X,k)
-\widetilde{D}^{<}(X,k)\right),
  \eeq
and the plasmon distribution functions
  ${\mathcal N}^{\mgl}(X,k)$ in the four-dimensional $k$  space through the
relations
 [cf. Eqs.~(\ref{SpF})]
   \beq
    \label{L:N}
 \widetilde{D}^{\mgl}(X,k)= -i \widetilde{a}^{}_{L}(X,k)
  {\mathcal N}^{\mgl}(X,k),
 \qquad
 {\mathcal N}^{>} -{\mathcal N}^{<}=1.
   \eeq
 Then it is easy to verify that
   \beq
     \label{a-L}
 \widetilde{a}^{}_{L}(X,k)=
 - \frac{4\left({\rm Im}\,\Pi^{+}\right)^3}
  {\left[\left({\vec{k}}^{\,2}- {\rm Re}\,\Pi^{+}\right)^2 +
   \left({\rm Im}\,\Pi^{+}\right)^2\right]^2}.
   \eeq
As in the case of transverse photons, the plasmon spectral function is
non-Lorentzian.

Let us turn back to the expression (\ref{dR-Long:Sc}) for the
 local radiating power. Substituting here the field correlation
 function from Eq.~(\ref{LCF:decomp}), we obtain two terms
which can be interpreted physically. The term arising from
 $\widetilde{D}^{\mgl}$ represents the contribution of the Compton effect
 on plasmons, which may
 be regarded as a conversion of longitudinal plasma waves into
 transverse electromagnetic waves. This relativistic effect was first studied
 by Galaitis and Tsytovich~\cite{GailTsytovich64} (see also~\cite{Tsytovich89})
and has been discussed as
 a frequency boosting mechanism for laboratory beam-laser
experiments~\cite{Benford83,Newman85}. The maximum frequency boost is
 $\omega^{}_{\text{max}}/\omega^{}_{e}\approx \gamma^2$, where
 $\omega^{}_{e}$ is the electron plasma frequency (\ref{omega-e}) and
 $\gamma$ is the Lorentz factor of the beam electrons. In laser-plasma
systems, the Compton effect on plasma waves may thus be relevant when
 the energies of relativistic electrons reach the GeV region.
 Note that the transition probability determined by
 Eqs.~(\ref{L:gen}) and (\ref{L:Scatt}) is very sensitive to the
collisional broadening of the electron quasiparticle spectral function and
the electron propagators. This feature of  Compton conversion in
plasmas deserves further investigation.

To gain some insight into radiative processes associated with the second
term in Eq.~(\ref{LCF:decomp}), we consider the
 longitudinal component
$\Pi(\und{1}\,\und{2})\equiv \Pi^{}_{00}(\und{1}\,\und{2})$
 of the polarization matrix given by the diagrams in
 Fig.~\ref{fig:SelfEn;Pol}. The leading contributions to
 $\Pi(\und{1}\,\und{2})$ come from those one-loop diagrams in which
the full electron and ion Green's functions are to be replaced by
 their quasiparticle parts. Then, in the local Wigner representation, we
 obtain
   \beq
      \label{Pi-L:full}
   \Pi^{<}(X,k)= \Pi^{<}_{\rm{el}}(X,k) + \Pi^{}_{\text{ion}}(X,k)
   \eeq
with the electron and ion polarization functions
  \begin{eqnarray}
    \label{Pi-L:el}
  & &
 \Pi^{<}_{\rm{el}}(X,k)=
  -ie^{2} \int \frac{d^4 p}{(2\pi)^4}\,
 {\rm tr}^{}_{D}\left\{\gamma^0
  \widetilde{G}^{<}_{}(X,p)\gamma^0\widetilde{G}^{>}_{}(X,p-k)
 \right\},\\[5pt]
    & &
     \label{Pi-L:ion}
 \Pi^{<}_{\text{ion}}(X,k)=-i\sum_{B}
  e^{2}_{B} \int \frac{d^4 p}{(2\pi)^4}\,
 {\rm tr}^{}_{S}\left\{
  \widetilde{\mathcal G}^{<}_{B}(X,p)\,
  \widetilde{\mathcal G}^{>}_{B}(X,p-k)
 \right\}.
  \end{eqnarray}
As shown above, the Green's functions can be
 expressed in terms of the quasiparticle spectral functions and the distribution
functions, but we will not write down the corresponding obvious formulas here.

The term $\Pi^{<}_{\rm{el}}(X,k)$ in the longitudinal polarization function
 (\ref{Pi-L:full})
 yields the contribution to the radiating power (\ref{dR-Long:Sc}) from
 relativistic electron
 scattering on off-resonant  fluctuations of the electron charge in the
 plasma. This effect is analogous to electron scattering on the current
 fluctuations discussed in the previous subsection. In the relativistic case
 both effects are important while the former dominates
 in the nonrelativistic limit. The ion term
 $\Pi^{<}_{\rm{ion}}(X,k)$  is associated with bremsstrahlung processes.
 As already stressed, the cross section for these processes reduces to
 the Bethe-Heitler cross section in vacuum QED if all medium corrections
 are removed.

 In this section we have considered the \textit{emission\/} of photons.
 The analysis of \textit{absorption\/} processes follows along
exactly the same lines by using Eqs.~(\ref{pi<s:final}) and
(\ref{Emitt-E:App2}) together with the symmetry relation
 $\pi^{>}_{s}(X,k)=\pi^{<}_{s}(X,-k)$.

 \section{Discussion and Outlook}
  \label{Sec:Discuss}
  One of the most important features of photon kinetics
   considered in this paper is the crucial role
 played by the off-shell parts of the particle and field correlation
 functions in the derivation of the photon emission rate. Note that in vacuum QED
 the separation of on-shell and off-shell states is in some sense
 trivial, since the on-shell states correspond to incoming or outgoing
 particles in a scattering process while the off-shell (virtual) states
 occur in the calculation of the S-matrix. In the kinetic theory, however,
  one is dealing with ensemble averaged correlation
  functions involving both  the resonant (quasiparticle) and
  off-shell parts. We have seen that, for a weakly coupled plasma,
 the structure of drift terms in the gradient-expanded
  KB equations provides a useful  guide to separate the quasiparticle and
   off-shell contributions to the correlation functions.
  Note that off-shell transport is also a problem of great
  interest for a proper dynamical
 treatment of stable particles and broad resonances in a dense nuclear
 medium~\cite{IvanKnollVoskr03}, but in that case it is hard to
 separate unambiguously the quasiparticle and off-shell contributions to correlation
functions  because of strong medium effects.

 Throughout the paper, we concentrated on the fundamental aspects of photon
kinetics in nonequilibrium relativistic plasmas. To compute explicitly
 the contributions from various scattering processes to the radiating
 power in the relativistic laser-plasma experiments,
  it is first of all necessary to have detailed information on the
  electron distribution function.
   As a first step
  one can use the distributions which are available at present
  from experiments and
  simulations for various laser powers~\cite{Tsung04,Pukhov03,Ren04,Antici08}.
 Other important quantities are the matrix self-energies $\Sigma^{\pm}$
which enter the propagators and the quasiparticle spectral function
  of relativistic electrons in the laser-plasma medium. The
   simple ansatz (\ref{Sig+-:Gamma}) can at best give an estimate
 of the contributions to the radiating power,
 so it would be desirable to have
 more elaborate approximations for the electron self-energies.

In closing, we would like to mention one specific feature
 of the laser-plasma medium which may have an appreciable influence on
 the QED processes.  When penetrating into the
plasma, the laser-driven relativistic electrons generate a return current
carried by the plasma electrons to maintain current
neutrality~\cite{Pukhov03}. For some time the common belief was that the
return current velocity $V^{}_{p}$ is nonrelativistic. One can easily see
this from the current
 neutralization condition $n^{}_{p}V^{}_{p}=n^{}_{b}V^{}_{b}$, where
  $n^{}_{b}$ and $n^{}_{p}$ are respectively the electron densities in the
  beam and in the bulk plasma. Indeed, since under most experimental conditions
   the dimensionless beam density $\alpha=n^{}_{b}/n^{}_{ p}$
varies from $10^{-3}$ (the dense  core plasma) to $10^{-1}$ (the plasma
corona), it is clear that  $V^{}_{p}\ll c$.
 However, recent simulations~\cite{Ren04} show that, at
high pulse intensities,  the complete current neutralization does not
occur and the plasma electrons in the beam region are relativistic. In
that case the medium has a relativistic multi-beam structure
 which can manifest itself through some peculiar features of plasma
  radiation. For instance, the hard photon production becomes possible
 in the direction opposite to
the laser-generated beam.

\section*{Acknowledgement}

This work was partially supported by the German Research Society (DFG)
under Grant SFB 652.

\setcounter{equation}{0}

\section*{Appendices}

 \appendix

\section{Energy Flux of  Radiation Field}
 \label{App:Flux}
 We start with the energy density operator of
 the radiation field in the Coulomb gauge and Heaviside's units.
Using the notation
  $x=(t,\vec{r})$ for space-time points, we have~\cite{Weinberg95}
      \beq
   \label{E:EM}
   \hat{\mathcal E}(x)=
   \frac{1}{2}\, \hat{\vec{P}}^{\, 2} +
     \frac{1}{2}\big(\vec{\nabla}\times \hat{\vec{A}}\,\big)^2,
  \eeq
 where the transverse field operators $\hat{\vec{A}}(x)$ and
 $\hat{\vec{P}}(x)=\partial\hat{\vec{A}}(x)/\partial t$ satisfy the canonical
 equal-time commutation relations
        \beq
      \label{Comm-rel}
  \big[ \hat{A}^{i}(t,\vec{r}^{}_{1}), \hat{P}^{j}(t,\vec{r}^{}_{2})
  \big]^{}_{-}= i\,\delta^{T}_{ij}\left(\vec{r}^{}_{1}
  -\vec{r}^{}_{2}\right).
   \eeq
The energy flux operator for the radiation field, $\hat{\vec{j}}(x)$, is
defined through the equation
 \beq
   \label{Flux:def}
  -i\left[\hat{\mathcal E}(x),\hat{H}^{}_{EM}(t)\right]^{}_{-}=
  -\vec{\nabla}\cdot \myhat{\vec{j}}(x)
 \eeq
with the Hamiltonian of the electromagnetic field
  \beq
    \label{H:EM}
 \hat{H}^{}_{EM}(t)=\int \hat{\mathcal E}(x)\, d^3\vec{r}\, .
  \eeq
Using Eq.~(\ref{Comm-rel}) to compute the commutator in
Eq.~(\ref{Flux:def}), we obtain
   \beq
   \label{Flux:op}
 \myhat{j}^{\,i}(x)= \frac{1}{2}\,
  \epsilon^{imn}
  \left[\hat{E}^{m}(x),\hat{B}^{n}(x)  \right]^{}_{+},
 \eeq
where $\epsilon^{ijk}$ is the completely antisymmetric unit tensor
($\epsilon^{123}=1$), and
 $[.\, , \, .]^{}_{+}$ stands for the anticommutator.
 We have also introduced the transverse electric field operator
$\hat{\vec{E}}(x)$ and the magnetic field operator $\hat{\vec{B}}(x)$:
     \beq
    \label{EB:op}
 \hat{\vec{E}}(x)= -\frac{\partial \hat{\vec{A}}(x)}{\partial t},
 \qquad
 \hat{\vec{B}}(x)=\vec{\nabla}\times \hat{\vec{A}}(x).
  \eeq
 The components of the average energy flux,
  $\vec{j}(x)=\langle\,\myhat{\vec{j}}(x)\rangle$, are represented
conveniently in the form
    \beq
      \label{Flux:ave}
   j^{\, i}(x)= \left(\vec{E}(x)\times \vec{B}(x)\right)^{i}+
    \frac{1}{2}\,\epsilon^{imn}
    \left\langle
    [\Delta \hat{E}^{m}(x),\Delta \hat{B}^{n}(x)]^{}_{+}
     \right\rangle,
    \eeq
where $\vec{E}$ and $\vec{B}$ are respectively the mean electric and
 magnetic fields, while $\Delta\hat{\vec{E}}=\hat{\vec{E}}-\vec{E}$ and
 $\Delta\hat{\vec{B}}=\hat{\vec{B}}-\vec{B}$ are the operators of the
 field fluctuations.
  Physically, the first term in
  Eq.~(\ref{Flux:ave}) is the energy flux associated with electromagnetic waves.
 The second term is the photon contribution which can be expressed in terms
 of the correlation functions $d^{<}_{ij}(X,k)$. With the aid of
 Eqs.~(\ref{EB:op}) the photon energy flux at a space-time point
 $X^{\mu}=(T,\vec{R})$ is transformed to
   \begin{eqnarray}
     \label{Flux:phot1}
     & &
     \hspace*{-20pt}
     j^{i}_{\text{phot}}(X)=
      -\frac{i}{2}\,\epsilon^{imn}\epsilon^{njl}
      \int \frac{d^4 k}{(2\pi)^4}
      \left\{
     \left(\frac{1}{2}\,\frac{\partial}{\partial T} +ik^0\right)
  \left(\frac{1}{2}\,\frac{\partial}{\partial R^{j}} +ik^{j}\right)
   d^{<}_{lm}(X,k)
      \right.
      \nonumber\\[5pt]
       & &
       \hspace*{100pt}
       {}+
       \left.
      \left(\frac{1}{2}\,\frac{\partial}{\partial T} - ik^0\right)
  \left(\frac{1}{2}\,\frac{\partial}{\partial R^{j}} - ik^{j}\right)
   d^{<}_{ml}(X,k)
       \right\}.
   \end{eqnarray}
This expression can be simplified by using conditions (\ref{Transv:W})
for $d^{<}_{ij}(X,k)$ and the identity
  $$
  \epsilon^{ijk}\epsilon^{i'j'k}=
  \delta^{}_{ii'}\,\delta^{}_{jj'}- \delta^{}_{ij'}\,\delta^{}_{ji'}\, .
  $$
 After some algebra which we omit, we find up to first-order
$X$-gradients
  \beq
    \label{Flux:phot2}
  j^{i}_{\text{phot}}(X)=i\int \frac{d^4 k}{(2\pi)^4}\,
   k^0 k^{i}\,\text{tr}\, d^{<}(X,k)+
   \frac{\partial}{\partial R^{j}}M^{}_{ij}(X),
  \eeq
where $\text{tr}\, d^{<}(X,k)=\sum_{j} d^{<}_{jj}(X,k)$, and
 \beq
   \label{M}
  M^{}_{ij}(X)= \frac{i}{2} \int \frac{d^4 k}{(2\pi)^4}\,
  k^0\left\{
    d^{<}_{ij}(X,k) - d^{<}_{ji}(X,k)
  \right\}.
 \eeq
Strictly speaking, Eq.~(\ref{Flux:phot2}) contains the \textit{full}
tensor $d^{<}_{ij}(X,k)$, and not just its transverse part with respect to
$\vec{k}$. However, as follows directly from Eq.~(\ref{Tens-Transv}), we
have up to first-order $X$-gradients
  \beq
     \label{Tr:T}
   \text{tr}\,T(X,k)= \text{tr}\,T^{\perp}(X,k)
  \eeq
for any $T^{}_{ij}(X,k)$ which satisfies conditions (\ref{Transv:W}). The
full tensor $d^{<}_{ij}(X,k)$ in Eq.~(\ref{Flux:phot2}) can thus be
replaced by its transverse part. Note that, in the slow variation case,
the last term in Eq.~(\ref{Flux:phot2}) is very small compared to the
first term. Moreover, it is easy to verify that $M^{}_{ij}=0$ in the
diagonal principal-axis representation (\ref{expan-pol}) for the
 transverse correlation functions $d^{<}_{ij}(X,k)$.
We thus obtain the expression
  \beq
     \label{Flux:d}
  j^{i}_{\text{phot}}(X)=i\sum_{s} \int
   \frac{d^4 k}{(2\pi)^4}\,k^0 k^{i}\,d^{<}_{s}(X,k)
  \eeq
which is valid up to first-order $X$-gradients.

\setcounter{equation}{0}

\section{Decomposition of Electron Correlation Functions}
 \label{App:Electrons}
We will show how the decomposition (\ref{G<>:dec}) follows from transport
equations for the electron correlation functions. To a degree our
consideration is similar to that given in Subsection~\ref{Subs:ResVirtPhot} for
photons.

The starting point are the Wigner transformed
KB equations (\ref{KB-12}). Keeping only first-order terms in $X$-gradients, we
obtain the set of equations (arguments $X$ and $p$ are omitted for
brevity)
   \begin{subequations}
   \label{G<>:W}
 \begin{eqnarray}
 \label{G<>:W1}
  & &
 \frac{1}{2}\left\{(g^{+})^{-1},G^{\mgl}\right\}
 - \frac{1}{2}\left\{\Sigma^{\mgl},G^{-}\right\}
  =
  i \left(\Sigma^{\mgl} G^{-} - (g^{+})^{-1} G^{\mgl}\right),
 \\[10pt]
  \label{G<>:W2}
 & &
 \frac{1}{2}\left\{G^{\mgl},(g^{-})^{-1}\right\}
 - \frac{1}{2}\left\{G^{+},\Sigma^{\mgl}\right\}
  =
 i \left( G^{+}\Sigma^{\mgl} - G^{\mgl}(g^{-})^{-1} \right),
 \end{eqnarray}
 \end{subequations}
where $g^{\pm}(X,p)$ are the local propagators~(\ref{g+-}). Note that,
generally speaking, on the right-hand sides of these equations the full
propagators $G^{\pm}(X,p)$ cannot be replaced by the local ones since
Eqs.~(\ref{G+-}) contain the gradient corrections.

The transport equations for $G^{\mgl}(X,p)$ are derived by taking the
difference of Eqs.~(\ref{G<>:W}). With expressions (\ref{g+-}) for the
local propagators and the relations
 \begin{equation}
  \label{G-Sig:ident}
  G^{>}-G^{<}=G^{+}- G^{-},
  \qquad
  \Sigma^{>}- \Sigma^{<}=\Sigma^{+}- \Sigma^{-},
 \end{equation}
a simple algebra gives
     \begin{eqnarray}
     \label{Kin:e}
     & &
     \hspace*{-15pt}
      \frac{1}{2} \left(
                \left\{(g^{+})^{-1}, G^{\mgl}\right\}
               - \left\{G^{\mgl},(g^{-})^{-1}\right\}
                  \right)
     + \frac{1}{2} \left(
                   \left\{g^{+},\Sigma^{\mgl}  \right\}
                   - \left\{\Sigma^{\mgl}, g^{-}\right\}
                  \right)
     \nonumber\\[5pt]
   & &
   {}=
   - \big[g,\Sigma^{\mgl}\big]^{}_{-}
   - \left[\,\mmynot{\mPi} -m -\sigma, G^{\mgl}\right]^{}_{-}
      + \frac{i}{2}\left(
   \left[ \Sigma^{>},G^{<} \right]^{}_{+}
    - \left[ \Sigma^{<},G^{>} \right]^{}_{+}
      \right),
  \end{eqnarray}
where $[A ,B]^{}_{\mp}=AB\mp BA$ is the commutator/anticommutator of spinor
matrices, and
   \begin{equation}
  \label{g-sig}
 g= \frac{1}{2}\left(G^{+} + G^{-} \right),
 \quad
 \sigma=\frac{1}{2}\left(\Sigma^{+} + \Sigma^{-} \right).
 \end{equation}
With regard to its spinor structure, Eq.~(\ref{Kin:e}) is a very
complicated $4\times 4$ matrix equation. In principle, upon multiplying
both sides of this equation by the Lorentz invariants\ $I,\gamma^{}_{\mu},
\gamma^{}_{5},\gamma^{}_{5}\gamma^{}_{\mu}, \sigma^{}_{\mu\nu}$, and then
taking the trace, one obtains coupled transport equations for the scalar,
vector, pseudo-scalar, axial-vector, and tensor components of the
correlation functions. It suffices for our purpose, however, to consider
only one of these transport equations, which is obtained from
Eq.~(\ref{Kin:e}) by taking the trace of both sides. A useful device for
manipulating the traces of the drift terms are the matrix identities
 which follow directly from the definition (\ref{Poiss}) of the
 Poisson bracket:
  \begin{equation}
    \label{Poiss:ident}
  {\rm tr}\left\{A,B\right\}=-\,{\rm tr}\left\{B,A\right\},
  \qquad
  {\rm tr}\left\{A,B\right\}=
  -\,{\rm tr}\left\{A^{-1}, ABA\right\}.
  \end{equation}
With the aid of these identities the traces of the drift terms in
Eq.~(\ref{Kin:e}) are rearranged as
 \begin{eqnarray*}
  & &
  \frac{1}{2}\, {\rm tr}^{}_{D}\left(
                \left\{(g^{+})^{-1}, G^{\mgl}\right\}
               - \left\{G^{\mgl},(g^{-})^{-1}\right\}
                  \right)=
       {\rm tr}^{}_{D}
       \left\{\,\mmynot{\mPi} - m -\sigma,G^{\mgl}\right\},
       \\[8pt]
   & &
  \frac{1}{2}\, {\rm tr}^{}_{D}
  \left( \left\{g^{+},\Sigma^{\mgl}  \right\}
    - \left\{\Sigma^{\mgl}, g^{-}\right\}\right)
  =
  \frac{1}{2}\,{\rm tr}^{}_{D}
  \left( \left\{g^{+} + g^{-}, \Sigma^{\mgl}\right\} \right)
 \\[5pt]
 & &
 \hspace*{60pt}
 {}= -\, \frac{1}{2}\, {\rm tr}^{}_{D}
 \left(\left\{(g^{+})^{-1}\! , g^{+}\Sigma^{\mgl} g^{+}\right\}
 + \left\{(g^{-})^{-1}\!, g^{-}\Sigma^{\mgl} g^{-}
 \right\}\right).
 \end{eqnarray*}
Then the trace of Eq.~(\ref{Kin:e}) can be written in the form
  \begin{eqnarray}
 \label{Kin:e1}
  & &
  \hspace*{-15pt}
  {\rm tr}^{}_{D}
  \left\{\,\mmynot{\mPi} -m -\sigma, G^{\mgl}-\frac{1}{2}
  \left( g^{+}\Sigma^{\mgl} g^{+} +
   g^{-}\Sigma^{\mgl} g^{-}\right)
   \right\}
    \nonumber\\[5pt]
    & &
   \hspace*{10pt}
 {}+\frac{1}{4}\,{\rm tr}^{}_{D}\left\{\Delta\Sigma, g^{+}\,\Sigma^{\mgl} g^{+}
 - g^{-}\,\Sigma^{\mgl} g^{-}
  \right\}
   = i\,{\rm tr}^{}_{D}\left(\Sigma^{>} G^{<} - \Sigma^{<}G^{>}
     \right),
 \end{eqnarray}
where $\Delta\Sigma=\Sigma^{+} -\Sigma^{-}$. For a weakly coupled plasma, the
first term on the left-hand side of Eq.~(\ref{Kin:e1}) dominates. It is seen
that this term contains just that part of $G^{\mgl}$ which is denoted as
$\widetilde{G}^{\mgl}$ in Eq.~(\ref{G<>:dec}). By analogy with photons, the
quantities $\widetilde{G}^{\mgl}$ may be identified as the quasiparticle
parts of the electron correlation functions, whereas the additional term in
 Eq.~(\ref{G<>:dec}) represents the off-shell parts.
 This interpretation is confirmed by
 the spectral properties of $\widetilde{G}^{\mgl}$ discussed
 in Section~\ref{Sec:Electrons}. It is also important to observe here that
 the off-shell parts do not contribute to the collision term on
 the right-hand of Eq.~(\ref{Kin:e1}) and hence we get a transport equation
 which involves only $\widetilde{G}^{\mgl}$:
  \begin{eqnarray}
   \label{Kin:e-quasi}
    & &
    \hspace*{-15pt}
   {\rm tr}^{}_{D}
  \left\{\,\mmynot{\mPi} -m -\sigma, \widetilde{G}^{\mgl}\right\}
  +\frac{1}{4}\,{\rm tr}^{}_{D}\left\{\Delta\Sigma, g^{+}\,\Sigma^{\mgl} g^{+}
 - g^{-}\,\Sigma^{\mgl} g^{-}
  \right\}
   \nonumber\\[5pt]
   & &
   \hspace*{190pt}
   {}=
   i\,{\rm tr}^{}_{D}\left(\Sigma^{>} \widetilde{G}^{<}
    - \Sigma^{<} \widetilde{G}^{>}\right).
  \end{eqnarray}
This equation is analogous to the transport equation (\ref{Kin2}) for
resonant photons.

\setcounter{equation}{0}

 \section{Two-Loop Contributions to Polarization Functions}
 \label{App:TwoLoop}

Here we examine the contributions to polarization functions, which are
generated by the second (two-loop) diagram for the polarization matrix
$\Pi^{}_{\mu\nu}(\und{1}\,\und{2})$ in Fig.~\ref{fig:SelfEn;Pol}.
     \begin{figure}[h]
 \centerline{\includegraphics[width=0.38\linewidth]{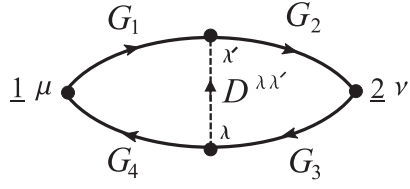}}
 \caption{\label{fig:Pi_two_loop} Two-loop diagrams for
  $i\Pi^{}_{\mu\nu}(\und{1}\,\und{2})$.
  }
 \end{figure}
 For convenience, we assign indices to the electron Green's functions
  as shown in Fig.~\ref{fig:Pi_two_loop}. Then, recalling the expression
   (\ref{Gam-e:bare}) for the bare four-vertex, we have
   \begin{eqnarray}
   \label{Pi:two-loop}
    & &
     \hspace*{-20pt}
   \left.\Pi^{}_{\mu\nu}(\und{1}\,\und{2})\right|^{}_{\text{2-loop}}=
   e^4 D^{\lambda\lambda'}(\und{6}'\,\und{3}')
   \nonumber\\[3pt]
    & &
   {}\times {\rm tr}^{}_{D}\left\{
   \delta^{}_{\mu\sigma^{}_{1}}(\und{1}'-\und{1})\gamma^{\sigma^{}_{1}}\,
              G^{}_{1}(\und{1}'\,\und{2}')\,
    \delta^{}_{\lambda'\sigma^{}_{2}}(\und{2}'-\und{3}')\gamma^{\sigma^{}_{2}}\,
              G^{}_{2}(\und{2}'\,\und{4}')\,
   \right.
   \nonumber\\[3pt]
    & &
    \hspace*{60pt}
    {}\times
    \left.
  \delta^{}_{\nu\sigma^{}_{3}}(\und{4}'-\und{2})\gamma^{\sigma^{}_{3}}\,
              G^{}_{3}(\und{4}'\,\und{5}')\,
  \delta^{}_{\lambda\sigma^{}_{4}}(\und{5}'-\und{6}')\gamma^{\sigma^{}_{4}}\,
              G^{}_{4}(\und{5}'\,\und{1}')
    \right\}.
   \end{eqnarray}
For the polarization functions
 $\Pi^{<}_{\mu\nu}(12)=\Pi^{}_{\mu\nu}(1^{}_{+}\,2^{}_{-})$, this formula
 generates
 several terms which involve different canonical components of $D^{\lambda\lambda'}$
 and $G^{}_{i}$. Let ${\mathcal P}^{<}_{\mu\nu}(12)$ be one of these terms
 with some space-time functions ${\mathcal D}^{\lambda\lambda'}(12)$ and
  ${\mathcal G}^{}_{i}(12)$. In the local Wigner form, we obtain
   (the fixed argument $X$ is omitted for brevity)
     \begin{eqnarray}
     \label{P-mn2}
     & &
     \hspace*{-20pt}
 {\mathcal P}^{<}_{\mu\nu}(k)=\frac{e^4}{(2\pi)^8}
  \int  d^4 k'\,
  \prod^{4}_{i=1} d^4 p^{}_{i}\,
   \delta^4(k- p^{}_{2}+ p^{}_{3})\,
 \delta^4(k'- p^{}_{3}+ p^{}_{4})\,
  {\mathcal D}^{\lambda\lambda'}(k')
   \nonumber\\[5pt]
    & &
    {}\times\delta^4(p^{}_{1}+p^{}_{3} -p^{}_{2} - p^{}_{4})\,
     {\rm tr}^{}_{D}
     \left\{
  \hat{\gamma}^{}_{\mu}(k)\,
   {\mathcal G}^{}_{1}(p^{}_{1})\,
  \hat{\gamma}^{}_{\lambda'}(k')\,
  {\mathcal G}^{}_{2}(p^{}_{2})
  \right.
   \nonumber\\[5pt]
  & &
    \hspace*{170pt}
    {}\times
     \left.
   \hat{\gamma}^{}_{\nu}(k)\,
    {\mathcal G}^{}_{3}(p^{}_{3})\,
 \hat{\gamma}^{}_{\lambda}(k')\,
  {\mathcal G}^{}_{4}(p^{}_{4})
     \right\},
  \end{eqnarray}
where we have used the notation (\ref{delta-k:mu}). Let us now compare
 Eq.~(\ref{P-mn2}) with the one-loop contribution to $\Pi^{<}_{\mu\nu}(X,k)$ that
corresponds to the first diagram for
 $\Pi^{}_{\mu\nu}(\und{1}\,\und{2})$ in Fig.~\ref{fig:SelfEn;Pol}:
  \begin{eqnarray}
   \label{Pi-mn1:W}
   & &
  \left.\Pi^{<}_{\mu\nu}(k)\right|^{}_{\text{1-loop}}=
 -i\,\frac{e^2}{(2\pi)^4}
  \int d^4 p^{}_{1}\,d^4 p^{}_{2}\,\delta^4(k+p^{}_{2}-p^{}_{1})
   \nonumber\\[5pt]
   & &
   \hspace*{140pt}
   {}\times {\rm tr}^{}_{D}
    \left\{
   \hat{\gamma}^{}_{\mu}(k)\,G^{<}(p^{}_{1})\,
 \hat{\gamma}^{}_{\nu}(k)\,G^{>}(p^{}_{2})
    \right\}.
  \end{eqnarray}
A simple analysis based on Eq.~(\ref{Pi:two-loop}) shows that each term
(\ref{P-mn2}) involves at least one electron correlation function $G^{<}$
 (${\mathcal G}^{}_{1}(p^{}_{1})$ or ${\mathcal G}^{}_{2}(p^{}_{2})$),
and at least one correlation function
 $G^{>}$  (${\mathcal G}^{}_{3}(p^{}_{3})$ or ${\mathcal G}^{}_{4}(p^{}_{4})$).
 As discussed in
 Section~\ref{Sec:Electrons}, the leading approximation for the two-loop
  contributions to the polarization functions is obtained by
 replacing $G^{\mgl}(p)$ by their quasiparticle parts
  $\widetilde{G}^{\mgl}(p)$ which are
   sharply peaked about the mass-shell. It is clear that, due to the
   four-dimensional delta functions, the terms (\ref{P-mn2}) with
   products $\widetilde{G}^{<}(p^{}_{1})\,\widetilde{G}^{>}(p^{}_{4})$
    and $\widetilde{G}^{<}(p^{}_{2})\,\widetilde{G}^{>}(p^{}_{3})$
     correspond to the same scattering process as the one-loop term
     (\ref{Pi-mn1:W}) in which $G^{\mgl}$ are replaced
     by $\widetilde{G}^{\mgl}$. For weakly coupled plasmas, such two-loop
     corrections may be neglected\footnote{Examples of the corresponding diagrams
     for the transverse polarization functions are given in
      Fig.~\ref{fig:One_loop_correct}.}. On the other hand,
       the terms (\ref{P-mn2}) with
    $\widetilde{G}^{<}(p^{}_{1})\,\widetilde{G}^{>}(p^{}_{3})$
    and $\widetilde{G}^{<}(p^{}_{2})\,\widetilde{G}^{>}(p^{}_{4})$
    describe new scattering processes and therefore must be retained. The
    corresponding diagrams for the transverse polarization functions are
    shown in Fig.~\ref{fig:pi_higher_order}.

 \setcounter{equation}{0}

\section{Electromagnetic Energy Production }
 \label{App:EnProd}
In the notation of Appendix~\ref{App:Flux}, the average electromagnetic
energy production in a unit volume due to the interaction with matter can
be written as
      \beq
   \label{E:EM-prod}
   \frac{\partial {\mathcal E}(x)}{\partial t}=
   -i\big\langle \big[\hat{\mathcal E}(x), H^{}_{\rm int }(t)\big]^{}_{-}\big\rangle,
  \eeq
 where ${\mathcal E}(x)=\langle \hat{\mathcal E}(x)\rangle$ and
     $$
   H^{}_{\rm int }(t)=
 - \int d^3\vec{r}\, \hat{\vec A}(x)\cdot \hat{\vec J}^{\,T}(x)
    $$
is the interaction Hamiltonian. Using the relations (\ref{Comm-rel}) to
calculate the commutator in Eq.~(\ref{E:EM-prod}) gives
  \beq
    \label{E:prod2}
\frac{\partial {\mathcal E}(x)}{\partial t}=
  \big\langle \hat{\vec{P}}(x)\cdot\hat{\vec J}^{\,T}(x)\big\rangle.
  \eeq
We now identify $x$ with the space-time point $X=(T,\vec{R})$ in the
kinetic picture and rewrite Eq.~(\ref{E:prod2}) as
  \beq
    \label{E:prod3}
 \frac{\partial {\mathcal E}(X)}{\partial T}=
 - \vec{E}(X)\cdot \vec{J}^{\,T}(X)
   + \left(\frac{\partial {\mathcal E}(X)}{\partial T}\right)^{}_{\rm phot},
  \eeq
where $\vec{E}(X)= -\,\partial \vec{A}(X)/\partial T$ is the mean electric
field, and $\vec{J}^{\,T}(X)$ is the mean transverse current density. The
second term in Eq.~(\ref{E:prod3}) is associated with photons:
      \begin{eqnarray}
    \label{Prod:E-phot}
    & &
    \hspace*{-20pt}
\left(\frac{\partial {\mathcal E}(X)}{\partial T}\right)^{}_{\rm phot}=
 \frac{1}{2}
 \int d^4 x^{}_{1}\,d^4 x^{}_{2}\,
  \delta^4(x^{}_{1}-X)\,\delta^4(x^{}_{2}-X)
   \nonumber\\[5pt]
    & &
    \hspace*{85pt}
   {}\times\frac{\partial}{\partial t^{}_{2}}
   \left[
  \big\langle \Delta \hat{J}^{\,T}_{i}(1)\,\Delta\hat{A}^{}_{i}(2) \big\rangle
   +
 \big\langle\Delta\hat{A}^{}_{i}(2)\,\Delta \hat{J}^{\,T}_{i}(1)\big\rangle
   \right],
  \end{eqnarray}
where
 $\Delta\hat{J}^{\,T}_{i}(1)=\hat{J}^{\,T}_{i}(x^{}_{1})-{J}^{\,T}_{i}(x^{}_{1})$
 and
$\Delta \hat{A}^{}_{i}(2)=\hat{A}^{}_{i}(x^{}_{2})-{A}^{}_{i}(x^{}_{2})$.
 The correlation functions in the above formula can be written in terms
 of the field Green's functions and the polarization functions by noting that
    $$
  \frac{1}{\langle S\rangle}\,
  \big\langle
 T^{}_{C}\big\{S\,\Delta\hat{J}^{\,T}_{Ii}(\und{1})\,
 \Delta\hat{A}^{}_{Ij}(\und{2})\big\}
  \big\rangle
  = i \,\frac{\delta J^{T}_{i}(\und{1})}{\delta
  J^{\,(\text{ext})j}(\und{2})}=
  - i\,\Pi^{}_{ij'}(\und{1}\,\und{1}')\,D^{}_{j'j}(\und{1}'\,\und{2})
  $$
as follows directly from Eqs.~(\ref{ave:C}) and (\ref{S-C}). In the
 physical limit one obtains
     \begin{eqnarray*}
    & &
    \big\langle \Delta \hat{J}^{\,T}_{i}(1)\,\Delta\hat{A}^{}_{i}(2) \big\rangle
   = -i\left[
    \pi^{>}_{ij}(11')\,d^{-}_{ji}(1'2)+
  \pi^{+}_{ij}(11')\,d^{>}_{ji}(1'2)
    \right],
    \\[5pt]
    & &
 \big\langle\Delta\hat{A}^{}_{i}(2)\,\Delta \hat{J}^{\,T}_{i}(1)\big\rangle
  =
 -i\left[
    \pi^{+}_{ij}(11')\,d^{<}_{ji}(1'2)+
  \pi^{<}_{ij}(11')\,d^{-}_{ji}(1'2)
    \right].
  \end{eqnarray*}
After inserting these expressions into Eq.~(\ref{Prod:E-phot}) we use the
rule (\ref{Transf:W}) to express all functions in terms of their
Wigner transforms, and then go over to the principal-axis representation
(\ref{expan-pol}). These manipulations
 give (on the right-hand side the $X$-dependence is not
 shown explicitly)
     \begin{eqnarray}
     \label{Prod:E-phot2}
     & &
     \hspace*{-20pt}
 \left(\frac{\partial {\mathcal E}(X)}{\partial T}\right)^{}_{\rm phot}=
  -\,\frac{i}{2}\sum_{s}
  \int \frac{d^4 k}{(2\pi)^4}
  \left(
   \frac{1}{2}\,\frac{\partial}{\partial T}+ ik^{}_{0}
  \right)
  \nonumber\\[5pt]
     & &
    \hspace*{50pt}{}
    \times
  \left[
   \pi^{+}_{s}(k)\big(d^{>}_{s}(k)+d^{<}_{s}(k)\big)+
  \big(\pi^{>}_{s}(k)+\pi^{<}_{s}(k)\big)d^{-}_{s}(k)
  \right].
  \end{eqnarray}
 Here the contributions from the Poisson brackets have
been omitted since they are small in the kinetic regime.
 For the same reason, the term with the derivative
$\partial/\partial T$ can be neglected since it gives a relatively small
correction to the energy density ${\mathcal E}(X)$, which is unimportant
for our present purposes. The remaining integral in
Eq.~(\ref{Prod:E-phot2}) represents the photon contribution to the local
electromagnetic energy balance.
  With the symmetry relations for the transverse functions
   \begin{eqnarray*}
    & &
    d^{\mgl}_{s}(X,-k)= d^{\mlg}_{s}(X,k),
    \qquad
   \pi^{\mgl}_{s}(X,-k)= \pi^{\mlg}_{s}(X,k),
    \\[5pt]
   & &
  d^{\pm}_{s}(X,-k)= d^{\mp}_{s}(X,k),
    \qquad
   \pi^{\pm}_{s}(X,-k)= \pi^{\mp}_{s}(X,k),
   \end{eqnarray*}
we get the expression in which the integration is over $k^{}_{0}>0$:
    $$
  \left(\frac{\partial {\mathcal E}(X)}{\partial T}\right)^{}_{\rm phot}=
 \sum_{s} \int\frac{d^4 k}{(2\pi)^4}\,\theta(k^{}_{0})
  k^{}_{0}\left[\pi^{>}_{s}(X,k)d^{<}_{s}(X,k)
   -\pi^{<}_{s}(X,k)d^{>}_{s}(X,k)   \right].
   $$
Recalling the decomposition (\ref{Res-d:def}), we see that here the field
correlation functions ${d}^{\mgl}_{s}(X,k)$ can be replaced by their
 resonant parts  $\widetilde{d}^{\,\mgl}_{s}(X,k)$.
 Then, using the relations~(\ref{SpF}) and (\ref{N(p)})
 leads to Eq.~(\ref{Emitt-E:App2}).

\setcounter{equation}{0}

\section{Photon Dispersion}
 \label{App:Dispers}
 The dispersion relation for photons follows from Eq.~(\ref{DispEq}). To solve
this equation, one needs an explicit expression for
 ${\rm Re}\,\pi^{+}_{s}(X,k)$. We start with the space-time functions
  $\pi^{\pm}_{ij}(12)$ which can be written in terms of the transverse
  polarization matrix $\Pi^{}_{ij}(\und{1}\,\und{2})$. Recalling the
  canonical notation (\ref{F:can}) for functions on the contour $C$, we have
   \begin{equation}
    \label{pi(12)}
    \pi^{\pm}_{ij}(12)=
    \pm\,\Pi^{}_{ij}(1^{}_{\pm}\,2^{}_{\pm})
     \mp \,\Pi^{}_{ij}(1^{}_{+}\,2^{}_{-}).
   \end{equation}
 Here it is sufficient to retain
only the contribution of the one-loop diagram
 (see Fig.~\ref{fig:SelfEn;Pol})
that dominates in the case of a weakly coupled plasma. Then the functions
(\ref{pi(12)}) are given by the space-time diagrams
  \begin{equation}
    \label{pi(12)-diagr}
 \pi^{\pm}_{ij}(12)=\,
  \raisebox{-26pt}{\includegraphics[scale=0.4]{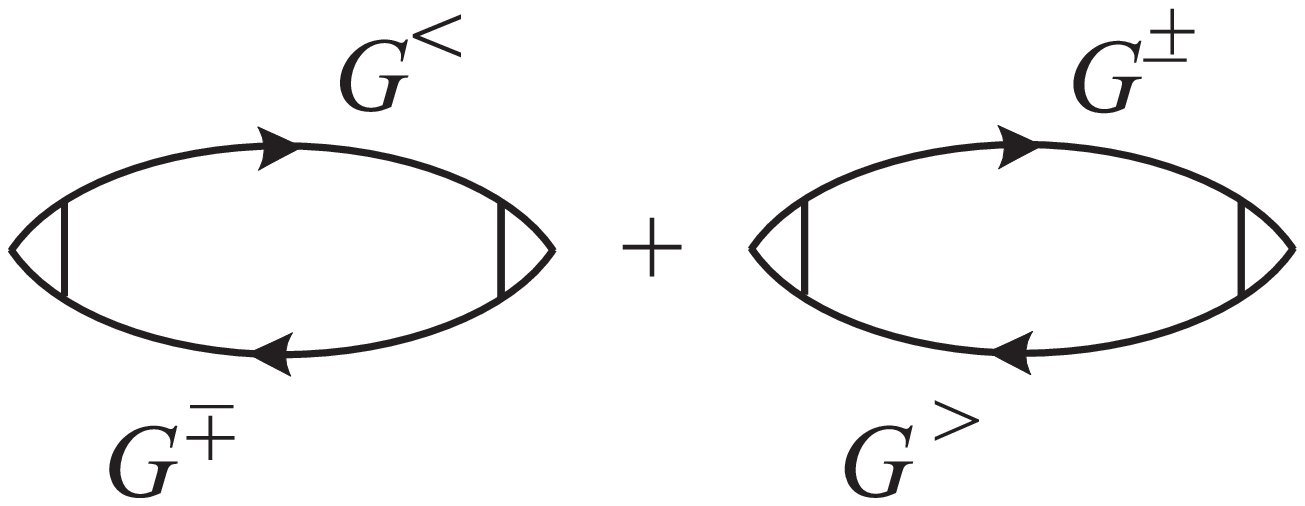}}
  \end{equation}
Going over to the local Wigner representation and using the first
 of Eqs.~(\ref{ReIm:+}), we obtain
  \begin{eqnarray}
    \label{Re-pi+}
     & &
     \hspace*{-15pt}
   {\rm Re}\,\pi^{+}_{ij}(X,k) = -ie^2
   \Delta^{\perp}_{ii'}(\vec{k})\, \Delta^{\perp}_{jj'}(\vec{k})
    \nonumber\\[5pt]
     & &
 \hspace*{20pt}
     {}\times
     \int \frac{d^4p}{(2\pi)^4}\, {\rm tr}^{}_{D}
     \left\{\gamma^{i'}\left[g(X,p+k) + g(X,p-k)\right]
     \gamma^{j'}\,G^{<}(X,p)
     \right\},
  \end{eqnarray}
where $\Delta^{\perp}_{ii'}(\vec{k})$ is the transverse projector
(\ref{projec}), and
  $$
  g(X,p)=\frac{1}{2}\left[ g^{+}(X,p) +  g^{-}(X,p)   \right].
  $$
Note that in  Eq.~(\ref{Re-pi+}) the off-shell part
of $G^{<}(X,p)$  must be dropped since it gives the contribution
of the same order as the two-loop diagram in Fig.~\ref{fig:SelfEn;Pol},
which  has been neglected. In other words, the full function
  $G^{<}(X,p)$ is to be replaced by its quasiparticle part
  $\widetilde{G}^{<}(X,p)$. To the leading approximation
  for a weakly coupled plasma, the local
  propagators $g^{\pm}(X,p)$ and the correlation functions
   $\widetilde{G}^{<}(X,p)$ may be taken in the collisionless form,
   Eqs.~(\ref{g+-:free}) and (\ref{G<>:free}).
   Then, by changing the integration variable $p$ into $p+eA(X)$, the polarization
   tensor (\ref{Re-pi+}) is expressed in terms of the gauge-invariant
   distribution functions (\ref{f-Ginv}). After some algebra, we
   have
     \begin{eqnarray}
       \label{Re-pi+:fin}
       & &
   {\rm Re}\,\pi^{+}_{ij}(X,k)=8\pi e^2
     \Delta^{\perp}_{ij}(\vec{k})
      \nonumber\\[5pt]
       & &
     {}\times\int \frac{d^4p}{(2\pi)^4}
      \left[1+ \frac{k^4}{4}\cdot
      \frac{\wp}{(p\cdot k)^2 - k^4/4}   \right]
      \eta(p^0)\,\delta\big(p^2 -m^2\big) f^{<}(X,p)
       \nonumber\\[8pt]
       & &
       -8\pi e^2 k^2\,
       \Delta^{\perp}_{ii'}(\vec{k})\Delta^{\perp}_{jj'}(\vec{k})
       \nonumber\\[5pt]
       & &
   {}\times\int \frac{d^4p}{(2\pi)^4}\,
  \frac{\wp}{(p\cdot k)^2 - k^4/4}\,
    p^{}_{i'}p^{}_{j'}\,  \eta(p^0)\,\delta\big(p^2 -m^2\big) f^{<}(X,p),
     \end{eqnarray}
 where $\wp$ denotes the principal  values of integrals.
 Two comments are relevant concerning the above expression. First, it should be
 noted that  $ {\rm Re}\,\pi^{+}_{ij}(X,k)$ contains a divergent vacuum
 term which is to be eliminated by applying the procedure of
 vacuum QED (for a discussion of this point see Bezzerides and DuBois
 \cite{BezDub72}). In what follows we
 ignore this vacuum term, assuming the physical mass and charge throughout.
 Second, for experimental conditions available at present, the positron
 contribution to $f^{<}(X,p)$ may be neglected in calculating
 $ {\rm Re}\,\pi^{+}_{ij}(X,k)$. Therefore, in
 Eq.~(\ref{Re-pi+:fin}) we shall make the replacement
   \begin{equation}
    \label{factor:e}
 \eta(p^0)\,\delta\big(p^2 -m^2\big) f^{<}(X,p)=
 \frac{\delta\big(p^0-E^{}_{p}\big)}{2E^{}_{p}}\,f(X,\vec{p}),
   \end{equation}
where $E^{}_{p}=\sqrt{\displaystyle|\vec{p}\,|^{\,2} +m^2}$, and
 $f(X,\vec{p})\equiv f^{}_{e^{-}}(X,\vec{p})$ is the electron distribution function.

 The next step is to find  eigenvectors and eigenvalues of the
 polarization tensor (\ref{Re-pi+:fin}). In general this is a rather
 complicated problem which requires
a knowledge of the nonequilibrium particle distribution functions. Note,
however, that our main interest is in the region of sufficiently high
frequencies $\omega=|k^0|$ where the dispersion curve for transverse photons
is close to the vacuum limit $\omega=|\vec{k}|$. Since in this region the
quantity $k^2=\omega^2 -\vec{k}^{\, 2}$
 is small compared to $\omega^2$, a simple and
 reasonable approximation for the polarization tensor can be obtained
 from Eq.~(\ref{Re-pi+:fin}) by setting $k^2=0$. Taking also Eq.~(\ref{factor:e})
 into account, we obtain a locally isotropic tensor
   \begin{equation}
    \label{Re-pi+:is}
  {\rm Re}\,\pi^{+}_{ij}(X,k)=\omega^{2}_{e}(X)
   \left(\delta^{}_{ij} - \frac{k^{}_{i}k^{}_{j}}{|\vec{k}|^2}\right)
   \end{equation}
with $\omega^{}_{e}(X)$  defined in Eq.~(\ref{omega-e}). In this case
 $ {\rm Re}\,\pi^{+}_{s}(X,k)=\omega^{2}_{e}(X)$, so that
 the dispersion equation (\ref{DispEq}) reduces to
 $k^2 - \omega^{2}_{e}(X)=0$, whence follows the expression (\ref{omega-s:isotr})
 for the effective photon frequencies.
If $\omega^2\gg\omega^{2}_{e}$, the anisotropic corrections to the polarization
tensor~(\ref{Re-pi+:fin}) are relatively small.
The leading anisotropic contribution comes from
the last term and is given by
  \begin{eqnarray}
     \label{Re-pi+:anis}
     & &
     \hspace*{-15pt}
     \left.
  {\rm Re}\,\pi^{+}_{ij}(X,k)
     \right|^{}_{\rm anisotr}=
     -4\pi e^2 k^2\,
     \Delta^{\perp}_{ii'}(\vec{k})\Delta^{\perp}_{jj'}(\vec{k})
  \nonumber\\[5pt]
   & &
    \hspace*{80pt}
   {}\times
   \int \frac{d^4 p}{(2\pi)^4}\,\frac{\wp}{(p\cdot k)^2}\,
    p^{}_{i'}p^{}_{j'}\,\frac{\delta\big(p^0 -E^{}_{p}\big)}{E^{}_{p}}\,
    f(X,\vec{p}),
  \end{eqnarray}
where we have used Eq.~(\ref{factor:e}).

 \setcounter{equation}{0}

 \section{Longitudinal Field Correlation Functions}
  \label{App:LFieldFunct}
We start with the equation of motion for the longitudinal field Green's
function $D(\und{1}\,\und{2})\equiv D^{00}(\und{1}\,\und{2})$
 on the time-loop contour $C$.
 Recalling Eqs.~(\ref{Dmn:eq}) and~(\ref{Block:Isotr}), we have
  \beq
     \label{D:eq}
  -\nabla^{2}_{1}D(\und{1}\,\und{2})= \delta(\und{1}-\und{2})
   + \Pi(\und{1}\,\und{1}')D(\und{1}'\,\und{2}).
  \eeq
The adjoint of this equation reads
  \beq
     \label{D:eq-adj}
  -\nabla^{2}_{2}D(\und{1}\,\und{2})= \delta(\und{1}-\und{2})
   + D(\und{1}\,\und{1}')\Pi(\und{1}'\,\und{2}).
  \eeq
Then, using the canonical form (\ref{F:can}) of $D(\und{1}\,\und{2})$ and
$\Pi(\und{1}\,\und{2})$, it is an easy matter to derive the
equations for
the retarded and advanced longitudinal ``propagators''
    \begin{subequations}
   \label{L:+-:ab}
   \begin{eqnarray}
     \label{L:+-:a}
    & &
    -\nabla^{2}_{1}D^{\pm}(12)= \delta(1-2)
     +\Pi^{\pm}(11')D^{\pm}(1'2),
     \\[5pt]
  \label{L:+-:b}
    & &
    -\nabla^{2}_{2}D^{\pm}(12)=\delta(1-2)
     +D^{\pm}(11')\Pi^{\pm}(1'2),
   \end{eqnarray}
   \end{subequations}
 and the KB
equations for the space-time correlation functions
  \begin{subequations}
   \label{KB:l<>:ab}
   \begin{eqnarray}
     \label{KB:l<>:a}
    & &
    -\nabla^{2}_{1}D^{\mgl}(12)=
     \Pi^{+}(11')D^{\mgl}(1'2)+\Pi^{\mgl}(11')D^{-}(1'2),
     \\[5pt]
  \label{KB:<>:b}
    & &
    -\nabla^{2}_{2}D^{\mgl}(12)=
     D^{\mgl}(11')\Pi^{-}(1'2)+D^{+}(11')\Pi^{\mgl}(1'2).
   \end{eqnarray}
   \end{subequations}
The analysis of the above equations proceeds exactly in parallel with that
 for the transverse field fluctuations. By going over to the Wigner representation
 (\ref{W-tr}) and keeping only first order terms in the $X$-gradients,
the sum and difference of Eqs.~(\ref{L:+-:ab}) become
   \beq
      \label{D+-:eq}
  \left({\vec{k}}^{\,2}- \Pi^{\pm}(X,k)\right)D^{\pm}(X,k)=1,
   \qquad
  \left\{\vec{k}^{\,2}- \Pi^{\pm}(X,k), D^{\pm}(X,k)  \right\}=0,
   \eeq
whence
 \beq
 \label{D+-}
 D^{\pm}(X,k)=\frac{1}{{\vec{k}}^{\,2} - \Pi^{\pm}(X,k)}.
 \eeq
In the Wigner representation, the KB equations (\ref{KB:l<>:ab}) are
manipulated to
   \begin{eqnarray}
    \label{L:Transp}
    & &
    \hspace*{-15pt}
    \left\{{\vec{k}}^{\,2} - {\rm Re}\,\Pi^{+},D^{\mgl}\right\}
    + \left\{{\rm Re}\,D^{+},\Pi^{\mgl}\right\}=
     i\left(\Pi^{>}D^{<} - \Pi^{<} D^{>}\right),
     \\[5pt]
     \label{L:M-shell}
    & &
      \hspace*{-15pt}
     \left\{
     {\rm Im}\,\Pi^{+},D^{\mgl}\right\}+
      \left\{
     {\rm Im}\,D^{+}, \Pi^{\mgl}\right\}=
     2\left( {\vec{k}}^{\,2}- {\rm Re}\,\Pi^{+}\right)
      \left( D^{\mgl} - |D^{+}|^2\,\Pi^{\mgl}\right),
   \end{eqnarray}
where the arguments $X$ and $k$ are omitted for brevity.
 Equation (\ref{L:Transp}) is derived by taking the difference of
 Eqs.~(\ref{KB:l<>:ab}) and may be regarded as the transport equation for
  longitudinal field fluctuations. It is quite similar in structure to the
  plasmon transport equation in nonrelativistic
  plasmas~\cite{Dub67}. Equation (\ref{L:M-shell}) is analogous to the mass-shell
  equation (\ref{MS}) for transverse field fluctuations.

\end{document}